\documentclass[iop,twocolappendix]{emulateapj}
\usepackage{natbib}
\usepackage{color}
\usepackage{mathptmx}
\usepackage{amsmath}
\usepackage{url}
\usepackage{multirow}
\usepackage{verbatim}
\usepackage[figuresright]{rotating}
\usepackage{afterpage}  
\usepackage[breaklinks=true]{hyperref}
\usepackage{refcount}
\def\lsim{\lower.5ex\hbox{$\; \buildrel < \over \sim \;$}}
\def\gsim{\lower.5ex\hbox{$\; \buildrel > \over \sim \;$}}
\def\msun{\,{\rm M_\odot}}
\def\zsun{\,{Z_\odot}}
\def\apj{ApJ}
\def\apjs{ApJS}
\def\araa{ARAA}
\def\mnras{MNRAS}

\def\aj{AJ}
\def\apjl{ApJL}

\def\aap{Astronomy \& Astrophysics}

\def\pasa{PASA}

\begin{document}

\shorttitle{{\it AGORA} Comparison. II: Isolated Disk}
\shortauthors{J. Kim et al. for the {\it AGORA} Collaboration}

\title{The {\it AGORA} High-Resolution Galaxy Simulations Comparison Project. II: Isolated Disk Test}

\author{Ji-hoon Kim$^{1, 2, 3, 4}$}
\author{Oscar Agertz$^{5, 6}$}
\author{Romain Teyssier$^{7}$}
\author{Michael J. Butler$^{8}$\textsuperscript{,\textdagger,\textdaggerdbl}}
\author{Daniel Ceverino$^{9}$\textsuperscript{,\textdagger,\textdaggerdbl}}
\author{Jun-Hwan Choi$^{10}$\textsuperscript{,\textdagger,\textdaggerdbl}}
\author{Robert Feldmann$^{7,11}$\textsuperscript{,\textdagger}}
\author{Ben W. Keller$^{12}$\textsuperscript{,\textdagger,\textdaggerdbl}}
\author{Alessandro Lupi$^{13}$\textsuperscript{,\textdagger,\textdaggerdbl}}
\author{Thomas Quinn$^{14}$\textsuperscript{,\textdagger,\textdaggerdbl}}
\author{Yves Revaz$^{15}$\textsuperscript{,\textdagger,\textdaggerdbl}}
\author{Spencer Wallace$^{14}$\textsuperscript{,\textdagger}}
\author{Nickolay Y. Gnedin$^{16,17,18}$\textsuperscript{,\textdaggerdbl}}
\author{Samuel N. Leitner$^{19}$\textsuperscript{,\textdaggerdbl}}
\author{Sijing Shen$^{20}$\textsuperscript{,\textdaggerdbl}}
\author{Britton D. Smith$^{21}$\textsuperscript{,\textdaggerdbl}}
\author{Robert Thompson$^{22}$\textsuperscript{,\textdaggerdbl}}
\author{Matthew J. Turk$^{23}$\textsuperscript{,\textdaggerdbl}} 
\author{Tom Abel$^{1, 2}$}
\author{Kenza S. Arraki$^{24}$}
\author{Samantha M. Benincasa$^{12}$}
\author{Sukanya Chakrabarti$^{25}$}
\author{Colin DeGraf$^{20}$}
\author{Avishai Dekel$^{26}$}
\author{Nathan J. Goldbaum$^{22}$} 
\author{Philip F. Hopkins$^{3}$}
\author{Cameron B. Hummels$^{3}$}
\author{Anatoly Klypin$^{24}$}
\author{Hui Li$^{27}$}
\author{Piero Madau$^{28, 13}$}
\author{Nir Mandelker$^{29, 26}$}
\author{Lucio Mayer$^{7}$} 
\author{Kentaro Nagamine$^{30, 31}$}
\author{Sarah Nickerson$^{7}$} 
\author{Brian W. O'Shea$^{32}$}
\author{Joel R. Primack$^{33}$}
\author{Santi Roca-F\`{a}brega$^{26}$}
\author{Vadim Semenov$^{17}$}
\author{Ikkoh Shimizu$^{30}$}
\author{Christine M. Simpson$^{34}$}
\author{Keita Todoroki$^{35}$}
\author{James W. Wadsley$^{12}$}
\author{John H. Wise$^{36}$ 
for the {\it AGORA} Collaboration$^{37}$}

\affil{$^{1}$Kavli Institute for Particle Astrophysics and Cosmology, SLAC National Accelerator Laboratory, Menlo Park, CA 94025, USA}
\affil{$^{2}$Department of Physics, Stanford University, Stanford, CA 94305, USA}
\affil{$^{3}$Department of Astronomy, California Institute of Technology, Pasadena, CA 91125, USA}
\affil{$^{4}$Einstein Fellow, \url{me@jihoonkim.org}}
\affil{$^{5}$Department of Physics, University of Surrey, Guildford, Surrey, GU2 7XH, United Kingdom}
\affil{$^{6}$Lund Observatory, Department of Astronomy and Theoretical Physics, Lund University, SE-22100 Lund, Sweden}
\affil{$^{7}$Centre for Theoretical Astrophysics and Cosmology, Institute for Computational Science, University of Zurich, Zurich, 8057, Switzerland} 
\affil{$^{8}$Max-Planck-Institut f\"{u}r Astronomie, D-69117 Heidelberg, Germany}
\affil{$^{9}$Zentrum f\"{u}r Astronomie der Universit\"{a}t Heidelberg, Institut f\"{u}r Theoretische Astrophysik, 69120 Heidelberg, Germany}
\affil{$^{10}$Department of Astronomy, University of Texas, Austin, TX 78712 , USA}
\affil{$^{11}$Department of Astronomy, University of California at Berkeley, Berkeley, CA 94720, USA}
\affil{$^{12}$Department of Physics and Astronomy, McMaster University, Hamilton, ON L8S 4M1, Canada}
\affil{$^{13}$Institut d'Astrophysique de Paris, Sorbonne Universites, UPMC Univ Paris 6 et CNRS, 75014 Paris, France}
\affil{$^{14}$Department of Astronomy, University of Washington, Seattle, WA 98195, USA}
\affil{$^{15}$Institute of Physics, Laboratoire d'Astrophysique, \'{E}cole Polytechnique F\'{e}d\'{e}rale de Lausanne, CH-1015 Lausanne, Switzerland}
\affil{$^{16}$Particle Astrophysics Center, Fermi National Accelerator Laboratory, Batavia, IL 60510, USA} 
\affil{$^{17}$Department of Astronomy and Astrophysics, University of Chicago, Chicago, IL 60637, USA}
\affil{$^{18}$Kavli Institute for Cosmological Physics, University of Chicago, Chicago, IL 60637, USA}
\affil{$^{19}$Department of Astronomy, University of Maryland, College Park, MD 20742, USA}
\affil{$^{20}$Kavli Institute for Cosmology, University of Cambridge, Cambridge, CB3 0HA, United Kingdom}
\affil{$^{21}$Institute for Astronomy, University of Edinburgh, Royal Observatory, Edinburgh, EH9 3HJ, United Kingdom}
\affil{$^{22}$National Center for Supercomputing Applications, University of Illinois, Urbana, IL 61801, USA}
\affil{$^{23}$School of Information Sciences, Department of Astronomy, University of Illinois, Urbana, IL 61801, USA}
\affil{$^{24}$Department of Astronomy, New Mexico State University, Las Cruces, NM 88001, USA}
\affil{$^{25}$School of Physics and Astronomy, Rochester Institute of Technology, Rochester, NY 14623, USA}
\affil{$^{26}$Center for Astrophysics and Planetary Science, Racah Institute of Physics, The Hebrew University, Jerusalem, 91904, Israel}
\affil{$^{27}$Department of Astronomy, University of Michigan, Ann Arbor, MI 48109, USA}
\affil{$^{28}$Department of Astronomy and Astrophysics, University of California at Santa Cruz, Santa Cruz, CA 95064, USA}
\affil{$^{29}$Department of Astronomy, Yale University, New Haven, CT 06520, USA}
\affil{$^{30}$Department of Earth and Space Science, Graduate School of Science, Osaka University, Toyonaka, Osaka, 560-0043, Japan}
\affil{$^{31}$Department of Physics and Astronomy, University of Nevada, Las Vegas, NV  89154, USA}
\affil{$^{32}$Department of Computational Mathematics, Science and Engineering, 
Department of Physics and Astronomy, National Superconducting Cyclotron Laboratory, Michigan State University, Lansing, MI 48824, USA}
\affil{$^{33}$Department of Physics, University of California at Santa Cruz, Santa Cruz, CA 95064, USA}
\affil{$^{34}$Heidelberger Institut f\"{u}r Theoretische Studien, 69118 Heidelberg, Germany}
\affil{$^{35}$Department of Physics and Astronomy, University of Kansas, Lawrence, KS 66045, USA}
\affil{$^{36}$Center for Relativistic Astrophysics, School of Physics, Georgia Institute of Technology, Atlanta, GA 30332, USA}
\affil{$^{37}$\url{http://www.AGORAsimulations.org/}}
\affil{\textsuperscript{\textdagger}These authors, in alphabetical order, contributed to the article by leading the effort within each code group to perform and analyze simulations.}
\affil{\textsuperscript{\textdaggerdbl}These authors, in alphabetical order, contributed to the article by developing {\sc Grackle} and implementing its interface with participating codes.}

\begin{abstract}
Using an isolated Milky Way-mass galaxy simulation, we compare results from 9 state-of-the-art gravito-hydrodynamics codes widely used in the numerical community. 
We utilize the infrastructure we have built for the {\it AGORA} High-resolution Galaxy Simulations Comparison Project.  
This includes the common disk initial conditions, common physics models (e.g., radiative cooling and UV background by the standardized package {\sc Grackle}) and common analysis toolkit {\tt yt}, all of which are publicly available. 
Subgrid physics models such as Jeans pressure floor, star formation, supernova feedback energy, and metal production are carefully constrained across code platforms.  
With numerical accuracy that resolves the disk scale height, we find that the codes overall agree well with one another in many dimensions including: gas and stellar surface densities, rotation curves, velocity dispersions, density and temperature distribution functions, disk vertical heights, stellar clumps, star formation rates, and Kennicutt-Schmidt relations.  
Quantities such as velocity dispersions are very robust (agreement within a few tens of percent at all radii) while measures like newly-formed stellar clump mass functions show more significant variation (difference by up to a factor of $\sim$3).  
Systematic differences exist, for example, between mesh-based and particle-based codes in the low density region, and between more diffusive and less diffusive schemes in the high density tail of the density distribution.
Yet intrinsic code differences are generally small compared to the variations in numerical implementations of the common subgrid physics such as supernova feedback.
Our experiment reassures that, if adequately designed in accordance with our proposed common parameters, results of a modern high-resolution galaxy formation simulation are more sensitive to input physics than to intrinsic differences in numerical schemes.  
\end{abstract}


\section{Introduction}\label{intro}

Decades of strenuous effort by computational astrophysicists have propelled numerical experiments to become one of the most widely used tools in theorizing how galaxies form in the Universe.  
Numerical experiments are often the only means to put our theory to a test, the result of which we can compare with observational data to validate the model's feasibility.  
Since the success of galaxy formation theory is predicated on robust numerical experiments, it is only reasonable that we apply the same scientific standard of reproducibility to galaxy formation simulations.  
In other words, it should be considered as a fundamental principle that researchers must not establish findings from a single numerical experiment as scientific knowledge.  
Only after the result is reproduced independently by other researchers and proven not to be an isolated incidence can we build any conclusive theory about how galaxies actually form in the Universe.  

However, the task of replicating galaxy simulations or, equivalently, comparing simulations between codes, has not received high priority.\footnote{Code comparisons in the astrophysical community have previously been undertaken, albeit with simplified physics in a different scale \citep[e.g.,][]{SBcluster, 2005ApJS..160....1O}, or focusing only on hydrodynamics solvers \citep[e.g.,][]{2007MNRAS.380..963A, 2008MNRAS.390.1267T}.}  
Instead, the task is considered complex and time-consuming because one needs to ensure that identical physics is used in an identical initial condition with identical runtime settings.
This is sometimes perceived as tedious and unrewarding for early-career researchers.   
In fact, the lack of reproducibility checks is not unique to the field of numerical galaxy formation \citep[e.g.,][]{aac4716, nature_survey}. 
And its cause is not simply an unwillingness of only computational astrophysicists, either \citep{10.3389/fpsyg.2015.01152}.  
Rather, addressing the system (or the lack thereof) which checks the reproducibility of simulations would require a collective action by the entire community.
It cannot be simply about asking individual researchers to release their data dumps, but it should be about building a system that incentivizes simulations published in a reproducible manner.
It should also be about assembling an infrastructure that reduces the cost of reproducibility checks, on which simulations are verified routinely and effortlessly \citep{Nosek1422, Begley02012015}.  

The {\it AGORA} High-resolution Galaxy Simulations Comparison Project ({\it Assembling Galaxies Of Resolved Anatomy}) is the collective response by the numerical galaxy formation community to such a challenge.  
Since its first meeting in 2012 at the University of California at Santa Cruz, the {\it AGORA} Collaboration has aimed to compare galaxy-scale numerical experiments on a variety of code platforms with state-of-the-art resolution.
Our shared goal is to ensure that physical assumptions are responsible for any success in galaxy formation simulations, rather than artifacts of particular implementations.
Through a multi-platform approach from the beginning, we strive to improve all our codes by ``increasing the level of realism and predictive power of galaxy simulations and the understanding of the feedback processes that regulate galaxy metabolism'' \citep{2014agora}, and by doing so to find solutions to long-standing problems in galaxy formation.  
Because the interplay between numerical resolution and subgrid modelings of stellar physics is crucial in galaxy-scale simulations, we require that simulations be designed with state-of-the-art resolution, $\lesssim$ 100 pc, which is currently allowed within realistic computational cost bounds. 

In the Project's flagship paper, \cite{2014agora}, we explained the philosophy behind the Project and detailed the publicly available Project infrastructure we have put together.  
We also described the proof-of-concept test, in which we field-tested our infrastructure with a dark matter-only cosmological zoom-in simulation, finding a robust convergence between participating codes.  
More than 140 researchers from over 60 academic institutions worldwide have since agreed to take part in the Collaboration, many of whom having been actively engaged in working groups and sub-projects.\footnote{See the Project website at \url{http://www.AGORAsimulations.org/} for more information on the Project including its membership, and its task-oriented and science-oriented working groups. \label{agora-website}}   
The cohort of numerical codes participating in the Project currently include, but are not limited to in future studies:  
the Lagrangian smoothed particle hydrodynamics codes \citep[SPH;][]{GingoldMonaghan1977, 1977AJ.....82.1013L, Monaghan1992} {\sc Changa}, {\sc Gadget}, {\sc Gasoline}, and {\sc Gear}, 
and the Eulerian adaptive mesh refinement codes \citep[AMR;][]{1984JCoPh..53..484B, BergerAMR} {\sc Art-I}, {\sc Art-II}, {\sc Enzo}, and {\sc Ramses}, 
and the mesh-free finite-volume Godunov code {\sc Gizmo} (see Section \ref{codes} for information on each code). 

\begin{table*}
\renewcommand{\arraystretch}{1.5}
\caption{Initial Condition Characteristics}
\centering
\begin{tabular}{c || c | c | c | c}
\hline\hline 
 & Dark matter halo & Stellar disk & Gas disk & Stellar bulge \\ 
\hline
Density profile & \cite{1997ApJ...490..493N} & Exponential & Exponential & \cite{1990ApJ...356..359H}\\
\hline
\multirow{2}{*}{Structural properties}  & $M_{200} = 1.074\times10^{12} \msun$, $v_{\rm c, 200} = 150 {\rm \,\,km\,s^{-1}}$, & $M_{\rm d,\star}=3.438\times 10^{10} \msun$, & $M_{\rm d,gas}=8.593\times 10^{9} \msun$, & $M_{\rm b,\star}=4.297\times 10^{9} \msun$, \\
 & $R_{200}=205.5$ kpc, $c=10$, $\lambda=0.04$  & $r_{\rm d}=3.432$ kpc, $z_{\rm d}=0.1\,r_{\rm d}$ & $f_{\rm gas}=0.2$  &   $M_{\rm b,\,\star}/M_{\rm d}=0.1$\\
\hline 
Number of particles  & $10^5$ & $10^5$ &  $10^5$ & $1.25\times10^4$ \\
\hline 
Particle mass  & $m_{\rm DM} = 1.254 \times 10^7 \msun$ & $m_{\rm \star,\, IC} = 3.437 \times 10^5 \msun$ &  ${m_{\rm gas, \,IC}} = 8.593\times10^4 \msun$ & $m_{\rm \star,\, IC} = 3.437 \times 10^5 \msun$ \\
\hline 
\end{tabular}
\label{table:IC}
\vspace*{5 mm}
\end{table*}

In this second report of our continuing endeavor, we use an isolated Milky Way-mass galaxy simulation to compare 9 widely used state-of-the-art gravito-hydrodynamics codes. 
As in all comparison studies in {\it AGORA}, the participating codes share the common initial condition (i.e., generated by {\sc Makedisk}; see Section \ref{IC}), common physics models \citep[e.g., radiative cooling and UV background provided by the standardized package {\sc Grackle}; see Section \ref{physics-sim-nosf};][]{Grackle-paper, Bryan2014, 2014agora},\footnote{The website is http://grackle.readthedocs.org/.\label{grackle-website}} and common analysis platform \citep[i.e, {\tt yt} toolkit;][]{yt}.\footnote{The website is http://yt-project.org/.\label{yt-website} }   
We adopt spatial resolution of 80 pc that resolves the scale height of the disk.  
This helps the codes to be less dependent on phenomenological prescriptions of sub-resolution processes which are inevitably introduced in low-resolution ($>$ kpc) simulations.  
As modern galaxy formation simulations with state-of-the-art resolution and physics prescriptions become more and more computationally expensive, it is timely that we compare high-resolution isolated disk simulations to check how successfully these galaxies are reproduced by their peers.\footnote{Comparisons of cosmological zoom-in simulations are also in the making to test the robustness of the code suite over 13.8 Gyr of evolution.  See the Project's flagship paper, \citep{2014agora}, for more information.} 
Readers should note that our intention is {\it not} to identify a ``correct'' or ``incorrect'' code, but to focus instead on juxtaposing the codes for physical insights and learn how much scatter one should expect among modern numerical tools in the field (see Section \ref{conclusion} for more discussion).

The remainder of this article is organized as follows.  
In Section \ref{IC} we explain the isolated disk initial condition used in the study.  
The common input physics and runtime parameters required in the participating codes are discussed in Section \ref{physics} and  \ref{params}, respectively.
Then Section \ref{codes} describes 9 hydrodynamics codes that participated in this comparison. 
Section \ref{results} presents the results of our comparison focusing on similarities and discrepancies discovered in various multi-dimensional analyses. 
Finally in Section \ref{conclusion} we summarize our findings and conclude the paper with remarks on future work.
We will also stress the importance of collaborative and reproducible research in the numerical galaxy formation community the {\it AGORA} Project strives to promote.

\section{Initial Condition}\label{IC}

In this section we describe the Milky Way-mass isolated initial condition (IC) we adopt in this study. 
While this IC is part of a set of disk ICs generated for {\it AGORA} simulations that were first introduced in Section 2.2 of the Project flagship paper \citep{2014agora}, we briefly explain its important structural properties for completeness.\footnote{The public Dropbox link is http://goo.gl/8JzbIJ.}    

\begin{figure*}
    \centering
    \includegraphics[width=1.04\textwidth]{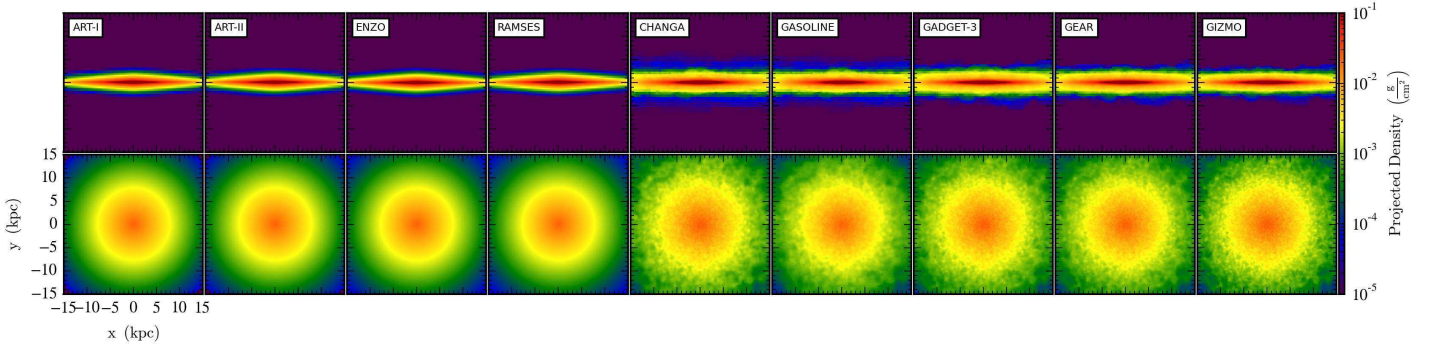}
    \caption{The 0 Myr snapshots of the isolated Milky Way-mass galaxy simulations by 9 participating codes.  Disk gas surface densities in a $30\,\,{\rm kpc}$ box, edge-on {\it (top)} and face-on {\it (bottom)}, produced with the common analysis toolkit {\tt yt}.  For visualizations of the particle-based codes hereafter (Figures \ref{fig:sigma_0_sim-nosf}-\ref{fig:sigma_500_sim-sff}, \ref{fig:temp_500_sim-nosf}-\ref{fig:temp_500_sim-sff}, \ref{fig:metal_500_sim-sff}, \ref{fig:elevation_500_sim-sff}, \ref{fig:resolution_500_sim-sff}) -- but not in any other analyses except these figures -- {\tt yt} uses an in-memory octree on which gas particles are deposited using smoothing kernels.  Comparing 0 Myr snapshots -- dumped immediately after each code reads in the IC -- is to check the exact identity of ICs interpreted by each code.  See Section \ref{results-gas-morph} for more information on this figure, and Section \ref{codes} for descriptions of participating codes in this comparison.  The full color version of this figure is available in the electronic edition.  The high-resolution versions of this figure and article are available at the Project website, http://www.AGORAsimulations.org/.
\label{fig:sigma_0_sim-nosf}}
\end{figure*}

The disk galaxy IC with properties characteristic of Milky Way-mass galaxies at redshift $z\sim1$ is generated with a privately shared version of {\sc Makedisk} \citep{2005MNRAS.361..776S}.\footnote{{\sc Makedisk} is an earlier realization of a code similar to {\sc Galic} \citep{2014MNRAS.444...62Y}.  {\sc Galic} is publicly available, and its website is http://www.h-its.org/tap-software-en/galic-code/.\label{makedisk-website}}$^{,}$\footnote{While the Milky Way's $f_{\rm gas}$ is $\sim$ 10\%, typical galaxies with the Milky Way stellar mass at $z\sim0$ have $f_{\rm gas}\sim$ 20\% \citep{2010MNRAS.403..683C}.  In this regard, one can say that we model a more typical galaxy at $z\sim0$ than the Milky Way.}
The IC has the following components (see also Table \ref{table:IC} and Figure \ref{fig:sigma_0_sim-nosf}): (1) a dark matter halo with $M_{200} = 1.074\times10^{12} \msun$, $R_{200} = 205.5$ kpc and circular velocity of $v_{\rm c, \,200} = 150 {\rm \,\,\,km\,\,s^{-1}}$ that follows the Navarro-Frenk-White \citep[NFW;][]{1997ApJ...490..493N} profile with concentration parameter $c=10$ and spin parameter $\lambda=0.04$, (2) an exponential disk with $M_{\rm d}=4.297\times 10^{10} \, \msun$, scale length $r_{\rm d}=3.432$ kpc and scale height $z_{\rm d}=0.1\,r_{\rm d}$ that is composed of $80\%$ stars and $20\%$ gas in mass (i.e., $f_{\rm gas}=M_{\rm d, \,gas}/M_{\rm d}=0.2$), (3) a stellar bulge with $M_{\rm b,\,\star}=4.297\times 10^{9} \, \msun$ that follows the Hernquist profile  \citep[$M_{\rm b,\,\star}/M_{\rm d} = 0.1$;][]{1990ApJ...356..359H}.  
The disk or bulge stars in the IC do not contribute to the feedback budget.

Among the three resolution choices provided in \cite{2014agora} here we employ a ``low-resolution'' IC which has $10^5$ particles each for the halo, the stellar disk, and the gas disk, and $1.25\times 10^4$ particles for the bulge. 
The initial gas temperature in the disk is set to $10^4$ K, not to the specific internal energy computed by {\sc Makedisk}.  
The initial metal fraction in the gas disk is 0.02041.\footnote{This fractional value 0.02041 corresponds to 1 $\zsun$ for {\sc Grackle} v2.0, but to 1.5761 $\zsun$ for {\sc Grackle} v2.1 or above.  It is because the solar metallicity unit $\zsun$ was updated from 0.02041 to 0.01295 in {\sc Grackle} v2.1.  Since cooling rates pre-tabulated by {\sc Cloudy} are at 1 $\zsun$, not at specific {\it metal fraction} value, the cooling rates in {\sc Grackle}'s equilibrium cooling mode will differ depending on which {\sc Grackle} version is adopted (see Section \ref{physics-sim-nosf} for more on {\sc Grackle}).  For example, in the current study, the codes using {\sc Grackle} v2.1 ({\sc Changa}, {\sc Gasoline}, {\sc Gadget-3}, and {\sc Gizmo}) show slightly enhanced cooling rates than the ones using {\sc Grackle} v2.0 or below ({\sc Art-I}, {\sc Art-II}, {\sc Enzo}, {\sc Ramses}, and {\sc Gear}).  Generally speaking, initial gas metallicity should be set up so that it is consistent with the chosen {\sc Grackle} version interfacing with the code.  We refer interested readers to the {\sc Grackle} v2.1 release note at https://goo.gl/BNRfwJ. \label{grackle-zsun}}
When the gas disk is initialized on mesh-based codes ({\sc Art-I}, {\sc Art-II}, {\sc Enzo}, and {\sc Ramses}), instead of using the particles provided by {\sc Makedisk}, we require that the participants use an analytic density profile of 
\begin{equation}
\rho_{\rm d, \,gas}(r,\,z)=\rho_0e^{-r/r_{\rm d}}\cdot e^{-|z|/z_{\rm d}}
\end{equation}
with $\rho_0= M_{\rm d, \, gas} / (4\pi r_{\rm d}^2 z_{\rm d})$, where $r$ is the cylindrical radius and  $z$ is the vertical height from the disk plane. 
To set up a disk in a centrifugal equilibrium we also ask that the participants utilize the rotational velocity profile binned from an actual gas particle distribution within $|z|<z_{\rm d}$.\footnote{This {\it actual} initial velocity profile, provided in {\tt vcirc\_SPH.dat} in our public Dropbox link, is different from the file {\tt vcirc.dat} produced by {\sc Makedisk} itself.  The difference is $\sim 5\%$ in the central few kpc.}
In mesh-based codes, we additionally include a uniformly low density gas halo with $n_{\rm H}=10^{-6}\,{\rm  cm}^{-3}$ and zero initial velocity, since they cannot have cells with zero density.  
The halo is initially set to $10^6$ K and zero metallicity.  
Note that this gaseous halo does not exist in particle-based codes (SPH codes or {\sc Gizmo}).  

It is worth to note one point about the common disk IC adopted here.  
As readers may find in Section \ref{results-sf}, in our experiment that includes radiative cooling, star formation and feedback ({\it ``Sim-SFF''}; see Section \ref{physics}), only $\sim 10^9 \msun$ additional  stars form in 500 Myr in all codes on average.  
When compared with the stellar and gas components present in the IC, this means only a $\sim 3\%$ increase in stellar mass, and a $\sim 12\%$ decrease in gas mass. 
As the discussion in Section \ref{results} should make clear, given this relatively small change in stellar mass, it is expected that the stellar feedback in our experiment will be very inefficient.

\vspace{3mm}

\section{Common Physics: Sim-noSF and Sim-SFF}\label{physics}

We now describe the common physics employed in our experiment including equilibrium gas cooling, metagalactic UV background, star formation, and energy and metal yields by supernovae.  
Note that the common physics adopted here is a variation of the common physics model recommended in all {\it AGORA} simulations by default; see Section 3 of the Project flagship paper \citep{2014agora}.  
For the present study all participating code groups are asked to run two simulations starting from the identical IC: (1) {\it ``Sim-noSF''} with radiative gas cooling but without star formation or feedback, and (2) {\it ``Sim-SFF''} with radiative cooling, star formation and feedback.  
In Section \ref{physics-sim-nosf}, we first list the gas physics that are common in both {\it Sim-noSF} and {\it Sim-SFF}. 
Then in Section \ref{physics-sim-sff}, the subgrid prescriptions of stellar physics for {\it Sim-SFF} are explained. 

\subsection{Gas Physics: Radiative Cooling, UV Background, and Pressure Floor}\label{physics-sim-nosf}

The rate at which the gas in our galaxy radiatively cools is determined by {\it AGORA}'s standard chemistry and cooling library {\sc Grackle} \citep{Grackle-paper, Bryan2014, 2014agora}.\footnotemark[\getrefnumber{grackle-website}]
For this study, the {\it equilibrium} cooling version of {\sc Grackle} is interfaced with each participating code, either via {\sc Grackle}'s original interface or via N. Gnedin's auxiliary API.\footnote{The website is https://bitbucket.org/gnedin/agora\_api/.}
In the chosen equilibrium cooling mode, {\sc Grackle} follows tabulated cooling rates pre-computed by the photoionization code {\sc Cloudy} \citep{2013RMxAA..49..137F}.\footnote{The website is http://www.nublado.org/.}
The pre-computed look-up table also includes metal cooling rates for solar abundances, 1 $\zsun$, as a function of gas number density and temperature.  
These metal cooling rates are then scaled linearly with metallicity which is followed in our simulations as a separate passive scalar.\footnote{See, however, footnote \ref{grackle-zsun} on how a different version of {\sc Grackle} may affect the cooling rates for the gas with the same {\it metal fraction} (but not the same {\it metallicity} interpreted by {\sc Cloudy}). \label{grackle-zsun-2}} 
We also adopt metagalactic UV background radiation at $z=0$ by \cite{HaardtMadau12} provided by {\sc Grackle}.  
For the difference between the chosen UV background model and previous calculations such as \cite{Haardt96} or \cite{Faucher09}, we refer the readers to Section 3.3 of \cite{2014agora}.  

Lastly, a non-thermal Jeans pressure floor is applied to stabilize the scales of the smoothing length (particle-based codes) or the finest cell (mesh-based codes) against unphysical collapse and to avoid artificial fragmentation due to unresolved pressure gradient \citep{1997ApJ...489L.179T, 2008ApJ...680.1083R}.  
In practice, it is achieved by enforcing that the local Jeans length $\lambda_{\rm Jeans}$ be sufficiently resolved with the finest resolution elements at all times.  
That is, 
\begin{equation}
\lambda_{\rm \,Jeans} = N_{\rm Jeans}\Delta x 
\label{eq:truelove}
\end{equation} 
where $\Delta x=80\,\,{\rm pc}$ is the adopted spatial resolution (finest cell size or softening length; see Section \ref{resolution}) and $N_{\rm Jeans} = 4$ is the Jeans number adopted from \cite{1997ApJ...489L.179T}.  
This gives the required pressure floor value as 
\begin{equation}
P_{\rm \,Jeans} = {1  \over \gamma \pi} N_{\rm Jeans}^2 G \rho_{\rm gas}^2  \Delta x^2 
\label{eq:floor}
\end{equation}
where $G$ is the gravitational constant, $\gamma=5/3$ is the adiabatic index, and $\rho_{\rm gas}$ is the gas density.  
Note that $N_{\rm Jeans}$ is not necessarily equal to the parameter controlling the pressure support in each code.  
For actual parameter choices for selected codes, see Appendix \ref{jeans-params}. 
For implementations using polytropes in {\sc Art-II} and {\sc Ramses}, see Sections \ref{art-ii} and \ref{ramses}, respectively.

\afterpage{ 
\clearpage
\global\pdfpageattr\expandafter{\the\pdfpageattr/Rotate 90}

\begin{sidewaysfigure}
    \includegraphics[width=1.0\textwidth]{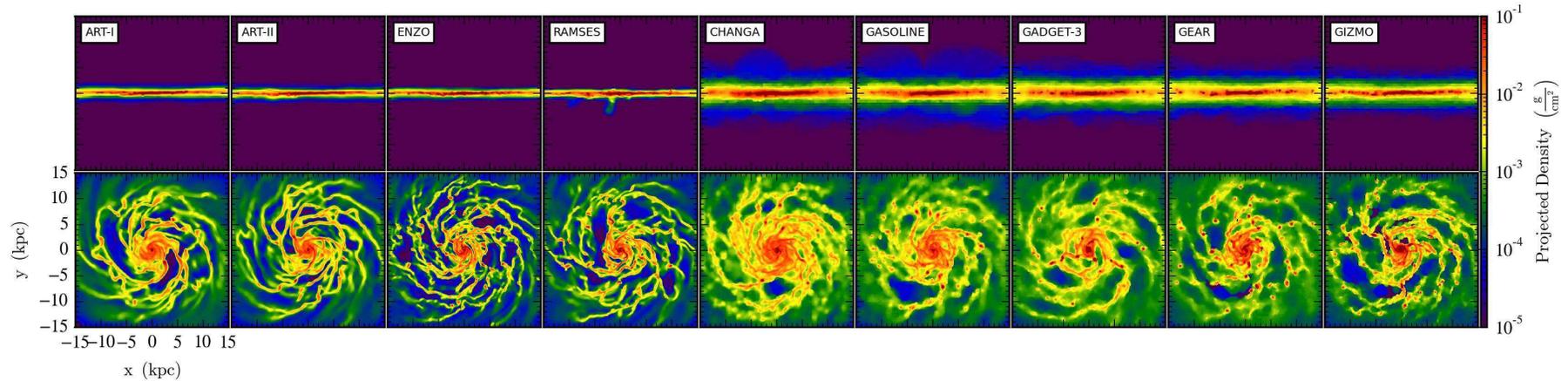}
    \caption{The 500 Myr composite of gas surface densities from {\it Sim-noSF} with radiative gas cooling but without star formation or supernova feedback.  Each frame is centered on the galactic center -- location of maximum gas density within 1 kpc from the center of gas mass.  For visualizations of the particle-based codes hereafter (Figures \ref{fig:sigma_0_sim-nosf}-\ref{fig:sigma_500_sim-sff}, \ref{fig:temp_500_sim-nosf}-\ref{fig:temp_500_sim-sff}, \ref{fig:metal_500_sim-sff}, \ref{fig:elevation_500_sim-sff}, \ref{fig:resolution_500_sim-sff}) -- but not in any other analyses except these figures -- {\tt yt} uses an in-memory octree on which gas particles are deposited using smoothing kernels.  See Section \ref{codes} for descriptions of participating codes in this comparison, and Section \ref{results-gas-morph} for a detailed explanation of this figure.  Compare with Figure \ref{fig:temp_500_sim-nosf}.  Simulations performed by:  Daniel Ceverino ({\sc Art-I}), Robert Feldmann ({\sc Art-II}), Mike Butler ({\sc Enzo}), Romain Teyssier ({\sc Ramses}), Spencer Wallace ({\sc Changa}), Ben Keller ({\sc Gasoline}), Jun-Hwan Choi ({\sc Gadget-3}), Yves Revaz ({\sc Gear}), and Alessandro Lupi ({\sc Gizmo}).  The full color version of this figure is available in the electronic edition.   The high-resolution versions of this figure and article are available at the Project website, http://www.AGORAsimulations.org/.  
	\label{fig:sigma_500_sim-nosf}}
\end{sidewaysfigure}

\begin{sidewaysfigure}
    \includegraphics[width=1.0\textwidth]{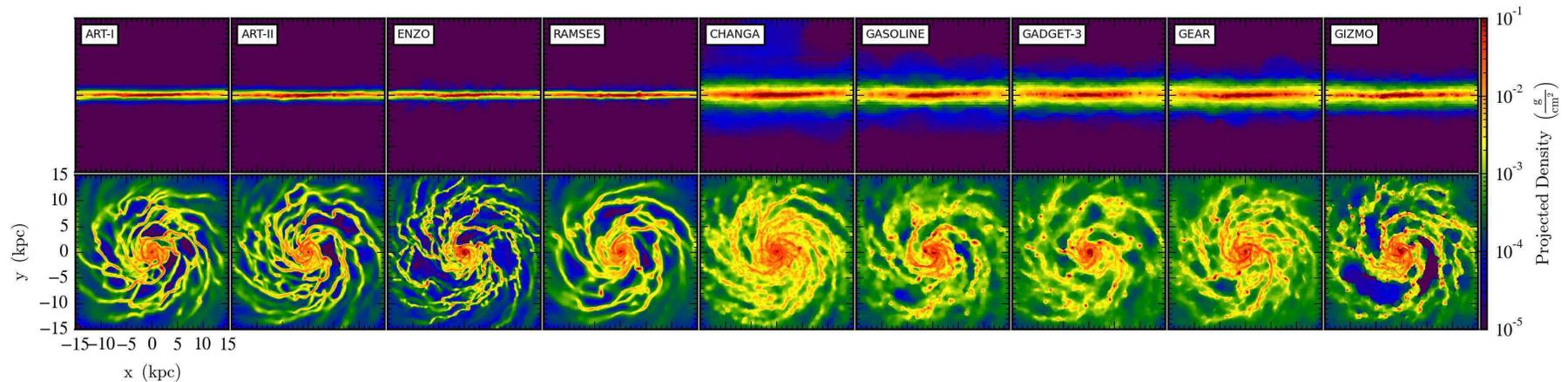}
    \caption{Same as Figure \ref{fig:sigma_500_sim-nosf} but for {\it Sim-SFF} with star formation and feedback.  See Section \ref{results-gas-morph} for a detailed explanation of this figure.  See also Section \ref{physics-sim-sff} for the common star formation prescription and the guideline for supernova feedback, and Section \ref{codes} for the exact deposit scheme of thermal feedback energy implemented in each code.  Compare with Figures \ref{fig:temp_500_sim-sff},  \ref{fig:star_with_clumps_fof_500_sim-sff}, \ref{fig:degraded_sigma_500_sim-sff}, \ref{fig:metal_500_sim-sff},  \ref{fig:elevation_500_sim-sff}, and \ref{fig:resolution_500_sim-sff}.   
	\label{fig:sigma_500_sim-sff}}
\end{sidewaysfigure}

\clearpage
}
\afterpage{\global\pdfpageattr\expandafter{\the\pdfpageattr/Rotate 0}}

\subsection{Stellar Physics: Star Formation, and Energy, Mass, Metal Yields From Core-Collapse Supernovae}\label{physics-sim-sff}

In addition to the gas physics described in the previous section, {\it Sim-SFF} incorporates subgrid models for star formation and supernova feedback.
First, a parcel of gas above the threshold $n_{\rm H, \,thres} = 10\,\, {\rm cm}^{-3} = \rho_{\rm gas,\, thres}/m_{\rm H} $ produces stars at a rate that follows the local Schmidt law as
\begin{equation}
{d\rho_\star \over dt} = {\epsilon_\star \rho_{\rm gas} \over t_{\rm ff}}
\label{eq:KS}
\end{equation}
where $\rho_\star$ is the stellar density, $t_{\rm ff}=(3\pi/(32 G\rho_{\rm gas}))^{1/2}$ is the local free-fall time, and $\epsilon_\star = 1\%$ is the star formation efficiency per free-fall time.  
We caution that the parameter $\epsilon_\star$ is not necessarily equal to the star formation efficiency parameter found in each code (e.g., for {\sc Changa} and {\sc Gasoline}, see Appendix \ref{sf-params}).
For a new star particle to spawn, it should have at least the mass of a gas particle in the IC, ${m_{\rm gas, \,IC}} = 8.593\times10^4 \msun$.
Note that $n_{\rm H, \,thres}$ adopted in this experiment is for this particular run only, and represents where the Jeans polytrope intersects with a typical $T - \rho$ equation of state in our disks (see Sections \ref{art-ii} and \ref{ramses}).\footnote{As noted in \cite{2014agora}, star formation prescription parameters such as $n_{\rm H, \,thres}$ or $\epsilon_\star$, the initial mass of star particles, and the stochasticity of star formation, are all highly dependent on numerical resolution.  An idealized test like the disk simulation presented here is essential to tune up such parameters for computationally expensive cosmological simulations.}

New star particles inject energy, mass, and metals back into the interstellar medium (ISM) through core-collapse (Type II) supernovae.
Assuming the {\it AGORA} standard \cite{Chabrier03} initial mass function (IMF) and that stars with masses between 8 and $40\,\msun$ explode as Type II supernovae, one Type II supernova occurs per every 91 $\msun$ stellar mass formed \citep[see Section 3.5 of][]{2014agora}.   
With the {\it AGORA} recommended fitting formulae Eqs. (4)-(6) of \cite{2014agora} and the assumed IMF, this single burst is found to release 2.63 $\msun$ of metals\footnote{Per unit stellar mass formed, the total fractional ejected metal masses (oxygen and iron combined) is $M_Z = 2.09 \,M_{\rm O} + 1.06 \,M_{\rm Fe} = 2.09\times0.0133 + 1.06\times0.0011 = 2.9\%$.}  
and 14.8 $\msun$ of gas (including metals). 
Per every 91 $\msun$ stellar mass, these metal and mass are instantaneously deposited into its surrounding after a delay time of 5 Myr, along with a net thermal energy of $10^{51}$ ergs.  

We note that exact deposit schemes for energy, mass, and metals are left at each participant's discretion.  
We do not intend to overly specify a single {\it common} deposit scheme which will need to be inevitably  different from one code to another (e.g., between mesh-based codes and particle-based codes), as we argued in Section 3.8 of the Project flagship paper \citep{2014agora}. 
Nevertheless, for all mesh-based codes ({\sc Art-I}, {\sc Art-II}, {\sc Enzo}, and {\sc Ramses}), the same strategy was chosen: thermal energy, mass, and metals are added to the cell where a 5 Myr old star particle sits at the time of explosion, and to this cell only.Ê
For particle-based codes ({\sc Changa}, {\sc Gasoline}, {\sc Gadget-3}, {\sc Gear}, and {\sc Gizmo}), each code's deposit scheme is discussed in detail in Section \ref{codes}. 
In future {\it AGORA} projects, we plan to calibrate different feedback schemes against observations and against one another.  
We refer the readers to Section \ref{conclusion} for more discussion on this future work.

\section{Common Runtime Parameters}\label{params}

Here we review the runtime parameters each group is required to adopt, such as gravitational softening and hydrodynamic smoothing lengths for particle-based codes and refinement thresholds for mesh-based codes.   

\subsection{Gravitational Softening Length and Finest Mesh Size}\label{resolution}

For all codes the gas mass resolution in hydrodynamics needs to be set as close as possible to ${m_{\rm gas,\,IC}} = 8.593\times10^4 \msun$. 
Assuming that we wish to resolve a self-gravitating clump with 64 of these resolution elements, the corresponding Jeans length scale becomes
\begin{equation}
\lambda_{\rm \,Jeans} = 2\left({64\,m_{\rm gas,\, IC} \over (4\pi/3)\,n_{\rm H, \,thres}}\right)^{1/3} = 348.7\,\,{\rm pc}\,,
\end{equation}
and therefore, from Eq. (\ref{eq:truelove}) we choose a spatial resolution of 80 pc.  
This value is used as the finest cell size  $\Delta x$ for mesh-based codes, and as the gravitational softening length $\epsilon_{\rm grav}$ for particle-based codes.  
For all particle-based codes taking part in the present study, gravity is softened according to the cubic spline kernel \citep[e.g., Eq. (A1) of][]{1989ApJS...70..419H}.  
For readers interested in the actual parameter choices, in Appendix \ref{sph-params} we examine the meanings of relevant parameters in different particle-based codes.  

\subsection{Minimum Hydrodynamical Smoothing Length}\label{smoothing}

For particle-based codes (including {\sc Gizmo}; see Section \ref{gizmo} and footnote \ref{gizmo-smoothing}), we require that the hydrodynamical smoothing lengths for collisional particles do not drop below 20\% of the gravitational softening lengths.
Unlike the gravitational softening kernel, exact smoothing kernel choices differ from code to code, and are detailed for each of the particle-based codes in Section \ref{codes}.
We also refer the readers interested in the actual parameter choices to Appendix \ref{sph-params} again.

\subsection{Refinement Strategy}\label{refinement}

We recommend to mesh-based code groups that a cell be split into 8 child cells once the cell contains more mass than ${m_{\rm gas,\,IC}} = 8.593\times10^4 \msun$ (1 gas particle mass in the IC of particle-based codes), or 8 collisionless particles (disk/bulge star particles in the IC with $m_{\rm \star,\, IC} = 3.437 \times 10^5 \msun$, or dark matter particles with $m_{\rm DM} = 1.254 \times 10^7 \msun$). 
This causes the grids to be refined in a fashion similar to the Lagrangian behavior of particle-based codes, and keeps the ratio of collisionless particle numbers to gas cells approximately unity on average.  
However, exact refinement strategies differ slightly from code to code, and are detailed for each of the mesh-based codes in Section \ref{codes}.\footnote{Given differences in refinement machineries among mesh-based codes it is impractical, if not impossible, to impose an exactly identical refinement criterion across all codes.  We instead adopt a trial-and-error approach within the guideline presented in Section \ref{refinement}, which resulted in all mesh-codes eventually converging to a similar overall grid structure. \label{ref-strategy}}   
We continue to refine the grids down to the resolution limit $\Delta x = $ 80 pc (see Section \ref{resolution}) where the non-thermal pressure floor kicks in (see Section \ref{physics-sim-nosf}).

\section{Participating Codes}\label{codes}

In this section we introduce the 9 gravito-hydrodynamics codes taking part in this test, focusing in particular on hydrodynamics solvers, refinement schemes for mesh-based codes ({\sc Art-I}, {\sc Art-II}, {\sc Enzo}, and {\sc Ramses}), and supernova feedback implementations for particle-based codes ({\sc Changa}, {\sc Gasoline}, {\sc Gadget-3}, {\sc Gear}, and {\sc Gizmo}).  
We leave out details that are commonly adopted across platforms such as gas cooling (Section \ref{physics-sim-nosf}) or star formation (Section \ref{physics-sim-sff}), or that were included in the {\it AGORA} flagship paper  such as gravitational dynamics \citep[Section 5 of ][]{2014agora}.  
We also point out that the codes involved in future {\it AGORA} studies are not necessarily limited to the ones described herein.

\begin{figure*}
\centering
\begin{minipage}[t]{.47\textwidth}
    \includegraphics[width=1.0\textwidth]{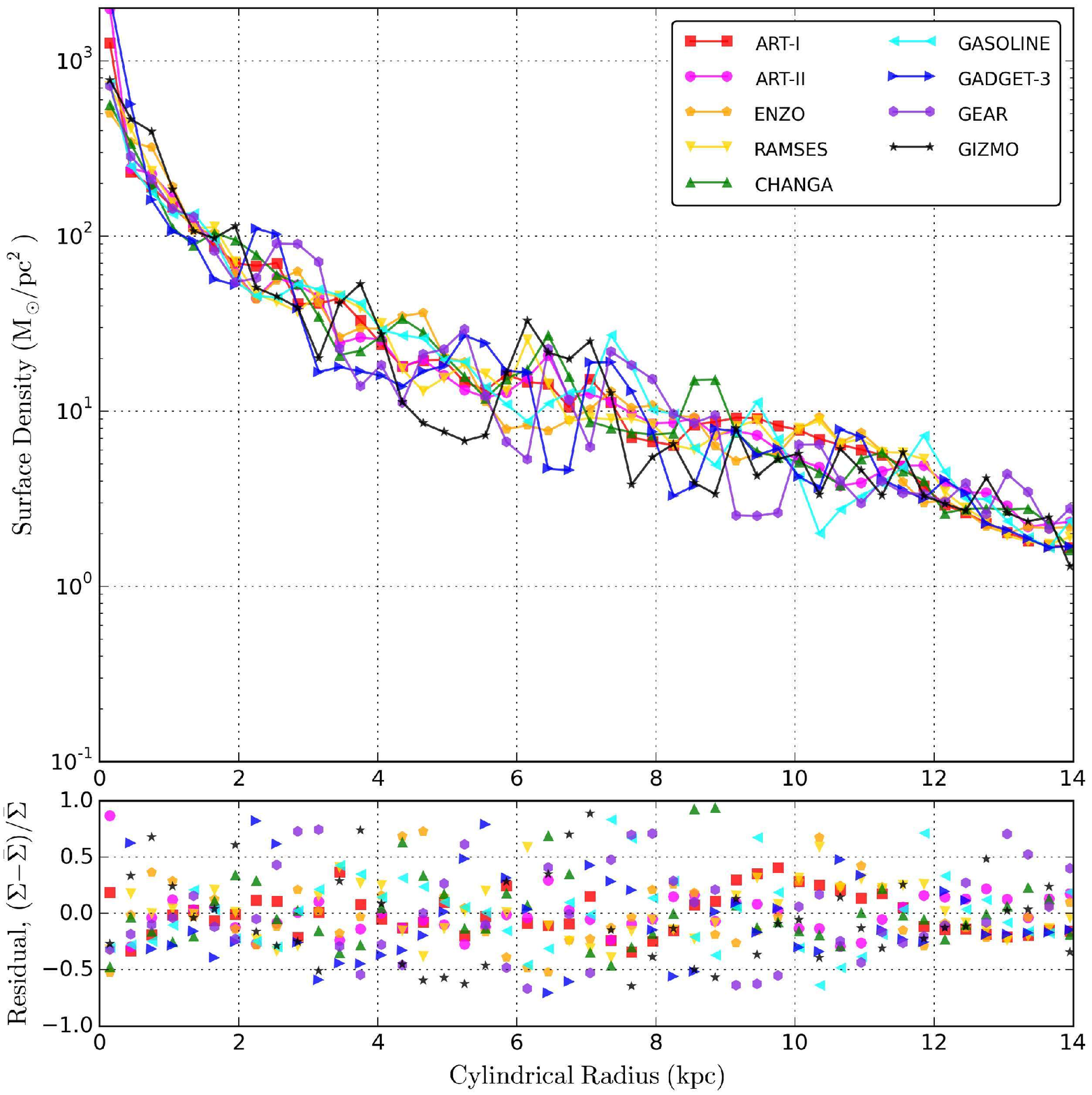}
    \caption{Cylindrically-binned gas surface density profiles at 500 Myr for {\it Sim-noSF} without star formation or feedback.  The cylindrical radius is from the galactic center -- location of maximum gas density within 1 kpc from the center of gas mass.   In all analyses for particle-based codes hereafter -- except the graphical visualizations such as Figures \ref{fig:sigma_0_sim-nosf}-\ref{fig:sigma_500_sim-sff} -- raw particle fields are used, not the interpolated or smoothed fields constructed in {\tt yt}.  Shown in the bottom panel is the fractional deviation from the mean of these profiles.  See Section \ref{results-gas-morph} for more information on this figure.  The $y$-axis range of the top panel is kept identical among Figures  \ref{fig:gas_surface_density_radial_500_sim-nosf}-\ref{fig:gas_surface_density_vertical_500_sim-sff} and \ref{fig:star_surface_density_radial_500_sim-sff} for easier comparison.  
\label{fig:gas_surface_density_radial_500_sim-nosf}}
\end{minipage}
\hfill
\begin{minipage}[t]{.47\textwidth}
    \includegraphics[width=1.0\textwidth]{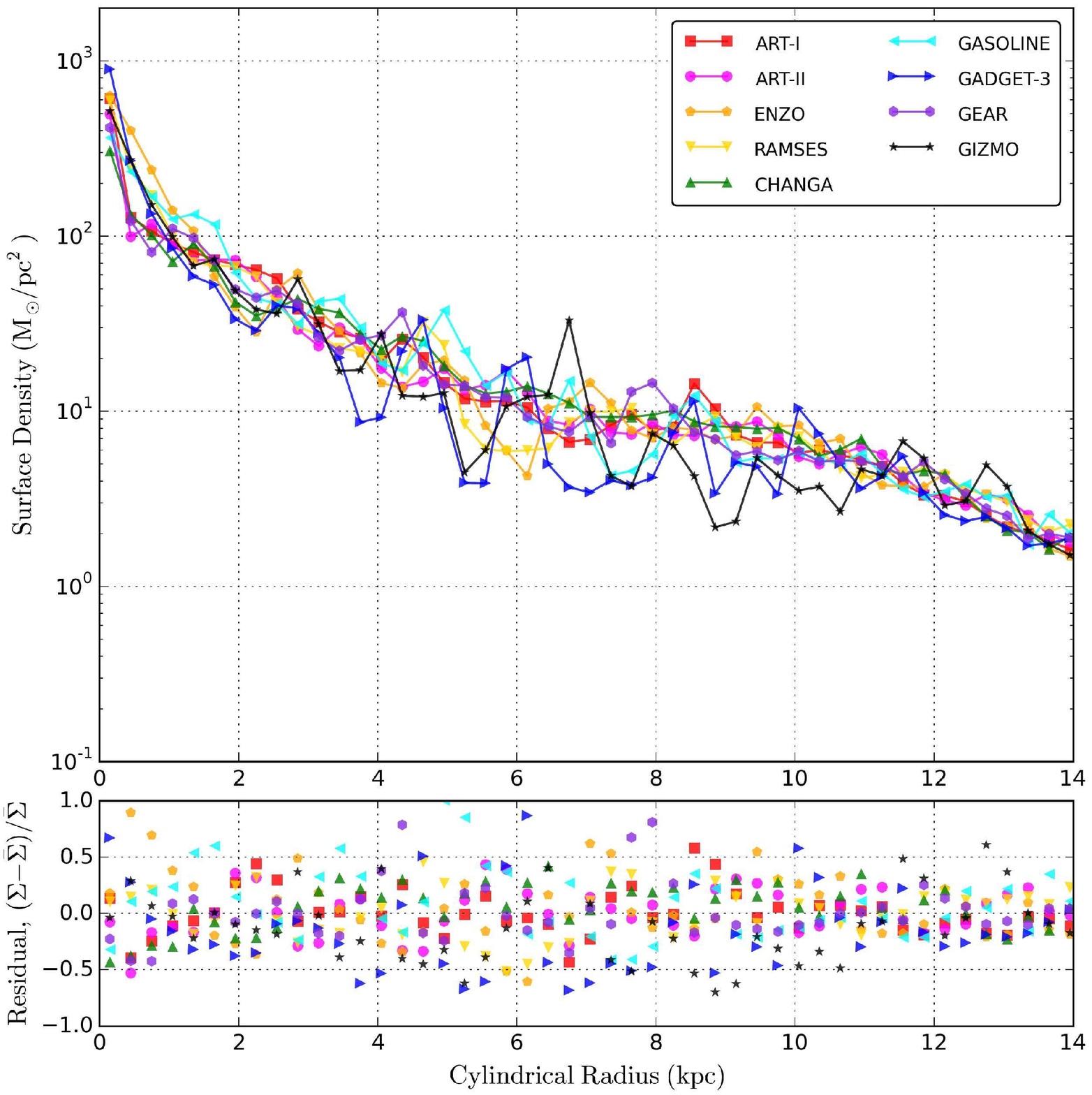}
    \caption{Same as Figure \ref{fig:gas_surface_density_radial_500_sim-nosf} but for  {\it Sim-SFF} with star formation and feedback.  Compare with Figures \ref{fig:star_surface_density_radial_500_sim-sff} and \ref{fig:sfr_surface_density_radial_500_sim-sff}.  
\label{fig:gas_surface_density_radial_500_sim-sff}}
\end{minipage}
\end{figure*}

\subsection{{\sc Art-I}}\label{art-i}

In {\sc Art-I}, differential equations of fluid dynamics are integrated using a shock-capturing Eulerian method described in \cite{Khokhlov1998}. 
It uses a 2nd-order accurate Godunov solver \citep{Godunov1959} that evaluates Eulerian fluxes by solving the Riemann problem at every cell interface \citep{Colella1985}. 
Left and right states of the Riemann problem are obtained by piecewise linear interpolation \citep{VanLeer1979}.
In contrast to other versions of {\sc Art} \citep[see Section 5.2.1 of][]{2014agora}, {\sc Art-I} with distinctive star formation and feedback recipes \citep[e.g.,][]{Ceverino2009, 2014MNRAS.442.1545C} have been developed by A. Klypin and collaborators.  

The octree-based, multi-level adaptive mesh allows users to control the grid structure at the individual cell level. 
For this comparison, the {\sc Art-I} group uses a $128^3$ root grid covering a $(1.304\,\,\,{\rm Mpc})^3$ box, then achieves an $\sim$ 80 pc cell size at maximum 7 levels of refinement. 
The mass thresholds above which a cell is adaptively refined into an oct of 8 child cells are ${m_{\rm gas,\,IC}} = 8.593 \times 10^4 \msun$  and $m_{\rm \star,\, IC} = 3.437 \times 10^5 \msun$ for gas and collisionless particles, respectively.\footnote{We note that {\sc Art-I} and {\sc Enzo} cannot refine cells by particle numbers, but only by particle masses.  By contrast, in the reported runs, {\sc Art-II} and {\sc Ramses} refine cells by particle numbers.  The refinement criteria are chosen to ensure an agreement among mesh-based codes in overall grid structures.  See also footnote \ref{ref-strategy}.  \label{art-i-enzo-ref}}
For the supernova feedback scheme to deposit the thermal energy adopted by all mesh-based codes, we refer the readers to Section \ref{physics-sim-sff}. 

\subsection{{\sc Art-II}}\label{art-ii}

{\sc Art-II} solves the gravito-hydrodynamics equations using a particle-mesh $+$ Eulerian AMR approach. 
{\sc Art-II} features MPI parallelization for distributed memory machines, flexible time-stepping hierarchy, and a variety of unique physics modules \citep[e.g.,][]{Gnedin2011, 2013ApJ...770...25A} developed by N. Gnedin, A. Kravtsov and collaborators. 

For the present study, starting from a uniform $128^3$ root grid covering $(1.311\,\,\,{\rm Mpc})^3$, cells are refined up to 7 additional levels to reach the finest size of 80 pc. 
Spherical regions of 4 (6, 10) root grid cells radius around the box center are always refined to at least 3 (2, 1) additional levels relative to the root grid.
The (de-)refinement procedure consists of three steps. 
First, cells are marked for refinement if the gas mass in the cell exceeds $0.6\,{m_{\rm gas,\,IC}} = 0.6\times{}8.593\times 10^4 \msun$, or if the cell contains 2 or more dark matter and/or star particles that were present in the IC. 
We then use a diffusion step to also mark neighboring cells for refinement and thus smooth the shape of the regions to be refined. 
By contrast, cells with gas masses below $0.2\,{m_{\rm gas,\,IC}}$ or without particles are marked for de-refinement provided they also satisfy a number of additional constraints. 
Finally, cells are refined (de-refined) by splitting them into 8 (by merging 8 child cells). 

The pressure floor implemented in {\sc Art-II} affects cells at the highest level of refinement by modifying the gas pressure values that enter the Riemann solver (i.e., not the actual pressure or temperature fields) with 
\begin{align}
P_{\rm cell} &= {\tt max}(P_{\rm\, Jeans}, \,\,P_{\rm gas}) \\
&= {\tt max}(n_{\rm H} k_{\rm B} T_{\rm\, Jeans}, \,\,P_{\rm gas})
\end{align}
where $P_{\rm cell}$ is the value entering the Riemann solver, $P_{\rm gas}$ is the gas pressure field in the simulation, $k_{\rm B}$ is the Boltzmann constant, $n_{\rm H} = \rho_{\rm gas}/m_{\rm H}$, and $T_{\rm Jeans} = T_{\rm J}(n_{\rm H}/n_{\rm H,\, J})$ with $T_{\rm J} = 1800\,\,{\rm K}$ and $n_{\rm H,\, J} = 8\,\, {\rm cm}^{-3}$.    
This polytrope choice is designed to match the common prescription Eq. (\ref{eq:floor}) with $N_{\rm Jeans} \simeq 4$. 
For the supernova feedback scheme to deposit the thermal energy adopted by all mesh-based codes, see Section \ref{physics-sim-sff}. 

\begin{figure*}
\centering
\begin{minipage}[t]{.47\textwidth}
    \includegraphics[width=1.0\textwidth]{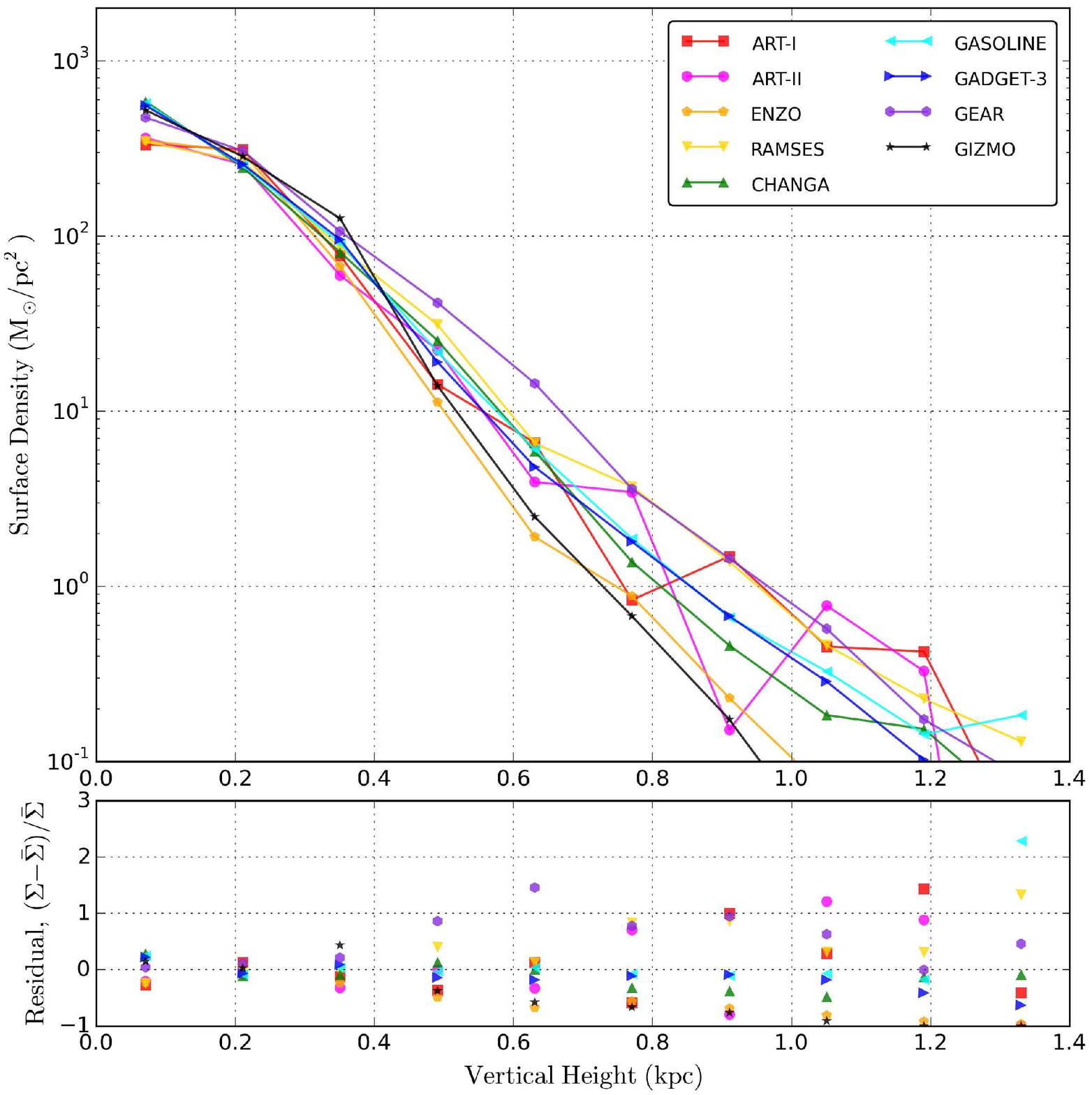}
    \caption{Vertically-binned gas surface density profiles at 500 Myr for {\it Sim-noSF} without star formation or feedback.  The height is the absolute vertical distance from the $x-y$ disk plane centered on the galactic center, $|z_{\rm i} - z_{\rm center}|$.  Shown in the bottom panel is the fractional deviation from the mean of these profiles.  The $y$-axis range of the top panel is kept identical among Figures  \ref{fig:gas_surface_density_radial_500_sim-nosf}-\ref{fig:gas_surface_density_vertical_500_sim-sff} and \ref{fig:star_surface_density_radial_500_sim-sff} for easier comparison.  
\label{fig:gas_surface_density_vertical_500_sim-nosf}}
\end{minipage}
\hfill
\begin{minipage}[t]{.47\textwidth}
    \includegraphics[width=1.0\textwidth]{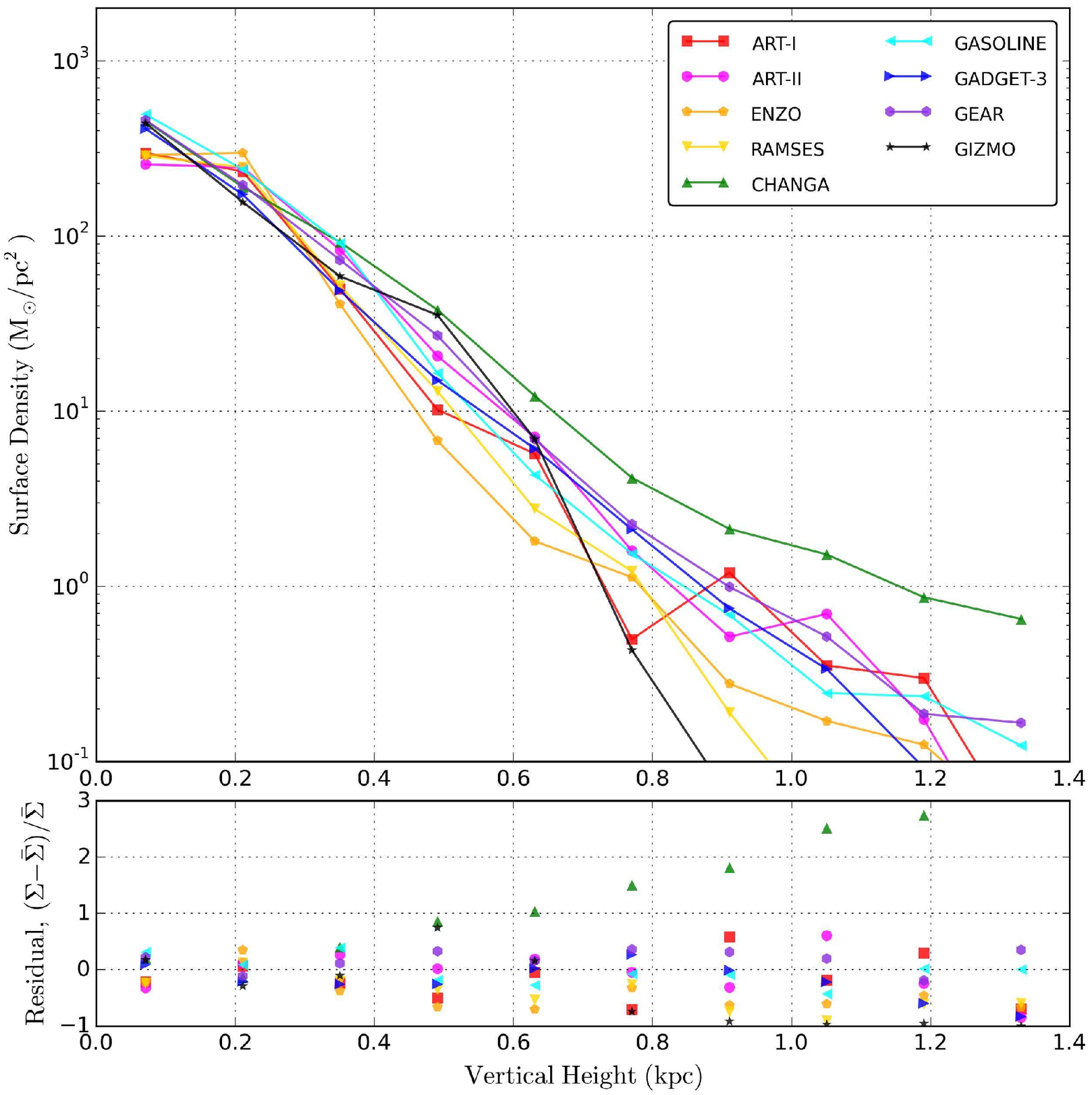}
    \caption{Same as Figure \ref{fig:gas_surface_density_vertical_500_sim-nosf} but for  {\it Sim-SFF} with star formation and feedback. 
\label{fig:gas_surface_density_vertical_500_sim-sff}}
\end{minipage}
\end{figure*}

\subsection{{\sc Enzo}}\label{enzo}

{\sc Enzo} is a block-structured adaptive mesh code, developed by an open-source, community-driven approach \citep{BryanNorman1997, OShea2004, Bryan2014}.\footnote{The website is http://enzo-project.org/.}Ê 
Among a variety of solver choices, for this comparison the 3rd-order accurate piecewise parabolic method (PPM) is selected to reconstruct the left and right states of the Godunov problem \citep{Colella84, Bryan95}, along with a Harten-Lax-van Leer with Contact (HLLC) Riemann solver \citep{1994ShWav...4...25T}.Ê 
A maximum 30\% of the required Courant-Friedrichs-Lewy (CFL) timestep is used to advance fluid elements; i.e., CFL safety factor $= 0.3$.
In addition to solving the conservation equations for mass, momentum and energy, the equation for internal energy is also solved in parallel, and the conservative or non-conservative formulation is adaptively selected based on a local estimate of the energy truncation errors.Ê
This ensures that the gas temperature remains physical, even in highly supersonic regions.

The {\sc Enzo} group uses a $64^3$ initial root grid covering a $(1.311\,\,\,{\rm Mpc})^3$ simulation box, then achieves 80 pc resolution with maximum 8 levels of refinement.  
The mass thresholds above which a cell is refined by factors of two in each axis are ${m_{\rm gas,\,IC}} = 8.593 \times 10^4 \msun$  and $8\,m_{\rm \star,\, IC} = 8\times 3.437 \times 10^5 \msun$ for gas and collisionless particles, respectively.\footnotemark[\getrefnumber{art-i-enzo-ref}]
The non-thermal pressure floor Eq. (\ref{eq:floor}) is used to modify the gas pressure inside the Riemann solver, but not to alter the actual gas energy field.  
For the supernova feedback scheme adopted by all mesh-based codes, we refer the readers to Section \ref{physics-sim-sff}. 

\subsection{{\sc Ramses}}\label{ramses}

{\sc Ramses} is an octree-based adaptive mesh code featuring an unsplit 2nd-order accurate Monotone Upstream-centered Scheme for Conservation Laws (MUSCL) Godunov scheme for the gaseous component \citep{ramses}.\footnote{The website is http://www.itp.uzh.ch/$\sim$teyssier/Site/RAMSES.html.}
For this comparison, {\sc Ramses} group uses a ideal gas equation of state with $\gamma=5/3$, along with the HLLC Riemann solver \citep{1994ShWav...4...25T} and the MinMod slope limiter \citep{1986AnRFM..18..337R}.
The CFL safety factor for controlling the time step is set to 0.5. 
The dual energy formalism adopted in {\sc Enzo} simulations (Section \ref{enzo}) is also used in {\sc Ramses} runs.

For this study, starting from a uniform $128^3$ root grid covering $(320\,\,\, {\rm kpc})^3$, cells are refined up to 5 additional levels to achieve an $\sim$ 80 pc cell size.
The refinement process works as follows. 
First, new refinement is triggered on a cell-by-cell basis if the baryonic mass (gas $+$ newly formed stars) exceeds ${m_{\rm gas,\,IC}} = 8.593 \times 10^4 \msun$, or if the number of dark matter and/or star particles that are present in the IC exceeds 8.\footnote{Readers should notice subtle differences here in refinement strategies between {\sc Ramses} and other mesh-based codes.  Newly formed stars are considered as part of the baryonic fluid, so they do not change the particle refinement based solely on collisionless particles in the IC.}  
We then mark additional cells by performing a mesh smoothing operation, expanding the initial area by one cell width in every direction.Ê
When new cells are created or old cells destroyed, density, momentum and internal energy are used as averaging and interpolating variables, thereby preventing a grid point with spurious temperature.Ê

In {\sc Ramses}, the gas pressure field includes the non-thermal pressure support term given by a temperature polytrope $T_{\rm Jeans} = \mu T_{\rm J}(n'_{\rm H}/n_{\rm H,\, J})$ with mean molecular weight $\mu$, $n'_{\rm H} = \rho_{\rm gas}X_{\rm H}/m_{\rm H}$, $T_{\rm J} = 1800\,\,{\rm K}$, $n_{\rm H,\, J} = 8\,\, {\rm cm}^{-3}$ and  $X_{\rm H} = 0.76$.\footnote{This means that in order to retrieve gas internal energy or temperature (e.g., Section \ref{results}), the pressure support term needs to be subtracted out from {\sc Ramses}'s pressure field, which is the only field being tracked.}    
As in {\sc Art-II}, this polytrope approximately matches the common pressure support prescription Eq. (\ref{eq:floor}).  
Newly created star particles in {\sc Ramses} have a fixed mass of $m_{\rm gas,\,IC}$, but they are spawned with a PoissonÊ probability distribution whose parameters are designed to mimic the local Schmidt law, Eq. (\ref{eq:KS}).
Lastly, for the common supernova feedback scheme adopted by all mesh-based codes, we refer the readers to Section \ref{physics-sim-sff}. 

\begin{figure*}
\centering
\begin{minipage}[t]{.47\textwidth}
    \includegraphics[width=1.0\textwidth]{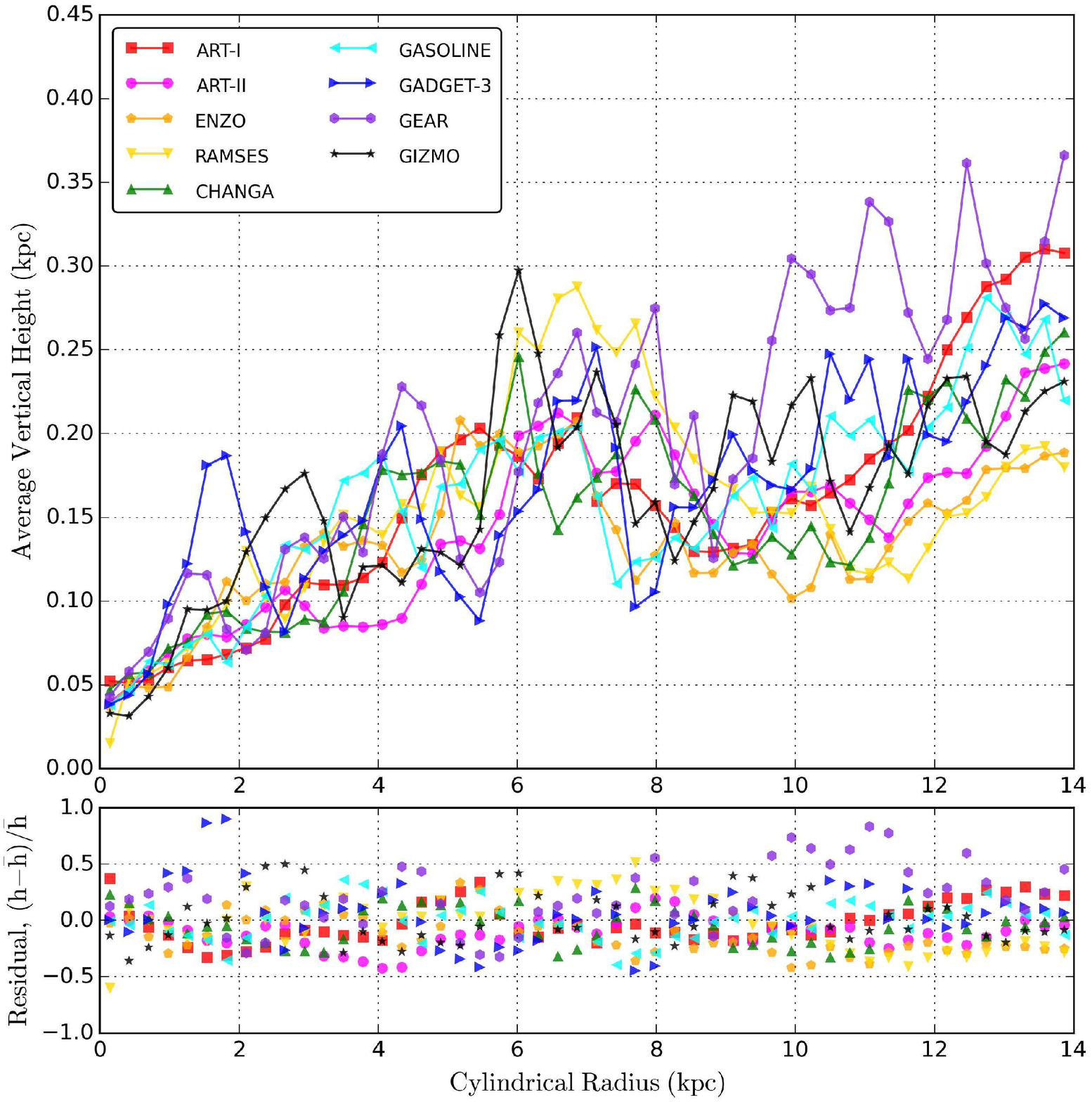}
    \caption{Cylindrically-binned, mass-weighted averages of gas vertical heights for {\it Sim-noSF} without star formation or feedback.  The height is the absolute vertical distance from the $x-y$ disk plane centered on the galactic center, $|z_{\rm i} - z_{\rm center}|$.  Shown in the bottom panel is the fractional deviation from the mean of these profiles.  See Section \ref{results-gas-morph} for an explanation on how this figure is made.  
\label{fig:rad_height_500_sim-nosf}}
\end{minipage}
\hfill
\begin{minipage}[t]{.47\textwidth}
    \includegraphics[width=1.0\textwidth]{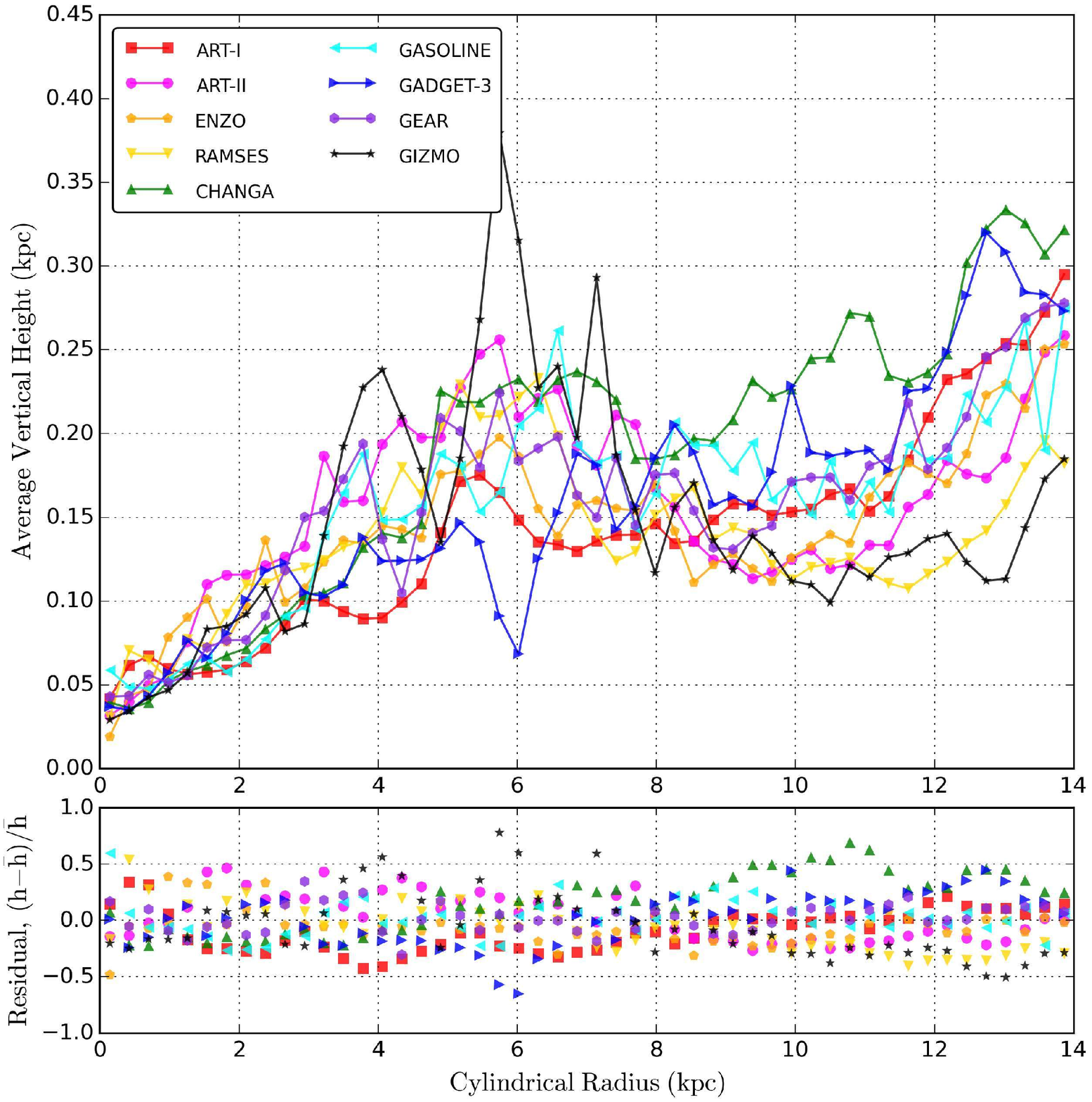}
    \caption{Same as Figure \ref{fig:rad_height_500_sim-nosf} but for  {\it Sim-SFF} with star formation and  feedback. 
\label{fig:rad_height_500_sim-sff}}
\end{minipage}
\end{figure*}

\subsection{{\sc Changa}}\label{changa}

{\sc Changa} is a reimplimentation of {\sc Gasoline} (see Section \ref{gasoline}) in the {\sc Charm++} runtime system.\footnote{The website is http://www-hpcc.astro.washington.edu/tools/changa.html.}  
{\sc Charm++} \citep{KaleKrishnan96}\footnote{The website is http://charm.cs.uiuc.edu/.} enables the overlap of computation and communication and provides adaptive load balancing infrastructure, allowing {\sc Changa} to scale to hundreds of thousands of processor cores \citep{Menon15}.
The hydrodynamics in {\sc Changa} closely follows that of {\sc Gasoline}.  
SPH forces are calculated using the method of \cite{RitchieThomas01}, and energy is diffused using the scheme of \cite{Shen10}, both of which providing a more accurate treatment of multi-phase ISM.
Timesteps are determined by the minimum of an acceleration and a CFL criterion.
Furthermore, the timesteps of neighbors are kept within a factor of 2 of each other as in \cite{Saitoh09} in order to accurately integrate highly supersonic flows.

For this work, a $k$-th nearest neighbor algorithm is used to find the $N_{\rm smooth}=$ 64 nearest neighbors which are smoothed with the Wendland C4 kernel \citep{DehnenAly12} to determine hydrodynamic properties.  
Unlike conventional versions of {\sc Changa} or {\sc Gasoline}, the supernovae thermal energy, mass, and metals are directly distributed to the 64 neighboring gas particles.\footnote{The feedback prescription used in this experiment needed a reimplementation of the feedback routine normally used in {\sc Changa} and {\sc Gasoline} \cite[e.g.,][]{Stinson06}.  In particular, in previous work, the supernovae rate determined by the stellar age and IMF is converted to an energy injection ``rate'' which is then incorporated into the thermal energy integration of neighboring gas particles.  By contrast, for this study, a supernova event occurs instantaneously, making the rate an ill-defined quantity in their existing energy integration machinery.}  
Gas particles that are neighbors of particles that will explode as a supernova in their next timestep are put on timesteps suitable for their post-supernova thermal energy, preventing them from being on a much smaller timestep required in the CFL condition.  
{\sc Grackle} cooling is implemented but it does not self-consistently account for the $PdV$ work or other external sources of energy, a requirement for {\sc Changa} and {\sc Gasoline}'s energy integration.  
Therefore, we split the energy integration into a half timestep of {\sc Grackle} cooling, then a full timestep of $PdV$ heating, and finally a second half timestep of cooling.  

\subsection{{\sc Gasoline}}\label{gasoline}

{\sc Gasoline} is a massively parallel SPH code, first described in \cite{gasoline}, that has subsequently been updated with modern SPH features.  
It contains a subgrid model for turbulent mixing of metals and energy \citep[e.g.,][]{Shen10}, a timestep limiter by \cite{Saitoh09} (see Section \ref{changa}), and a geometric density estimator for SPH force expressions \citep[see Section 2.4 of][for a latest detailed description of the code and its performance]{2014MNRAS.442.3013K}.  

For the current work, the {\sc Gasoline} group uses a Wendland C4 smoothing kernel \citep{DehnenAly12} with $N_{\rm smooth}=200$ neighbors.\footnote{Note the difference in $N_{\rm smooth}$ from {\sc Changa} in Section \ref{changa}.  \cite{DehnenAly12} showed that this kernel can use larger neighbor numbers without the pairing instability which may effectively remove resolution, and that doing this improves performance on a number of basic hydrodynamics tests.} 
The same feedback scheme as {\sc Changa}'s (Section \ref{changa}) is implemented, smoothed with the Wendland C4 kernel over 64 neighbors (not $N_{\rm smooth}= 200$) to better match the amount of mass heated by feedback events with other particle-based codes.
Gas particles that receive feedback compute their required CFL timestep at the timestep prior to receiving feedback, which helps to prevent numerical instability and overcooling.  
{\sc Grackle} cooling is implemented by applying a half timestep of cooling, then a full timestep of external $PdV$ heating, followed by a final half timestep of cooling, as in {\sc Changa} (Section \ref{changa}).  

\begin{figure*}
\centering
\begin{minipage}[t]{.47\textwidth}
    \includegraphics[width=1.0\textwidth]{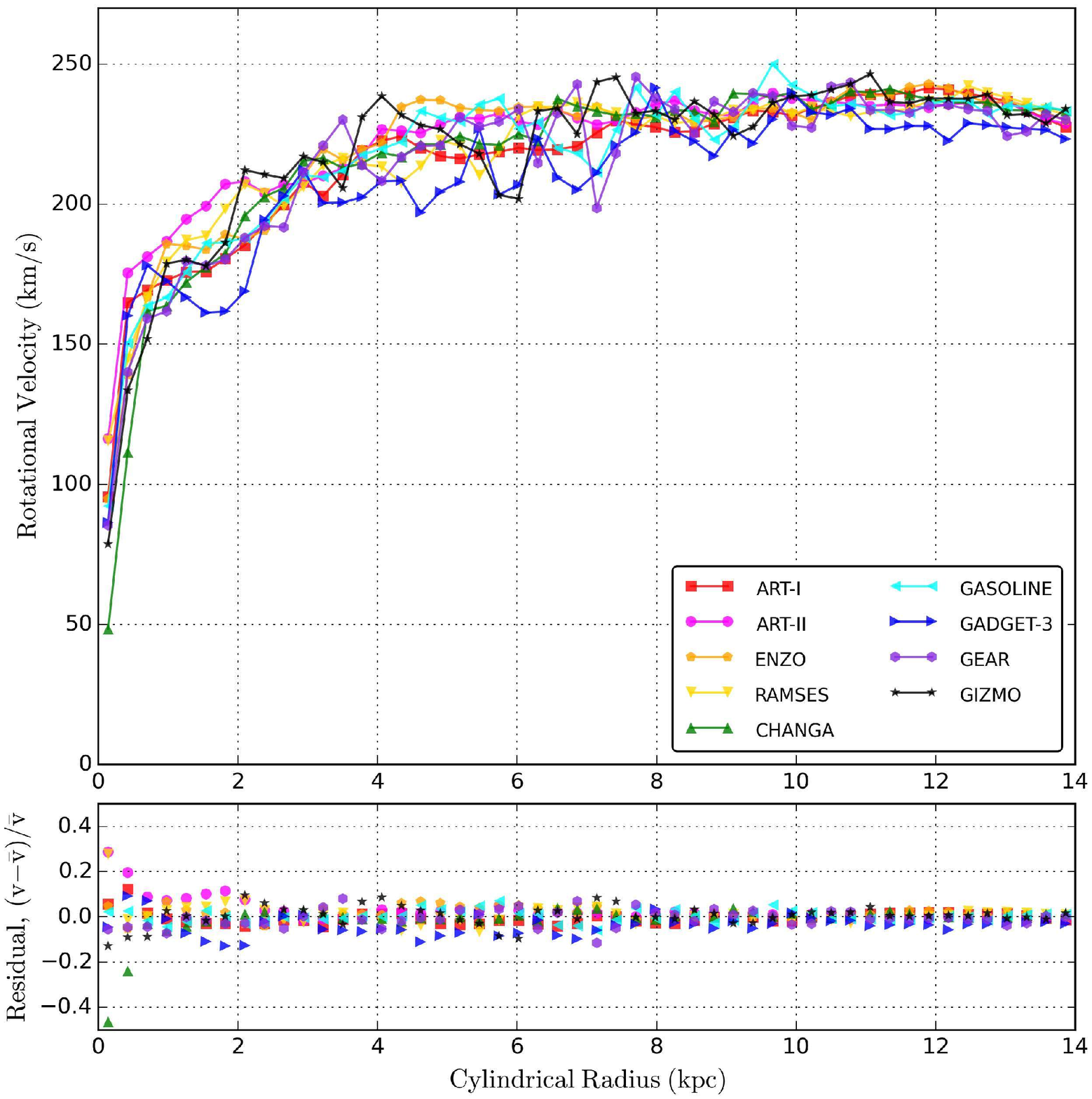}
    \caption{Gas rotation velocity curves at 500 Myr for {\it Sim-noSF} without star formation or feedback. The cylindrical radius and rotational velocity are with respect to the galactic center -- location of maximum gas density within 1 kpc from the center of gas mass.   Shown in the bottom panel is the fractional deviation from the mean of these profiles.  See Section \ref{results-gas-kin} for a detailed explanation on how this figure is made.    
\label{fig:pos_vel_500_sim-nosf}}
\end{minipage}
\hfill
\begin{minipage}[t]{.47\textwidth}
    \includegraphics[width=1.0\textwidth]{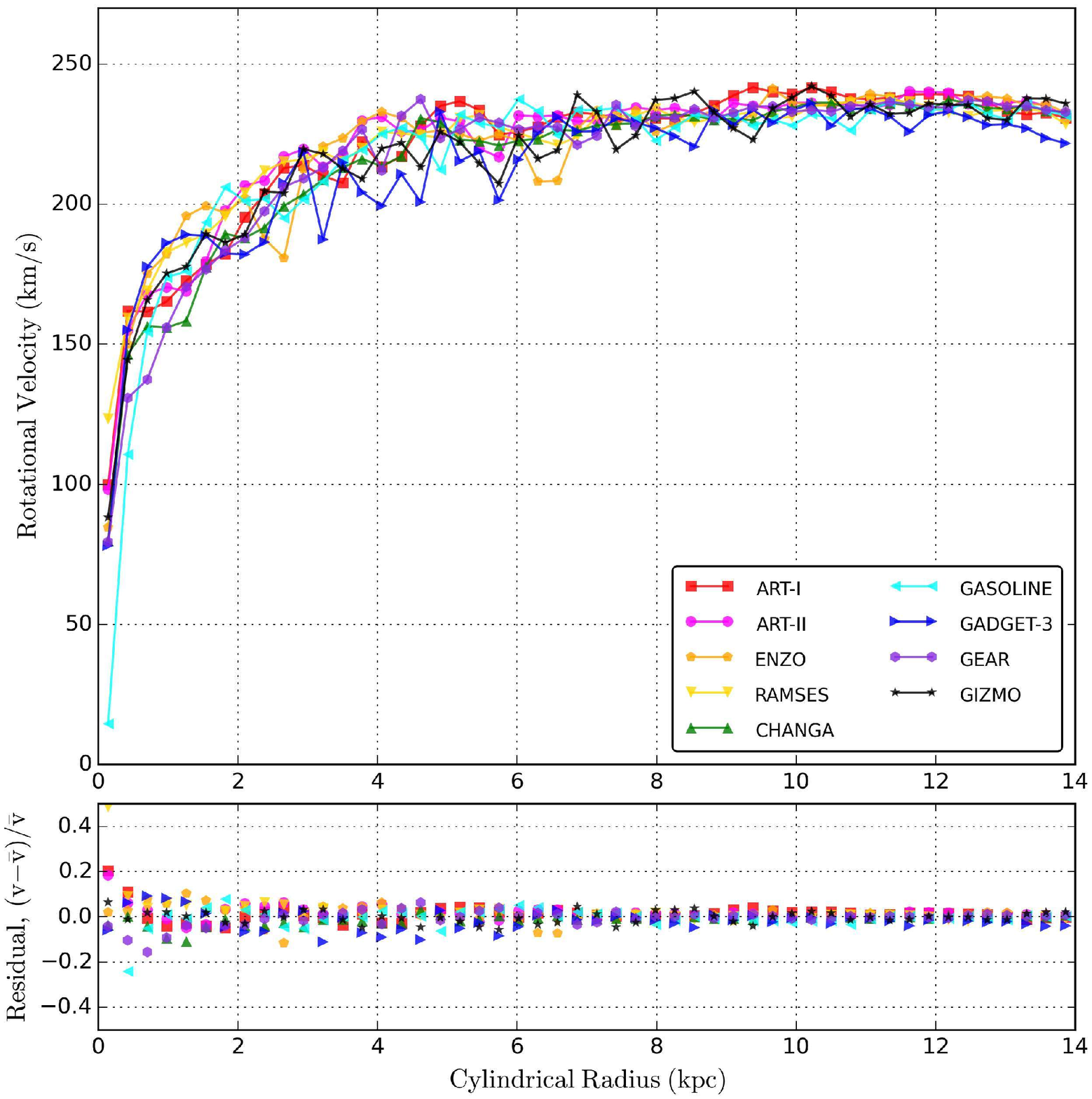}
    \caption{Same as Figure \ref{fig:pos_vel_500_sim-nosf} but for  {\it Sim-SFF} with star formation and feedback.  Compare with Figure \ref{fig:star_pos_vel_500_sim-sff}.  
\label{fig:pos_vel_500_sim-sff}}
\end{minipage}
\end{figure*}

\subsection{{\sc Gadget-3}}\label{gadget}

{\sc Gadget-3} is an updated version of {\sc Gadget-2}, a cosmological tree-particle-mesh (TPM) SPH code that was originally developed by V. Springel \citep{Springel01, gadget2}.\footnote{The website is http://www.h-its.org/tap-software-en/gadget-code/ or http://www.mpa-garching.mpg.de/gadget/.} Ê
{\sc Gadget-3} has important updates from {\sc Gadget-2}, such as domain decomposition and dynamic tree reconstruction which may slightly alter the $N$-body dynamics.  
The {\sc Gadget-3} code used in this comparison is a modified version of the original {\sc Gadget-3}  by K. Nagamine and his collaborators, which includes pressure-entropy formulation by \cite{hopkins2013}, time-dependent artificial viscosity, variable smoothing lengths, among others \citep[e.g.,][]{2012MNRAS.419.1280C, 2014ApJ...780..145T, 2016arXiv160907547A}.  

For the present study, the {\sc Gadget-3} group adopts a quintic spline smoothing kernel \citep{1996PASA...13...97M} with $N_{\rm ngb}=64$.  
The implementation of supernova feedback is based on an updated version of \cite{2014MsT..........1T} that largely follows a Sedov-Taylor blast wave method outlined in \citet[][but not their cooling shutoff model]{Stinson06, 2013MNRAS.428..129S}.  
The exact model used in the current work is fully described in \cite{2016arXiv160907547A}, but, in brief, the implementation comprises the following steps. 
Every time a star particle explodes, we compute the ``shock radius'' based on \cite{1974ApJ...188..501C} and \cite{1977ApJ...218..148M}, and then find the gas particles within the radius.  
We then inject thermal energy and metal yields into the identified gas particles within the shock radius, weighted by the SPH spline kernel.
Finally, we note that the results of this version of {\sc Gadget-3} are not representative of all the {\sc Gadget-3} codes in the community, because some of the results are strongly dependent on the detailed implementations of baryonic physics, such as star formation and feedback.

\subsection{{\sc Gear}}\label{gear}

{\sc Gear} is a self-consistent, fully parallelized, chemo-dynamical tree SPH code \citep{revaz2012} which is built on the publicly available {\sc Gadget-2} code \citep[see Section \ref{gadget};][]{gadget2}.
The simulations reported here are run with the improvements dicussed in \citet{revaz2016}, including the pressure-entropy formulation proposed by \cite{hopkins2013}, individual and adaptive time-stepping schemes \citep{durier2012}, artificial viscosity \citep{monaghan83} supplemented with the Balsara switch \citep[$\,f_{ij}$ from][]{balsara1995}, and particle-based time-dependent viscosity coefficient \citep{rosswog2000}.

For this study, the standard cubic spline smoothing kernel \citep{1985A&A...149..135M} in {\sc Gadget-2} is used with $N_{\rm ngb}=50$.
The feedback energy, mass, and metal injection into the ISM is implemented following the standard SPH scheme.  
The implementation comprises the following steps.   
Every time a star particle explodes, we first find the nearest gas particles, according to the {\it weighted} number of neighbors as defined in \cite{springel2002}.
A {\it desired} number of neighbors $N_{\rm ngb}=50$ is used. 
Then we inject thermal energy and yields into the neighboring gas particles, weighted by the SPH spline kernel.
{\sc Grackle} cooling is performed after the kick step, once gas particles have eventually received supernova feedback energy and once the size of the next timestep is known. 
The adiabatic cooling/heating is first applied, and then the radiative one provided by {\sc Grackle}.

\begin{figure*}
\centering
\begin{minipage}[t]{.47\textwidth}
    \includegraphics[width=1.0\textwidth]{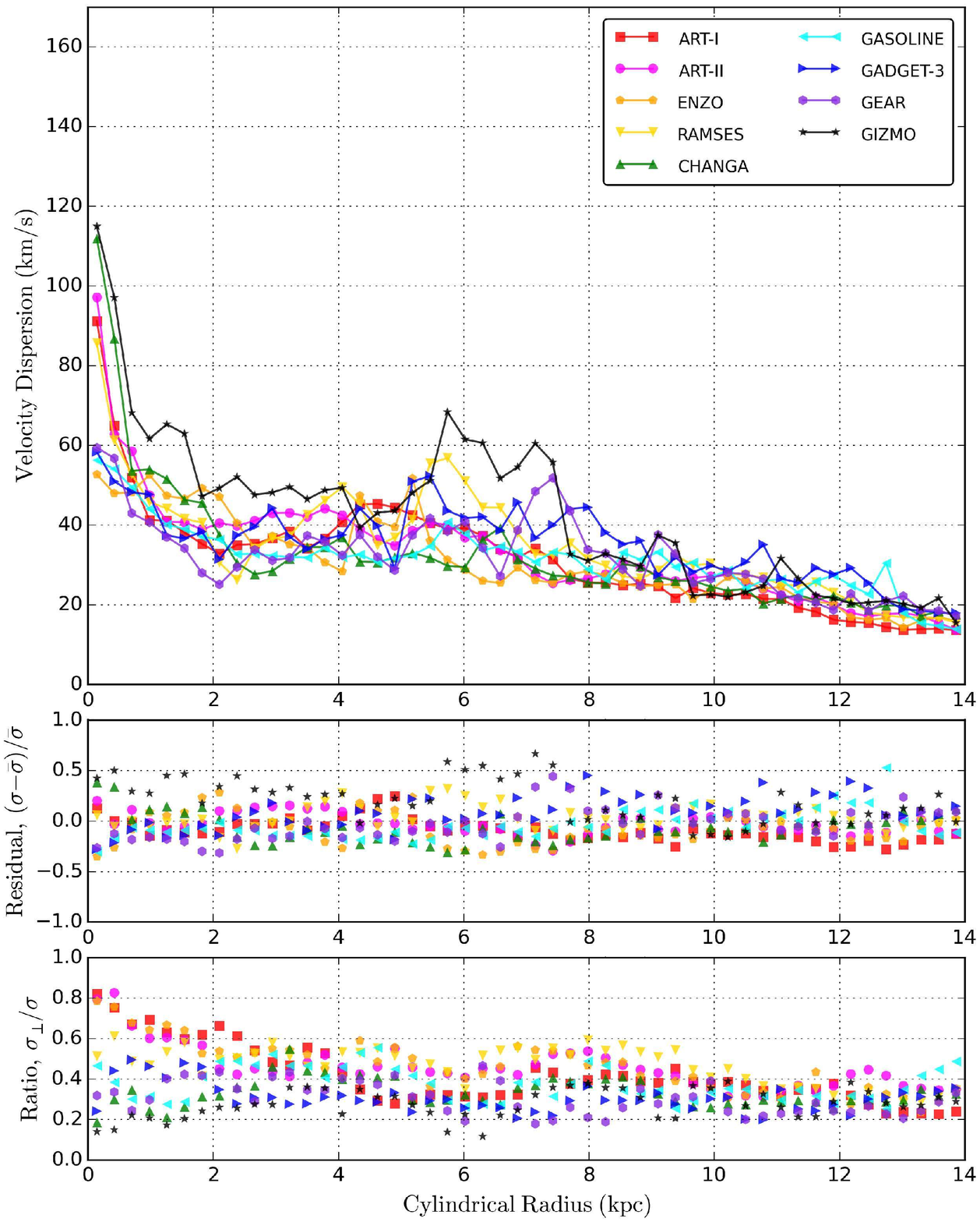}
    \caption{Gas velocity dispersion curves at 500 Myr for {\it Sim-noSF} without star formation or feedback.  The velocity dispersion is the square root of mass-weighted averages of $(v_{\rm i} - v_{\rm rot}(r))^2$.   Shown in the middle panel is the fractional deviation from the mean of these profiles.  In the bottom panel we plot the ratio of {\it vertical} velocity dispersion ($z$-direction) to {\it total} velocity dispersion.  See Section \ref{results-gas-kin} for a detailed explanation on how this figure is made.  The $y$-axis range of the top panel is kept identical among Figures  \ref{fig:pos_disp_500_sim-nosf}-\ref{fig:pos_disp_500_sim-sff} and \ref{fig:star_pos_disp_500_sim-sff} for easier comparison.  
\label{fig:pos_disp_500_sim-nosf}}
\end{minipage}
\hfill
\begin{minipage}[t]{.47\textwidth}
    \includegraphics[width=1.0\textwidth]{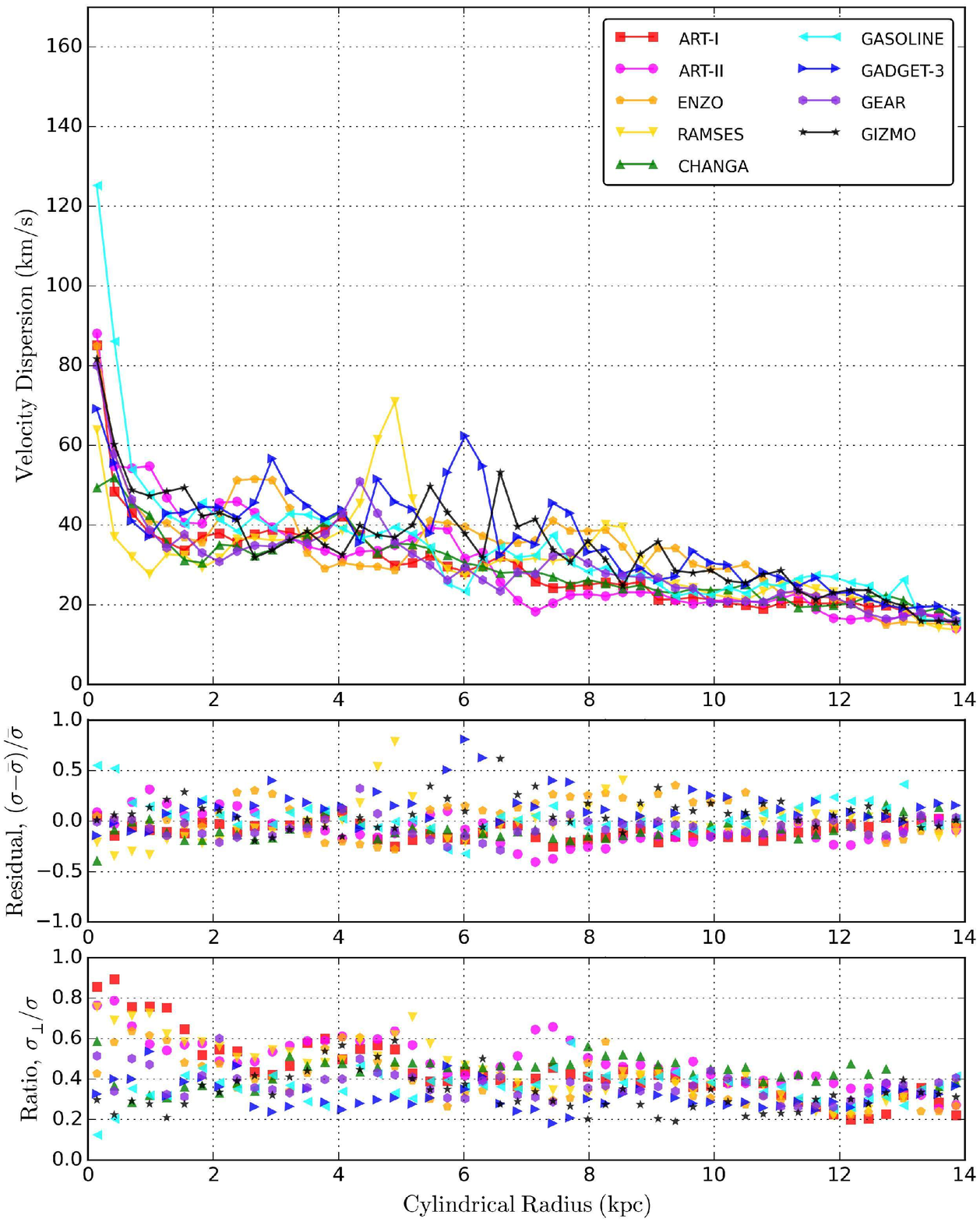}
    \caption{Same as Figure \ref{fig:pos_disp_500_sim-nosf} but for  {\it Sim-SFF} with star formation and  feedback.  Compare with Figure \ref{fig:star_pos_disp_500_sim-sff}.  
\label{fig:pos_disp_500_sim-sff}}
\end{minipage}
\end{figure*}

\subsection{{\sc Gizmo}}\label{gizmo}

{\sc Gizmo} \citep{hopkins2015} is a new mesh-free Godunov code based on discrete tracers, aimed at capturing the advantages of both Lagrangian and Eulerian techniques.\footnote{The website is http://www.tapir.caltech.edu/$\sim$phopkins/Site/GIZMO.html.} 
The numerical scheme implemented in {\sc Gizmo}, initially proposed by \cite{Lanson2008}, follows the implementation of \cite{2011MNRAS.414..129G} and relies on the discretization of the Euler equations of hydrodynamics among a set of discrete tracers.Ê
Unlike in the moving mesh technique, where the volume is partitioned by a Voronoi tessellation, {\sc Gizmo} distributes the volume fraction assigned to the tracers through a kernel function. 
For the current work, {\sc Gadget}'s standard cubic spline smoothing kernel is used with $N_{\rm ngb}=32$.
Note, unlike SPH codes, these tracers only represent unstructured cells, sharing an ``effective face'' with the neighboring cells.\footnote{It is worth noting that the kernel size in {\sc Gizmo} (what is called the ``smoothing length'' in SPH) does not play any role in the dynamics, but is simply related to each cell's ``effective volume.''  For $N_{\rm ngb}=32$, a radius enclosing approximately this many neighbors is used to estimate the effective volume per particle at 2nd-order accuracy, with most of the ``effective volume'' coming from the region within a single inter-particle separation length around the tracer. \label{gizmo-smoothing}} 
The Riemann problem is then solved across these faces using a Godunov method as in mesh-based codes, to accurately resolve shocks without artificial dissipation terms. 
Unlike mesh-based codes, these cells are not fixed in space and time, resulting in the scheme's Lagrangian behavior with intrinsically adaptive resolution.
When the time evolution of the common face between two cells is considered, we use the 2nd-order accurate Meshless Finite Mass method \citep[MFM; described in][]{hopkins2015}. 
{\sc Gizmo}'s time-stepping scheme is fully adaptive, and closely follows {\sc Gadget-3} or {\sc Arepo} \citep{arepo}. 
It also includes a timestep limiter by \cite{Saitoh09} (see Section \ref{changa}).

{\sc Gizmo}'s gravity solver is based on the tree algorithm inherited from {\sc Gadget-3}, itself descending from {\sc Gadget-2} \citep[see Section \ref{gadget};][]{gadget2}. 
Gravitational softenings in {\sc Gizmo} can be fixed or fully adaptive, but in the reported runs fixed softening length is used matching SPH codes.
To model supernova feedback, the {\sc Gizmo} simulations shown here adopts a similar feedback strategy used in {\sc Gear} (Section \ref{gear}).
That is, energy, mass and metals are distributed among the neighboring gas particles/cells in a kernel-weighted fashion, but with $N_{\rm ngb}=32$.
For star particles, timesteps are constrained to prevent supernovae from exploding in the timestep when the stars formed.
Lastly, we caution that different feedback implementations using {\sc Gizmo} in the literature adopt different algorithms to distribute supernova feedback energy \citep[e.g.,][by the {\it FIRE} Collaboration]{2014MNRAS.445..581H}.

\begin{figure*}
    \centering
    \includegraphics[width=1.04\textwidth]{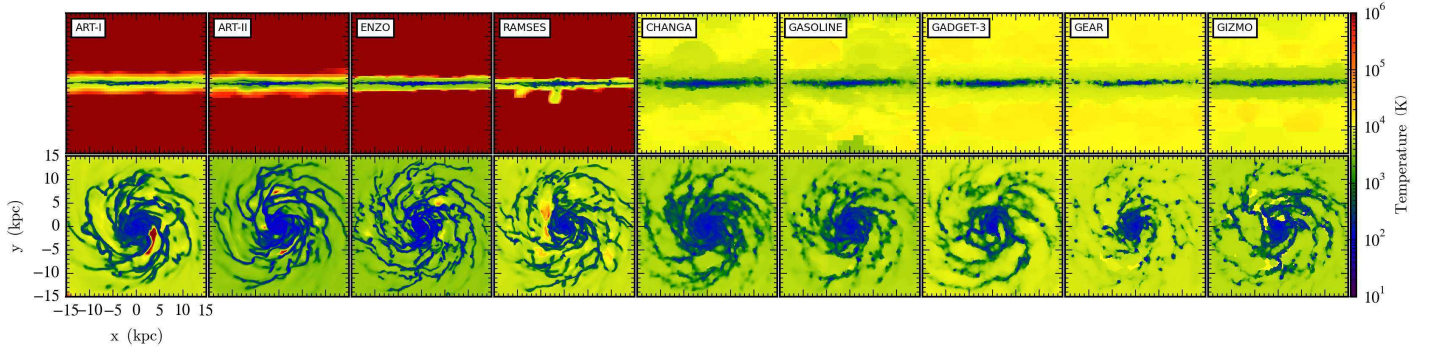}
    \caption{The 500 Myr composite of density-square-weighted gas temperature projections, edge-on {\it (top)} and face-on {\it (bottom)}, for {\it Sim-noSF} with radiative gas cooling but without star formation or supernova feedback.  See Section \ref{codes} for descriptions of participating codes in this comparison, and Section \ref{results-gas-thermal} for a detailed explanation of this figure.  Compare with Figure \ref{fig:sigma_500_sim-nosf}.   The full color version of this figure is available in the electronic edition.  The high-resolution versions of this figure and article are available at the Project website, http://www.AGORAsimulations.org/.
\label{fig:temp_500_sim-nosf}}
\end{figure*}

\begin{figure*}
    \centering
    \includegraphics[width=1.04\textwidth]{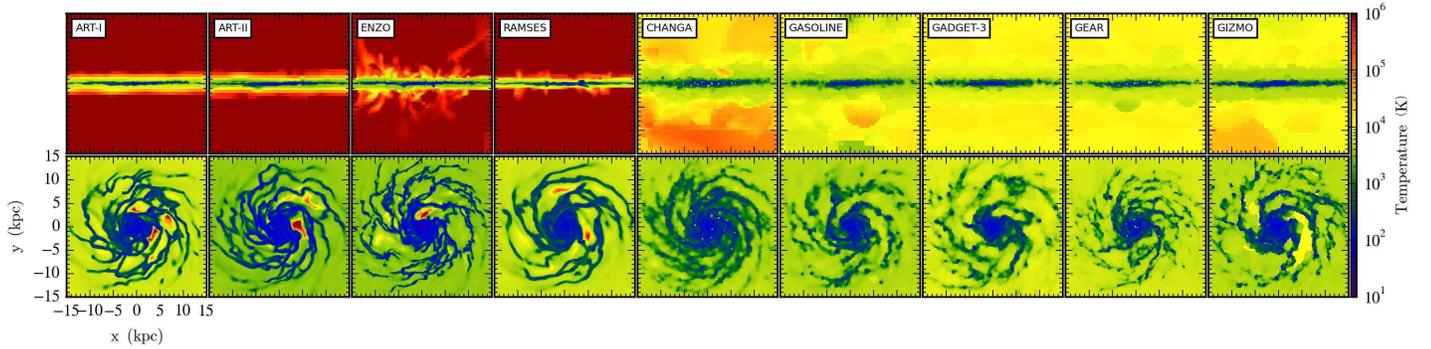}
    \caption{Same as Figure \ref{fig:temp_500_sim-nosf} but for  {\it Sim-SFF} with star formation and feedback.  See Section \ref{results-gas-thermal} for a detailed explanation of this figure.  See also Section \ref{physics-sim-sff} for the common star formation prescription and the guideline for supernova feedback, and Section \ref{codes} for the exact deposit scheme of thermal feedback energy implemented in each code.  Compare with Figure \ref{fig:sigma_500_sim-sff}.   
\label{fig:temp_500_sim-sff}}
\end{figure*}

\vspace{1mm}

\section{Results}\label{results}

In this section, we lay out the results of the first isolated disk galaxy comparison by the {\it AGORA} Collaboration.  
We focus on similarities and discrepancies discovered by comparing 500 Myr snapshots of the participating simulations in two setups: {\it Sim-noSF} and {\it Sim-SFF}. 
As defined in Section \ref{physics}, {\it Sim-noSF} refers to a run with radiative gas cooling but without star formation or feedback, and {\it Sim-SFF} refers to a run with radiative cooling, star formation and supernova feedback. 

In our simulation analyses, a key role has been played by the {\it AGORA} recommended community-driven analysis platform {\tt yt} \citep{yt, TurkSmith11, Turk13}.\footnotemark[\getrefnumber{yt-website}]  
It natively processes data from all 9 participating simulation codes discussed in this paper, plus many other modern astrophysics codes such as {\sc Athena} \citep{2008ApJS..178..137S}, {\sc Flash} \citep{2000ApJS..131..273F}, {\sc Gadget-3-sphs} \citep{2012MNRAS.422.3037R}, {\sc Nyx} \citep{AlmBell12}, and {\sc Orion} \citep{1998ApJ...495..821T}, to name a few.
Interested readers may try a unified, publicly available {\tt yt} script employed in the present analyses that has been developed throughout the progress of this study.\footnote{\label{agora-analysis-script-website}The website is http://bitbucket.org/mornkr/agora-analysis-script/.}$^{,}$\footnote{{\tt yt} version 3.3 or later is required for the script to reproduce our analyses.  For the figures and plots in Section \ref{results}, the {\tt yt-dev} changeset d7f213e1752e is used.}
We also plan to make datasets used in the present study publicly available in the near future (see Section \ref{conclusion} for more information).  

\subsection{Gas Disk Morphology}\label{results-gas-morph}

We first examine the morphology of gas disks evolved in each of the codes in our experiments.  
In Figures \ref{fig:sigma_0_sim-nosf} to \ref{fig:sigma_500_sim-sff}, we compile 9 panels that exhibit the results of the isolated disk galaxy simulations, first with radiative gas cooling but without star formation or feedback, {\it Sim-noSF}, and second with star formation and feedback, {\it Sim-SFF}, by the 9 participating codes. 
Each panel displays the disk gas surface density in a $30\,\,{\rm kpc}$ box centered on the location of maximum gas density within 1 kpc from the center of gas mass.  
This centering criterion is adopted in all subsequent figures and plots.   
For visualizations of the particle-based codes hereafter (Figures \ref{fig:sigma_0_sim-nosf}-\ref{fig:sigma_500_sim-sff}, \ref{fig:temp_500_sim-nosf}-\ref{fig:temp_500_sim-sff}, \ref{fig:metal_500_sim-sff}, \ref{fig:elevation_500_sim-sff}, \ref{fig:resolution_500_sim-sff}) -- but not in any other analyses except these figures -- {\tt yt} uses an in-memory octree on which gas particles are deposited using smoothing kernels.  
The resolution of this octree governs the resolution of produced images.  
If more than 8 particles are in an oct, that oct is refined into 64 child octs (i.e., {\tt yt} parameters ${\tt n\_ref} = 8$ and ${\tt over\_refine\_factor}=2$), providing compatible or better image resolution than a typical SPH visualization.  
The densities are assigned to the octree in a scatter step. 
That is, we first calculate a particle's smoothing length, and then add the particle's density contribution to all cell centers of the octree cells that are within the particle's smoothing sphere. 

We asked every code to output the state of the simulation immediately after it was initialized -- so-called ``0 Myr snapshot'' -- to allow ourselves to  directly compare whether the initial condition (IC) generation was successful and consistent among codes. 
This exercise has been strenuously carried out for all the analyses items presented in Sections \ref{results-gas-morph}-\ref{results-gas-thermal} and \ref{results-etc}, enabling us to correct inconsistently initialized simulations early in the study.    
One such example, the surface density comparison of 0 Myr snapshots, is shown in Figure \ref{fig:sigma_0_sim-nosf}. 
A clear distinction in gas disk initialization between mesh-based and particle-based codes can be seen in this figure. 
To model the gas disk, SPH particles or {\sc Gizmo}'s discrete tracers are generated by drawing random numbers from the distribution function given by an analytic density profile. 
This by definition results in Poisson noise in the disk surface density shown here.
Readers can also observe slight differences between mesh-based and particle-based codes in how the density field is represented in their calculations. 
By the nature of reconstructing the density from the positions of particles, the particle-based codes may smooth out the strong density contrast in the IC at the edge of the initial gas disk.  

\begin{figure*}
    \centering
    \includegraphics[width=0.83\textwidth]{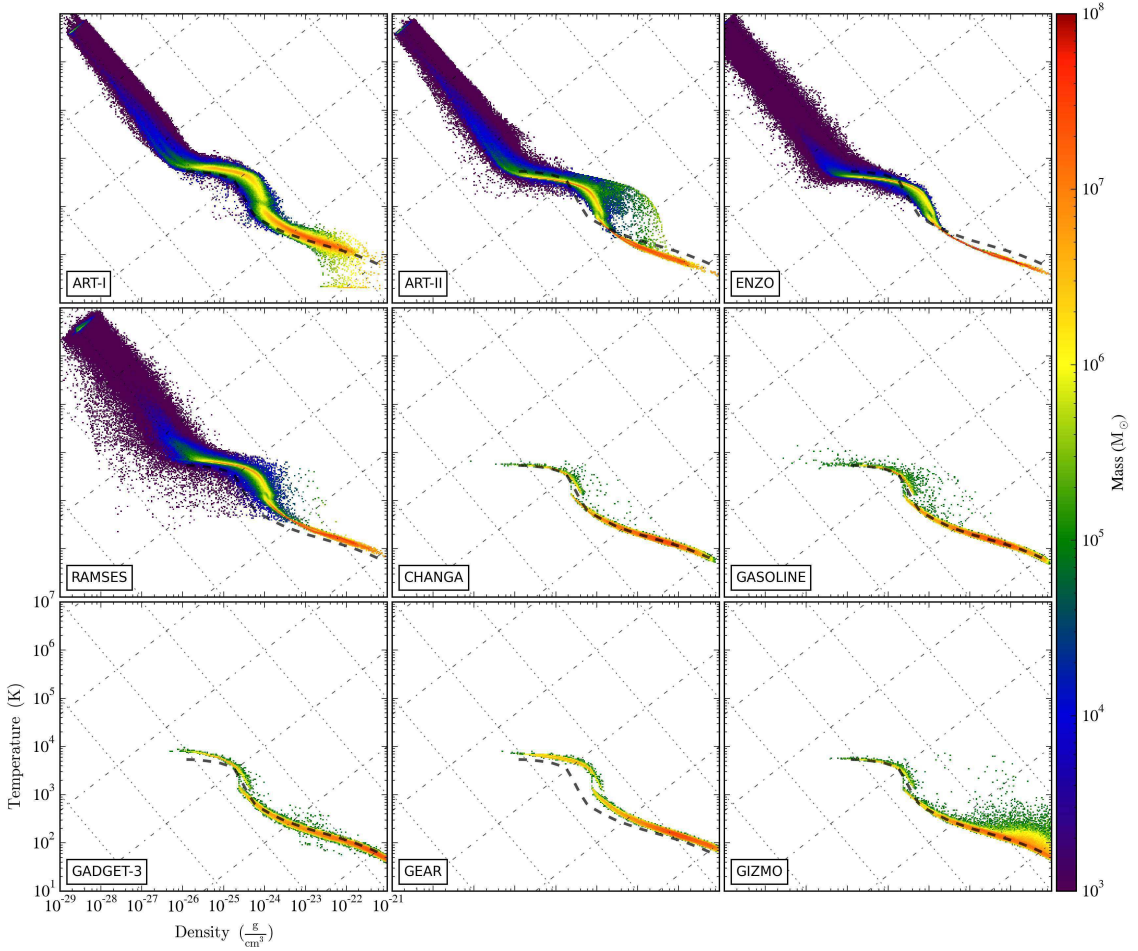}
    \caption{The 500 Myr composite of 2-dimensional probability distribution function of density and temperature for the gas within 15 kpc from the galactic center in {\it Sim-noSF} without star formation or feedback.  Colors represent the total gas mass in each 2-dimensional bin.  A gaseous halo -- low density, high temperature gas in the upper left corner of each panel -- exists only in mesh-based codes, but not in particle-based codes (SPH codes or {\sc Gizmo}).  To guide readers' eyes, we use a thick dashed line in each panel to plot the mean temperature in each density bin for {\sc Changa}.  The thin dotted diagonal lines denote the slope of constant pressure process, and the thin dot-dashed diagonal lines that of constant entropy process.  Note that different versions of {\sc Grackle} are interfaced with different codes ({\sc Grackle} v2.1 in {\sc Changa}, {\sc Gasoline}, {\sc Gadget-3} and {\sc Gizmo}, versus {\sc Grackle} v2.0 or below in {\sc Art-I}, {\sc Art-II}, {\sc Enzo}, {\sc Ramses} and {\sc Gear}).  See Section \ref{results-gas-thermal} for more information on this figure.  The full color version of this figure is available in the electronic edition.  The high-resolution versions of this figure and article are available at the Project website, http://www.AGORAsimulations.org/.
\label{fig:pdf_500_sim-nosf}}
\end{figure*}

Interestingly, in Figures \ref{fig:sigma_500_sim-nosf} and \ref{fig:sigma_500_sim-sff} at 500 Myr, there are other subtle differences noticeable between mesh-based and particle-based codes as well as within these sub-groups. 
While the peak densities and filamentary structures in the disks are very similar across all codes, it is noticeable that the mesh-based codes typically show lower densities in the inter-arm regions of the disks. 
The typical densities in those inter-arm regions -- while not containing much mass -- may differ by as much as an order of magnitude between mesh-based and particle-based codes (see also Section \ref{results-etc} and Figure \ref{fig:resolution_500_sim-sff} for a related discussion on spatial resolution). 
Another distinguishing aspect among the participating codes is the number of dense clumps formed. 
This is true in both the simulations with and without star formation and feedback (see also Section \ref{results-star-morph} and Figures \ref{fig:star_with_clumps_fof_500_sim-sff} and \ref{fig:star_clump_stats_fof_500_sim-sff} for a related comparison of newly-formed stellar clumps). 

Figures \ref{fig:gas_surface_density_radial_500_sim-nosf} and \ref{fig:gas_surface_density_radial_500_sim-sff} are the cylindrically-binned gas surface density profiles for {\it Sim-noSF} and {\it Sim-SFF}, respectively.  
The cylindrical radius is defined as the distance from the galactic center. 
Raw particle fields are used for profiles of the particle-based codes, not the interpolated or smoothed fields constructed in {\tt yt}.  
In other words, the total mass in each cylindrical annulus is divided by its area so that each gas particle contributes only to a bin in which its center falls.
As noted earlier, the gas surface densities show a high degree of correspondence across all codes in both {\it Sim-noSF} and {\it Sim-SFF}.  
All 9 profiles agree very well within a factor of a few at all radii (averaged fractional deviation for 2 $< r <$ 10 kpc in {\it Sim-SFF} is 32.2\% or 0.121 dex),\footnote{Defined as $ N_i^{\,-1} \sum\limits_{2< r_i <10} \left[ N_{\rm code}^{\,-1}  \sum\limits_{\rm code} \left\{ (f_{i, \, {\rm code}}/ \, \overline{f_i}) - 1 \right\}^2 \right]^{1/2} $. \label{fractional_deviation_radius}} and can be approximated by exponential profiles at radii $>$ 1.5 kpc.
Note that due to the gas consumed by star formation, the gas surface density slightly decreases from Figure \ref{fig:gas_surface_density_radial_500_sim-nosf} to Figure \ref{fig:gas_surface_density_radial_500_sim-sff}, the latter of which could be compared with  Figures \ref{fig:star_surface_density_radial_500_sim-sff} (newly-formed stellar surface density) and \ref{fig:sfr_surface_density_radial_500_sim-sff} (star formation rate surface density). 
Aside from this small decrease in density, there is little impact of supernova feedback from contrasting {\it Sim-noSF} and {\it Sim-SFF}.  
The inefficiency of stellar feedback is partly related to a only small increase in stellar mass in the first 500 Myr of evolution (see Sections \ref{IC} and \ref{results-sf} for more discussion).  

\begin{figure*}
    \centering
    \includegraphics[width=0.83\textwidth]{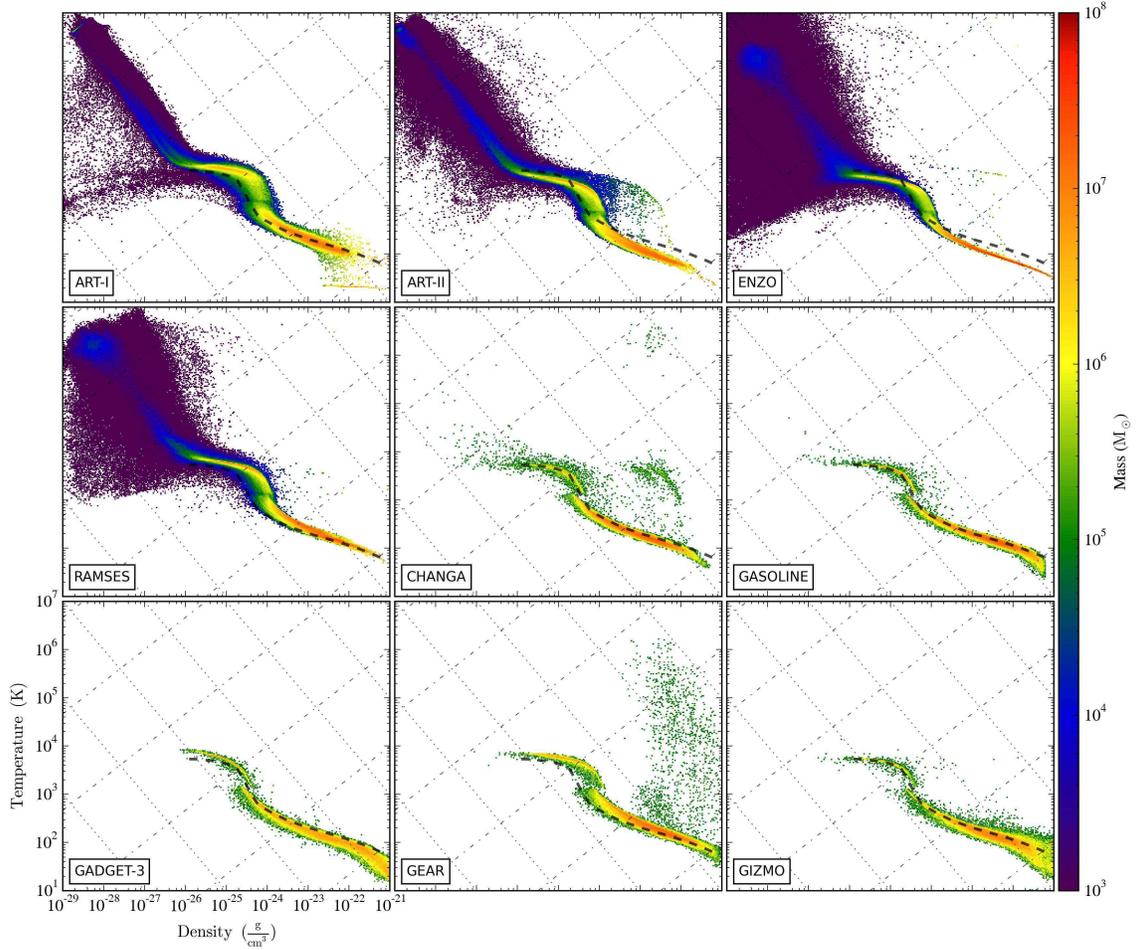}
    \caption{Same as Figure \ref{fig:pdf_500_sim-nosf} but for {\it Sim-SFF} with star formation and feedback.   We use a thick dashed line in each panel to plot the mean temperature in each density bin in {\sc Changa}'s {\it Sim-noSF} run (same as in Figure \ref{fig:pdf_500_sim-nosf}).  See Section \ref{results-gas-thermal} and the caption of Figure \ref{fig:pdf_500_sim-nosf} for a detailed explanation of this figure.  Compare with Figure \ref{fig:metal_pdf_500_sim-sff}. 
\label{fig:pdf_500_sim-sff}}
\end{figure*}

Displayed in Figures \ref{fig:gas_surface_density_vertical_500_sim-nosf} and \ref{fig:gas_surface_density_vertical_500_sim-sff} are the vertically-binned gas surface density profiles for {\it Sim-noSF} and {\it Sim-SFF}, respectively.  
The height is defined as the absolute vertical distance from the $x-y$ disk plane centered on the galactic center, i.e., $|z_{\rm i} - z_{\rm center}|$. 
Again, for particle-based codes raw particle fields are used to produce the profiles, not the interpolated or smoothed fields in {\tt yt}.  
Note a smaller range in $x$-axes than in the previous figure (only one tenth of Figures \ref{fig:gas_surface_density_radial_500_sim-nosf} and \ref{fig:gas_surface_density_radial_500_sim-sff}). 
The surface densities begin to diverge above $\sim$0.6 kpc from the disk plane, but below that substantial agreement appears (averaged fractional deviation for $z <$ 0.6 kpc in {\it Sim-SFF} is 30.4 \% or 0.115 dex).
There is no systematic difference between mesh-based and particle-based codes, similar to what we find in radial density profiles, Figures \ref{fig:gas_surface_density_radial_500_sim-nosf} and \ref{fig:gas_surface_density_radial_500_sim-sff}. 

Figures \ref{fig:rad_height_500_sim-nosf} and \ref{fig:rad_height_500_sim-sff} show the cylindrically-binned, mass-weighted average of gas vertical heights for {\it Sim-noSF} and {\it Sim-SFF}, respectively.  
As defined before, the height is an absolute vertical distance from the disk plane.  
Thus, each line in Figures \ref{fig:rad_height_500_sim-nosf} and \ref{fig:rad_height_500_sim-sff} represents the averaged $|z_{\rm i} - z_{\rm center}|$ as a function of cylindrical radius. 
Combined with Figures \ref{fig:gas_surface_density_vertical_500_sim-nosf} and \ref{fig:gas_surface_density_vertical_500_sim-sff}, these plots provide insight into the thicknesses of the gas disks in each of the codes.  
Again, no systematic difference is found between mesh-based and particle-based codes. 

\subsection{Gas Disk Kinematics}\label{results-gas-kin}

Here we examine the kinematics of gas disks that are evolved using each of the participating codes.
First, in Figures \ref{fig:pos_vel_500_sim-nosf} and \ref{fig:pos_vel_500_sim-sff} we show the gas rotation velocity curves for {\it Sim-noSF} and {\it Sim-SFF}, respectively.  
The curve reveals a mass-weighted average of gas rotational velocity of gas cells/particles, as a function of cylindrical radius. 
The cylindrical radius and rotational velocity are defined with respect to the galactic center (see Section \ref{results-gas-morph} for our adopted centering scheme).  
For mesh-based codes, only the dense enough gas cells ($\rho_{\rm gas} > 10^{-25}\,{\rm g\,cm^{-3}}$) are considered in order to minimize the contribution of the gaseous halo which is present only in mesh-based codes (see Section \ref{IC} for more information about the halo gas distribution in our IC).  
From these two figures it is clear that there exists a very good agreement on gas kinematics in all the codes, as good as within a few percent at $\sim$ 10 kpc for both {\it Sim-noSF} and {\it Sim-SFF} (averaged fractional deviation for 2 $< r <$ 10 kpc in {\it Sim-SFF} is 2.8\% or 0.012 dex). 
Discrepancies in the central region ($<$ 1.5 kpc) are a result of determining the galactic center that may constantly shift its position. 

\begin{figure*}
\centering
\begin{minipage}[t]{.47\textwidth}
    \includegraphics[width=1.0\textwidth]{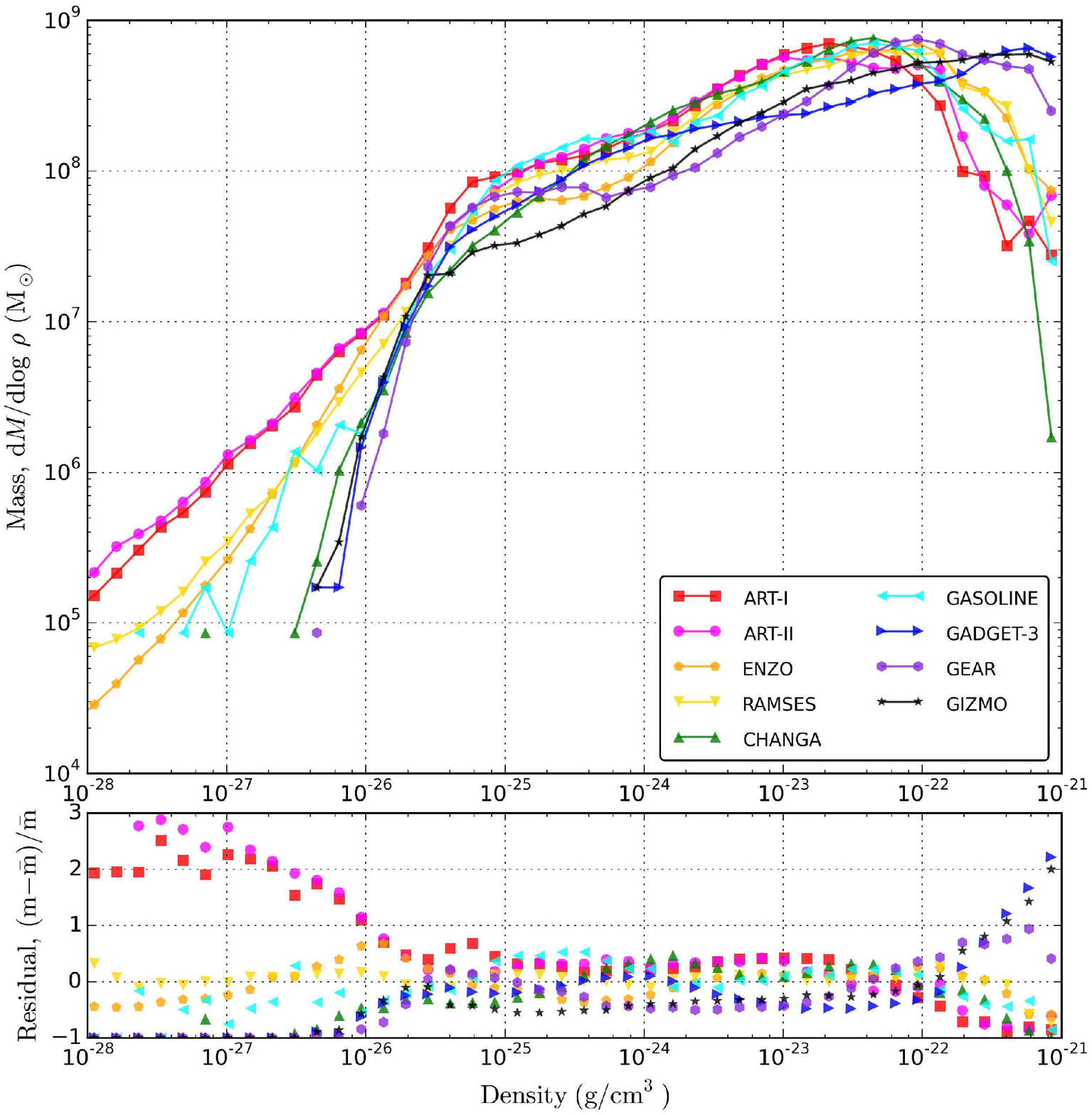}
    \caption{Gas density probability distribution function at 500 Myr for {\it Sim-noSF} without star formation or feedback.  Note that a gaseous halo -- low density tails towards the left side of this plot -- exists only in mesh-based codes, but not in particle-based codes (SPH codes or {\sc Gizmo}).   Shown in the bottom panel is the fractional deviation from the mean of these profiles.  See Section \ref{results-gas-thermal} for more information on this figure. 
\label{fig:density_df_500_sim-nosf}}
\end{minipage}
\hfill
\begin{minipage}[t]{.47\textwidth}
    \includegraphics[width=1.0\textwidth]{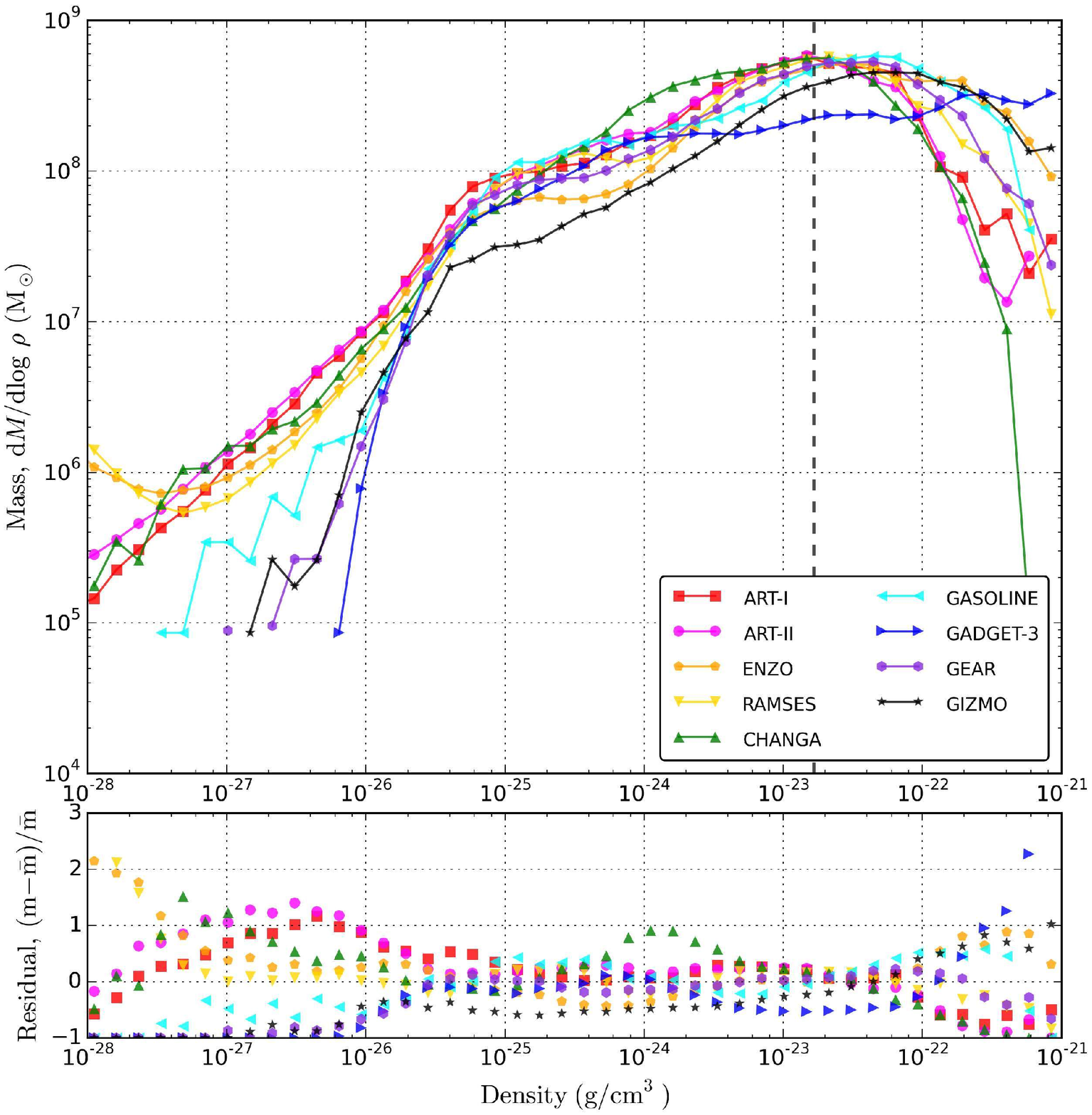}
    \caption{Same as Figure \ref{fig:density_df_500_sim-nosf} but for {\it Sim-SFF} with star formation and feedback.  The thick dashed line denotes the star formation threshold density, $n_{\rm H, \,thres} = 10\,\, {\rm cm}^{-3}$.  
\label{fig:density_df_500_sim-sff}}
\end{minipage}
\end{figure*}

Figures \ref{fig:pos_disp_500_sim-nosf} and \ref{fig:pos_disp_500_sim-sff} reveal the gas velocity dispersion curves for {\it Sim-noSF} and {\it Sim-SFF}, respectively.  
The velocity dispersion quantifies the residual velocity components of gas other than the rotational velocity found in Figures \ref{fig:pos_vel_500_sim-nosf} and \ref{fig:pos_vel_500_sim-sff}. 
In other words, each line in Figures \ref{fig:pos_disp_500_sim-nosf} and \ref{fig:pos_disp_500_sim-sff} denotes a square root of a mass-weighted average of $(v_{\rm i} - v_{\rm rot}(r))^2$, as a function of  cylindrical radius.   
As in Figures \ref{fig:pos_vel_500_sim-nosf} and \ref{fig:pos_vel_500_sim-sff}, for mesh-based codes only dense enough cells are used to compute the dispersion. 
Again a good agreement is found in the velocity dispersion between all codes within a few tens of percent for both {\it Sim-noSF} and {\it Sim-SFF} (averaged fractional deviation for 2 $< r <$ 10 kpc in {\it Sim-SFF} is 17.8\% or 0.071 dex).
Larger variations in the central region ($<$ 1.5 kpc) are partly due to the center determination, just like in rotation velocity curves, Figures \ref{fig:pos_vel_500_sim-nosf} and \ref{fig:pos_vel_500_sim-sff}. 
The discrepancies are also produced by vertical movement of gas in the inner disk, which is captured in mesh-based codes ({\sc Art-I}, {\sc Art-II}, {\sc Enzo}, and {\sc Ramses}) but not as well in particle-based codes ({\sc Changa}, {\sc Gasoline}, {\sc Gadget-3}, {\sc Gear}, and {\sc Gizmo}) given our very particular choice of 80 pc resolution.
This can be clearly seen in the bottom panels of Figures \ref{fig:pos_disp_500_sim-nosf} and \ref{fig:pos_disp_500_sim-sff}, in which we plot the ratio of {\it vertical} velocity dispersion ($z$-direction) to {\it total} velocity dispersion to illustrate the contribution by vertical movement of gas.
From these figures, we find that a significant portion of gas velocity dispersion in the inner disk measured in mesh-based codes are driven by vertical movement, but not in particle-based codes (see Section \ref{results-etc} and Figure \ref{fig:resolution_500_sim-sff} for a related discussion on spatial resolution). 

\subsection{Thermal Structure of the Interstellar Medium}\label{results-gas-thermal}

In Figures \ref{fig:temp_500_sim-nosf} and \ref{fig:temp_500_sim-sff}, we compile 9 panels of the density-square-weighted gas temperature projections for {\it Sim-noSF} and {\it Sim-SFF}, respectively.\footnote{Temperature information may not be readily available in some codes in which only internal energy ($T/\mu$; {\sc Gadget-3} and {\sc Gear}) or pressure ($\rho T/\mu$; {\sc Ramses}) is tracked instead.  In this case, we use the $T-\mu$ table derived from {\sc Grackle} v2.0 to acquire the temperature.}  
Each panel is for the central $30\,\,{\rm kpc}$ box (see Section \ref{results-gas-morph} for our adopted centering scheme).  
Readers should note that, by design only mesh-based codes ({\sc Art-I}, {\sc Art-II}, {\sc Enzo}, and {\sc Ramses}) initially include a gaseous halo with low density and high temperature ($T_{\rm gas} = 10^6\,\, {\rm K}$; colored dark red), but not the particle-based codes ({\sc Changa}, {\sc Gasoline}, {\sc Gadget-3}, {\sc Gear}, and {\sc Gizmo}; see Section \ref{IC}).  
All of these codes show similar features without star formation and feedback in Figure \ref{fig:temp_500_sim-nosf}. 
We continue a related discussion on {\it Sim-noSF} using Figures \ref{fig:pdf_500_sim-nosf} and \ref{fig:density_df_500_sim-nosf} later in this section.

\begin{figure}
    \centering
    \includegraphics[width=0.47\textwidth]{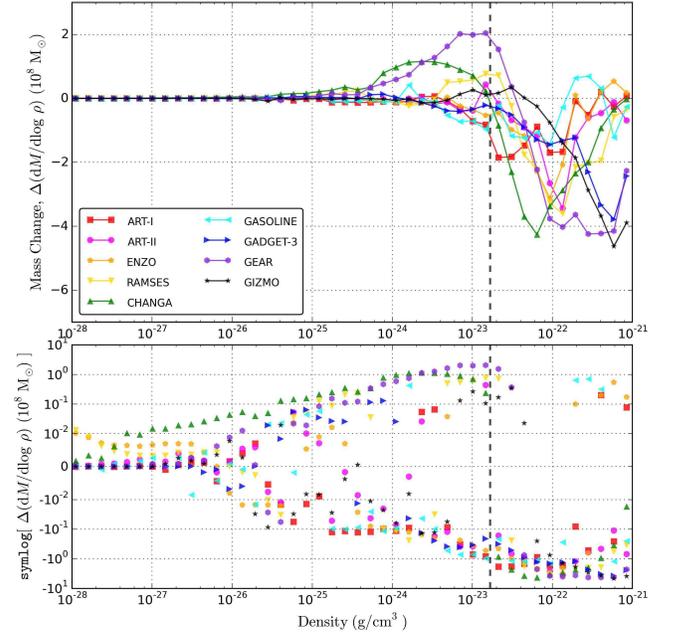}
    \caption{Code-by-code mass change in each density bin from Figure \ref{fig:density_df_500_sim-nosf} to  \ref{fig:density_df_500_sim-sff}.  This plot measures the effect of star formation and feedback by subtracting the density probability distribution function of {\it Sim-noSF} from that of {\it Sim-SFF}.  The $y$-axis spans from $-7\times10^8\,\msun$ to $\,+3\times10^8\,\msun$ in a linear scale.  Shown in the bottom panel is the sign-preserving logarithm of the mass difference in order to make smaller changes to stand out.\protect\footnotemark[\getrefnumber{symlog-definition}]  The thick dashed line denotes the star formation threshold density, $n_{\rm H, \,thres} = 10\,\, {\rm cm}^{-3}$.  See Section \ref{results-gas-thermal} for more information on this figure. 
\label{fig:density_df_change_500_sim-sff}}
\end{figure}

\begin{figure*}
    \centering
    \includegraphics[width=1.04\textwidth]{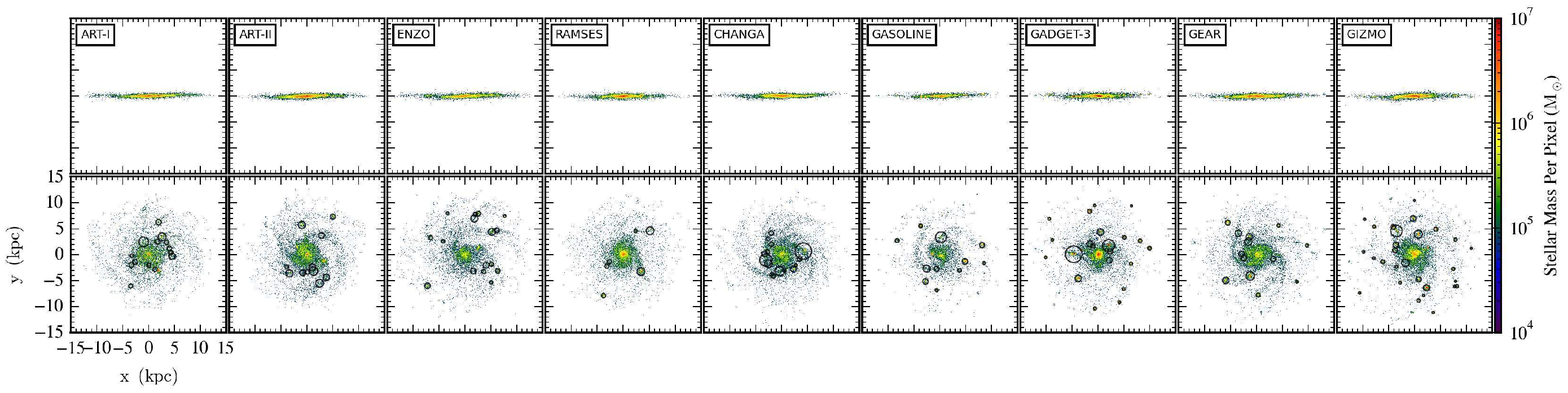}
    \caption{The 500 Myr composite of newly-formed star particle distributions, edge-on {\it (top)} and face-on {\it (bottom)}, for {\it Sim-SFF} with star formation and feedback.  Only the newly-formed star particles are drawn, not the disk or bulge stars that were present in the IC.  Colors represent the total newly-formed stellar mass in each 2-dimensional bin.  We also highlight the clumps of newly-formed star particles identified by the Friends-of-Friends (FOF) algorithm with circles.  Clumps with masses below $2.6 \times 10^6 \msun$ and the most massive clump found by FOF (always associated with the stellar bulge at the galactic center) are excluded.  See Section \ref{results-star-morph} for a detailed explanation of this figure.  Compare with Figures \ref{fig:sigma_500_sim-sff}, \ref{fig:temp_500_sim-sff}, and \ref{fig:resolution_500_sim-sff}.  The full color version of this figure is available in the electronic edition.  The high-resolution versions of this figure and article are available at the Project website, http://www.AGORAsimulations.org/. 
\label{fig:star_with_clumps_fof_500_sim-sff}}
\end{figure*}

But slight differences in supernova feedback implementation -- even within a common guideline -- may cause different features in the temperature map, Figure \ref{fig:temp_500_sim-sff}. 
For example, small hot bubbles are visible in {\sc Changa} and {\sc Gear}, but not in {\sc Gasoline} or {\sc Gadget-3} in which the common supernova feedback schemes are implemented slightly differently.\footnote{Both {\sc Changa} and {\sc Gasoline} spread supernova feedback energy to 64 neighboring particles, but they use different neighbor numbers for smoothing in hydrodynamics: $N_{\rm smooth}=$ 64 in {\sc Changa} versus 200 in {\sc Gasoline} (see Sections \ref{changa} and \ref{gasoline}).  In addition, unlike {\sc Gear}, {\sc Gadget-3} adopts the Sedov-Taylor blast wave method (see Section \ref{gadget}).  A slight change in the details of shock radius estimation is shown to cause a large difference.} 
We also note the chimneys of hot gas departing from the disk in {\sc Enzo} and {\sc Ramses}, clearly visible in the edge-on maps of Figure \ref{fig:temp_500_sim-sff}.  
This gas ejecta from the disk as a result of supernova feedback is not as evident in {\sc Art-I}, {\sc Art-II}, or particle-based codes, although some hot bubbles are seen in {\sc Changa} and {\sc Gear}.
We caution that the spatial resolution employed in this paper is only 80 pc.  
It is not as high as the resolutions in some of the modern zoom-in cosmological simulations, and may not be enough for particle-based codes to resolve the chimney-like structure above and below the disk (see Section \ref{results-etc} and Figure \ref{fig:resolution_500_sim-sff} for a related discussion on spatial resolution). 
We refer the readers to a continued discussion on {\it Sim-SFF} using Figures \ref{fig:pdf_500_sim-sff} and \ref{fig:density_df_500_sim-sff} later in this section.\footnote{In Figures \ref{fig:temp_500_sim-sff} and \ref{fig:metal_500_sim-sff} readers may notice hemispherical shapes of size $\sim$ 5 kpc  in some particle-based codes, e.g., {\sc Gasoline} or {\sc Gear}.  These are not expanding blast waves driven by supernova feedback, but a visualization effect due to {\tt yt}'s smoothing kernel for SPH particles or {\sc Gizmo}'s discrete tracers.  This feature is also clearly seen in Figure \ref{fig:metal_500_sim-sff}, and barely seen in Figure \ref{fig:sigma_500_sim-sff}.  Higher-resolution simulations would minimize this artifact. \label{hemispherical-artifact}}  

To better understand what we find in Figures \ref{fig:temp_500_sim-nosf} and \ref{fig:temp_500_sim-sff}, we show the 2-dimensional probability distribution functions (PDFs) of gas density and temperature in Figures \ref{fig:pdf_500_sim-nosf} and \ref{fig:pdf_500_sim-sff}  for {\it Sim-noSF} and {\it Sim-SFF}, respectively.  
We consider gas within 15 kpc from the galactic center.  
Colors represent the total gas mass in each 2-dimensional bin. 
As explained in Section \ref{results-gas-morph}, raw particle fields are used for the PDFs of particle-based codes, not the smoothed fields constructed by {\tt yt}.
Note that a gaseous halo, represented by low density, high temperature gas in the upper left corner of each panel, is designed to exist only in mesh-based codes, but is absent in particle-based codes (SPH codes and {\sc Gizmo}; see Section \ref{IC}).  
It is therefore {\it not} the intended scope of this paper to compare this hot halo or circumgalactic medium between codes.
To guide readers' eyes, we plot the mean temperature in each density bin from {\sc Changa}'s {\it Sim-noSF} run with a thick dashed line in each panel (in both Figures \ref{fig:pdf_500_sim-nosf} and \ref{fig:pdf_500_sim-sff}; in the range of $[10^{-26}, 10^{-21}]\,{\rm g\,cm^{-3}}$).  
{\sc Changa}'s mean profile is close to the ``mean of means'' of these 9 codes, thus helps to compare the PDFs' relative positions between codes.  
The thin dotted diagonal lines denote the slope of constant pressure process, and the thin dot-dashed diagonal lines that of constant entropy process. 

\begin{figure}
    \centering
    \includegraphics[width=0.47\textwidth]{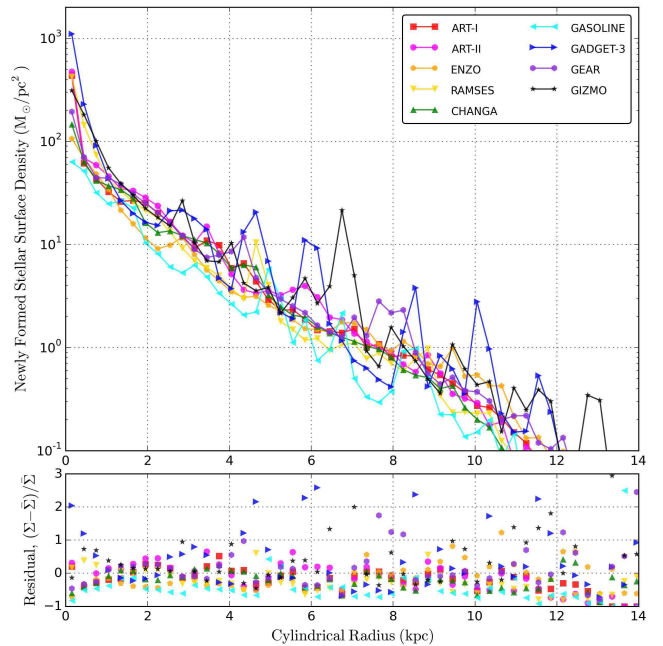}
    \caption{Cylindrically-binned newly-formed stellar surface density profiles at 500 Myr for {\it Sim-SFF} with star formation and feedback.  Only the newly-formed star particles are considered, not the disk or bulge stars that were present in the IC.  Shown in the bottom panel is the fractional deviation from the mean of these profiles.  See Section \ref{results-star-morph} for a detailed explanation of this figure. Compare with Figures \ref{fig:gas_surface_density_radial_500_sim-sff} and  \ref{fig:sfr_surface_density_radial_500_sim-sff}.  The $y$-axis range of the top panel is kept identical among Figures  \ref{fig:gas_surface_density_radial_500_sim-nosf}-\ref{fig:gas_surface_density_vertical_500_sim-sff} and \ref{fig:star_surface_density_radial_500_sim-sff} for easier comparison.  
\label{fig:star_surface_density_radial_500_sim-sff}}
\end{figure}

\begin{figure*}
    \centering
    \includegraphics[width=0.72\textwidth]{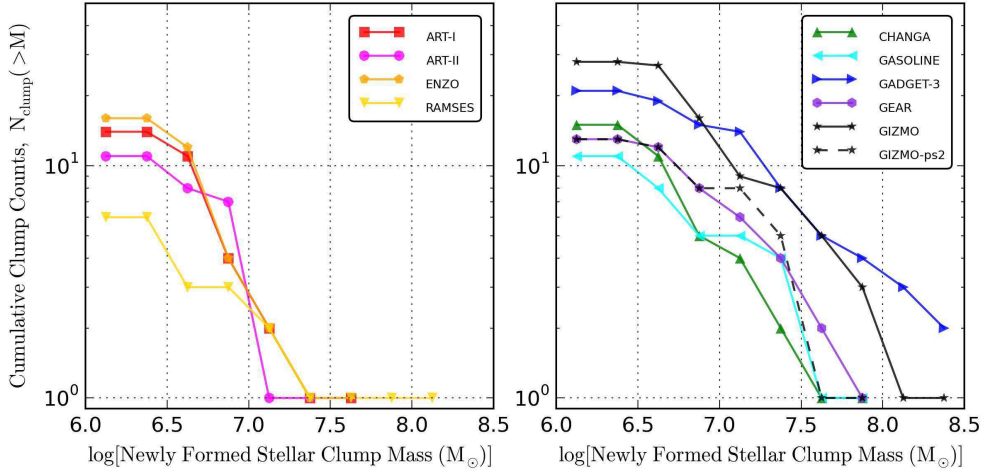}
    \caption{The cumulative mass function of newly-formed stellar clumps at 500 Myr for {\it Sim-SFF}.  Clumps with masses below $2.6 \times 10^6 \msun$ and the most massive clump found by FOF (always associated with the stellar bulge at the galactic center) are excluded.  See Section \ref{results-star-morph} for more information on this figure, including an additional test {\sc Gizmo-ps2} for which {\sc Gizmo}'s Jeans pressure support is increased by a factor of 2.
\label{fig:star_clump_stats_fof_500_sim-sff}}
\end{figure*}

Overall, all codes exhibit similar behaviors in Figure \ref{fig:pdf_500_sim-nosf} when without star formation and feedback, just like the broad similarity observed in Figure \ref{fig:temp_500_sim-nosf}. 
A clear branch of gas is visible in all codes extending towards higher density, lower temperature, owing to the common treatment of cooling by the {\sc Grackle} library (see also Figure \ref{fig:density_df_500_sim-nosf}, and a related discussion later in this section).  
But, as noted in Section \ref{physics-sim-nosf}, due to varying {\sc Grackle} versions participating codes are interfaced with ({\sc Grackle} v2.1 in {\sc Changa}, {\sc Gasoline}, {\sc Gadget-3} and {\sc Gizmo}, versus {\sc Grackle} v2.0 or below in {\sc Art-I}, {\sc Art-II}, {\sc Enzo}, {\sc Ramses} and {\sc Gear}), the cooling rates differ slightly from code to code even at the same density, temperature, and metal fraction.  
See Section \ref{IC} (footnote \ref{grackle-zsun}) or Section \ref{physics-sim-nosf} (footnote \ref{grackle-zsun-2}) for more information.  
Any remaining discrepancy is attributable to the difference in how each code sub-cycles its cooling module in the hydrodynamics calculation.  

Now we compare the density$-$temperature PDFs when star formation and supernova feedback are included in Figure \ref{fig:pdf_500_sim-sff}.  
All codes successfully lower the fraction of low temperature, high density gas by forming stars and then injecting their feedback energy (see also Figures \ref{fig:density_df_500_sim-nosf}-\ref{fig:density_df_change_500_sim-sff}, and a related quantitative  discussion later in this section).  
However, notable differences exist between codes as to how gas reacts to the supernova feedback.   
For example, in {\sc Changa} and {\sc Gear} some gas is leaving the aforementioned high density branch towards higher temperature (up to $10^6$ K) due to supernova feedback, but not in {\sc Gasoline}, {\sc Gadget-3} and {\sc Gizmo}.  
The high temperature, high density gas seen in {\sc Changa} and {\sc Gear} is associated with the small hot bubbles discussed in Figure \ref{fig:temp_500_sim-sff}.
As explained earlier, these discrepancies  in particle-based codes are attributed to different numerical implementations of the common feedback physics. 
Also noticeable is the hot gas being ejected from the disk as a result of feedback particularly in {\sc Enzo} and {\sc Ramses}, seen as a broader distribution of hot gas in Figure \ref{fig:pdf_500_sim-sff} compared to Figure \ref{fig:pdf_500_sim-nosf}.  

For a more quantitative  comparison of the ISM thermal structure, shown in Figures \ref{fig:density_df_500_sim-nosf} and \ref{fig:density_df_500_sim-sff} are the gas mass distributions along the density axis (density probability distribution function) for {\it Sim-noSF} and {\it Sim-SFF}, respectively, simply derived from Figures  \ref{fig:pdf_500_sim-nosf} and \ref{fig:pdf_500_sim-sff} above.  
Note again that a gaseous halo, represented by low density tails towards the left side of these plots, is by design included only in the mesh-based codes, but not in the particle-based codes. 
In Figure \ref{fig:density_df_500_sim-nosf}, when without star formation and feedback all codes show similar distributions within a factor of a few difference in the density range $[10^{-25}, 10^{-22}]\,{\rm g\,cm^{-3}}$.  
A notable deviation is that three particle-based codes, {\sc Gadget-3}, {\sc Gear} and {\sc Gizmo}, hold more mass at density above $10^{-22}\,{\rm g\,cm^{-3}}$ than the rest of the codes do. 
However, in Figure \ref{fig:density_df_500_sim-sff}, now in a more realistic setup including star formation and feedback, no clear systematic difference exists between mesh-based and particle-based codes. 
While the agreement in the range $[10^{-25}, 10^{-22}]\,{\rm g\,cm^{-3}}$ is again within a factor of a few (averaged fractional deviation in $10^{-25} < \rho < 10^{-22}\,{\rm g\,cm^{-3}}$ is 28.6\% or 0.109 dex), the codes diverge from one another up to more than an order of magnitude at density above $10^{-22}\,{\rm g\,cm^{-3}}$, translating into differences in clumping properties (Figure \ref{fig:star_clump_stats_fof_500_sim-sff}) and star formation rates (Figure \ref{fig:sfr_500_sim-sff}). 

Then, Figure \ref{fig:density_df_change_500_sim-sff} plots the code-by-code mass change  in each density bin from Figure \ref{fig:density_df_500_sim-nosf} to Figure \ref{fig:density_df_500_sim-sff}. 
This plot aims to measure the effect of star formation and supernova feedback by subtracting the density probability distribution of {\it Sim-noSF} from that of {\it Sim-SFF}.
As noted in our discussion of Figure \ref{fig:pdf_500_sim-sff}, {\it Sim-SFF} lowers the fraction of high density gas by forming stars at above $n_{\rm H, \,thres} = 10\,\, {\rm cm}^{-3} = \rho_{\rm gas,\, thres}/m_{\rm H}$ (see Section \ref{physics-sim-sff}) and then injecting their feedback energy. 
This impact is more evident in the bottom panel of Figure \ref{fig:density_df_change_500_sim-sff}, where we show the sign-preserving logarithm, {\tt symlog()}, of the mass change, making smaller changes more discernible.\footnote{${\rm symlog}(x) =  {\rm sgn}(x)\times {\rm log}|x|$. \label{symlog-definition}}
Without exception, gas masses at density above $\rho_{\rm gas,\, thres} = 1.67 \times 10^{-23}\,{\rm g\,cm^{-3}}$ (thick dashed line) are reduced by star formation and feedback, even if inefficiently. 
The gas is either consumed by star formation or redistributed by thermal supernova feedback to a less dense region below $\rho_{\rm gas,\, thres}$ away from star-forming sites.   

\begin{figure*}
\centering
\begin{minipage}[t]{.47\textwidth}
    \includegraphics[width=1.0\textwidth]{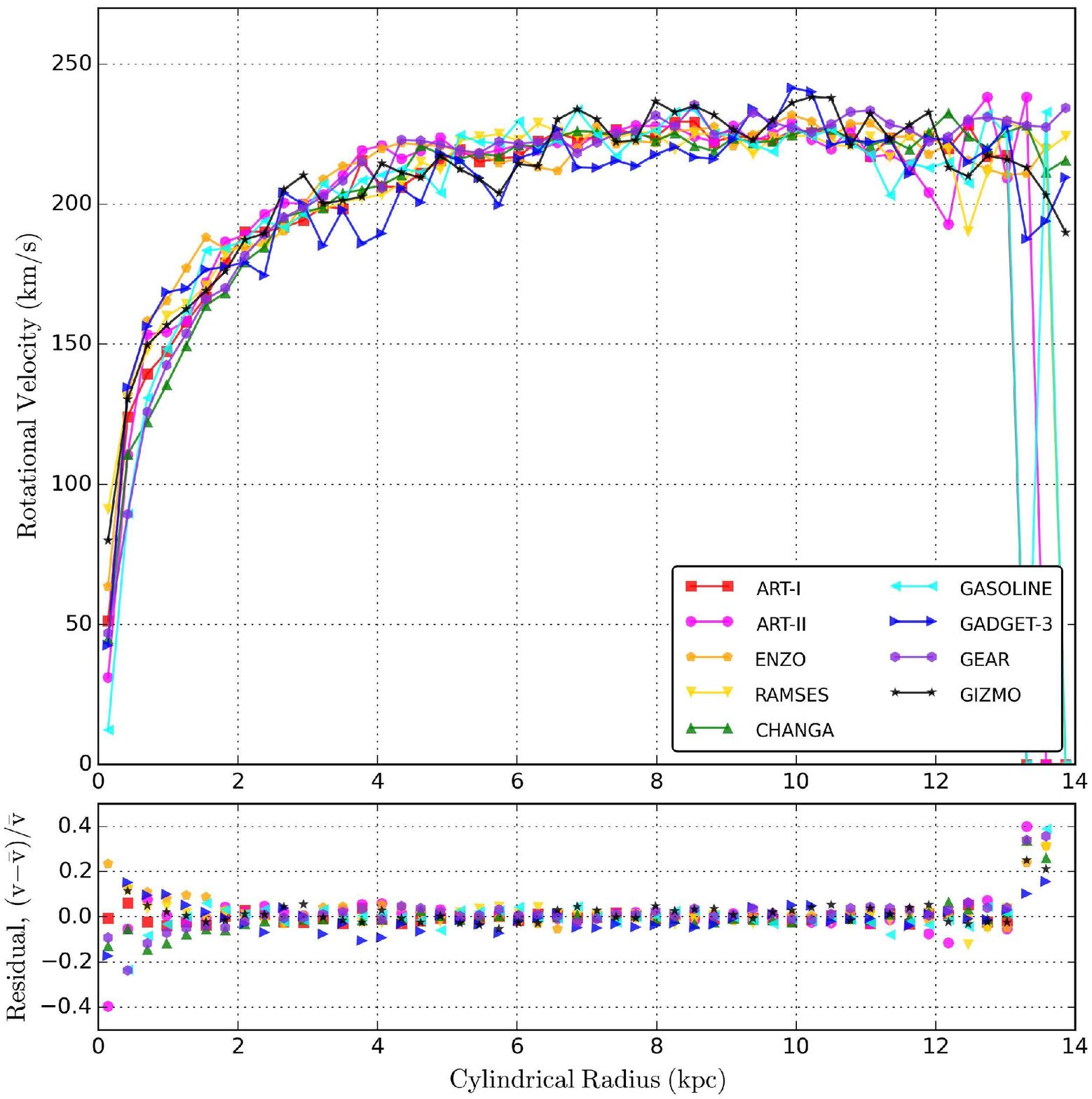}
    \caption{Stellar rotation velocity curves at 500 Myr for {\it Sim-SFF} with star formation and feedback. The cylindrical radius and rotational velocity are with respect to the galactic center -- location of maximum gas density within 1 kpc from the center of gas mass.  Only the newly-formed star particles are considered.  See Section \ref{results-star-kin} for a detailed explanation on how this figure is made.  Compare with Figure \ref{fig:pos_vel_500_sim-sff}.  The $y$-axis range of the top panel is kept identical among Figures  \ref{fig:pos_vel_500_sim-nosf}-\ref{fig:pos_vel_500_sim-sff} and \ref{fig:star_pos_vel_500_sim-sff} for easier comparison.  
\label{fig:star_pos_vel_500_sim-sff}}
\end{minipage}
\hfill
\begin{minipage}[t]{.47\textwidth}
    \includegraphics[width=1.0\textwidth]{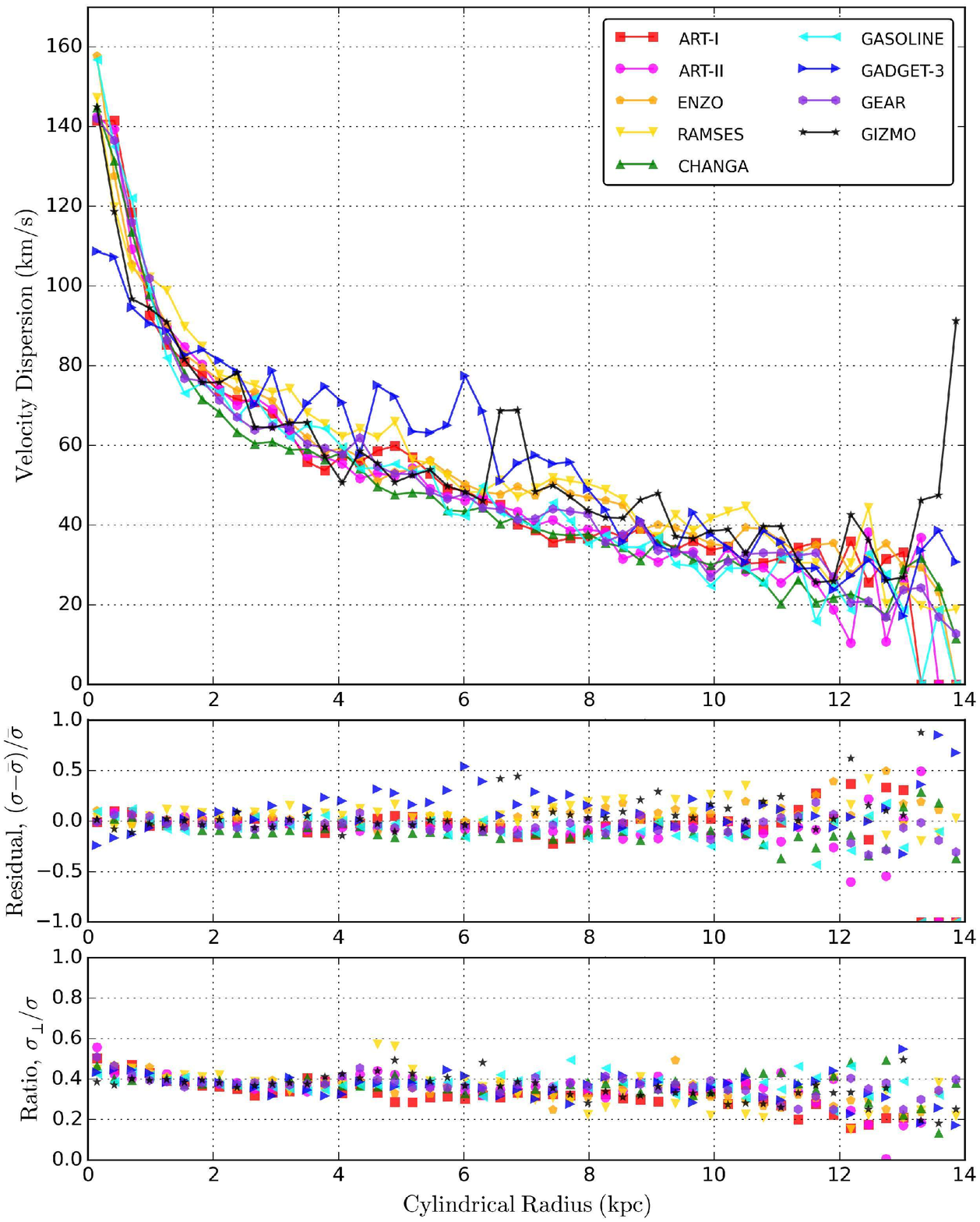}
    \caption{Stellar velocity dispersion curves at 500 Myr for {\it Sim-SFF}.  Shown in the middle panel is the fractional deviation from the mean of these profiles.  In the bottom panel we plot the ratio of {\it vertical} velocity dispersion ($z$-direction) to {\it total} velocity dispersion.  Compare with Figure \ref{fig:pos_disp_500_sim-sff}. The $y$-axis range of the top panel is kept identical among Figures  \ref{fig:pos_disp_500_sim-nosf}-\ref{fig:pos_disp_500_sim-sff} and \ref{fig:star_pos_disp_500_sim-sff} for easier comparison.  
\label{fig:star_pos_disp_500_sim-sff}}
\end{minipage}
\end{figure*}

\subsection{Stellar Disk Morphology}\label{results-star-morph}

In this section, we study the morphology of stellar disks formed in {\it Sim-SFF}. 
Figure \ref{fig:star_with_clumps_fof_500_sim-sff} shows the distributions of newly-formed star particles in {\it Sim-SFF} for each of the 9 codes. 
Only the newly-formed star particles are drawn, not the disk or bulge stars that were present in the IC. 
Star particles present in the IC are excluded from all ``stellar'' particle analyses hereafter. 
Each frame is centered on the galactic center, defined in Section \ref{results-gas-morph} as the location of peak gas density within 1 kpc from the center of gas mass. 
This center almost always coincides approximately with the center of the most massive stellar clump.    
Colors represent the total newly-formed stellar masses in each 2-dimensional bin. 
The bottom rows also highlights clumps of newly-formed stars (more discussion on stellar clumps later in this section).

\begin{figure*}
\centering
\begin{minipage}[t]{.47\textwidth}
    \includegraphics[width=1.0\textwidth]{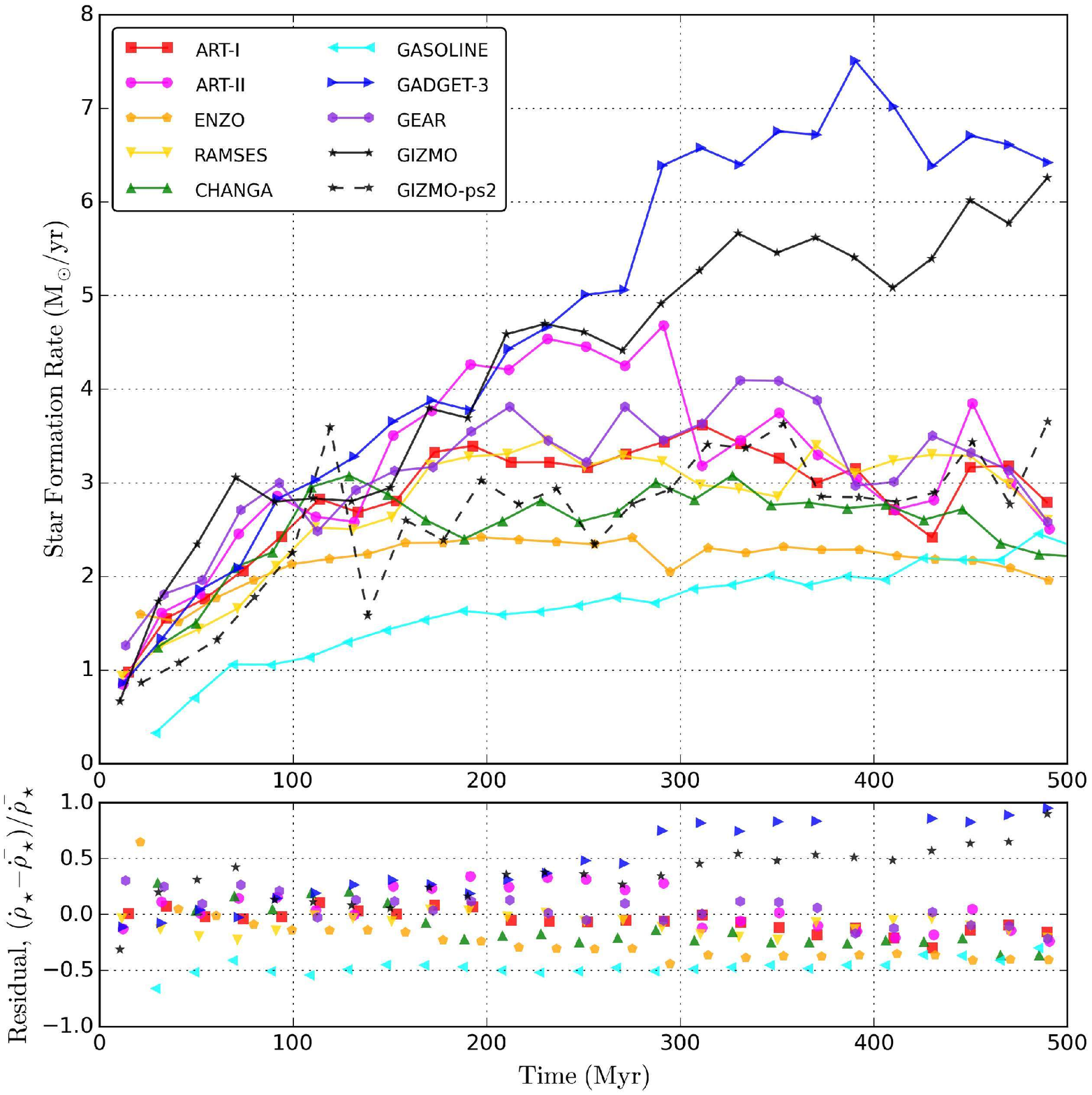}
    \caption{Galaxy-wide star formation rates by 500 Myr for {\it Sim-SFF}.    Shown in the bottom panel is the fractional deviation from the mean.  See Section \ref{results-sf} for a detailed explanation on this figure, including an additional test {\sc Gizmo-ps2} for which {\sc Gizmo}'s Jeans pressure support is increased by a factor of 2.
\label{fig:sfr_500_sim-sff}}
\end{minipage}
\hfill
\begin{minipage}[t]{.47\textwidth}
    \includegraphics[width=1.0\textwidth]{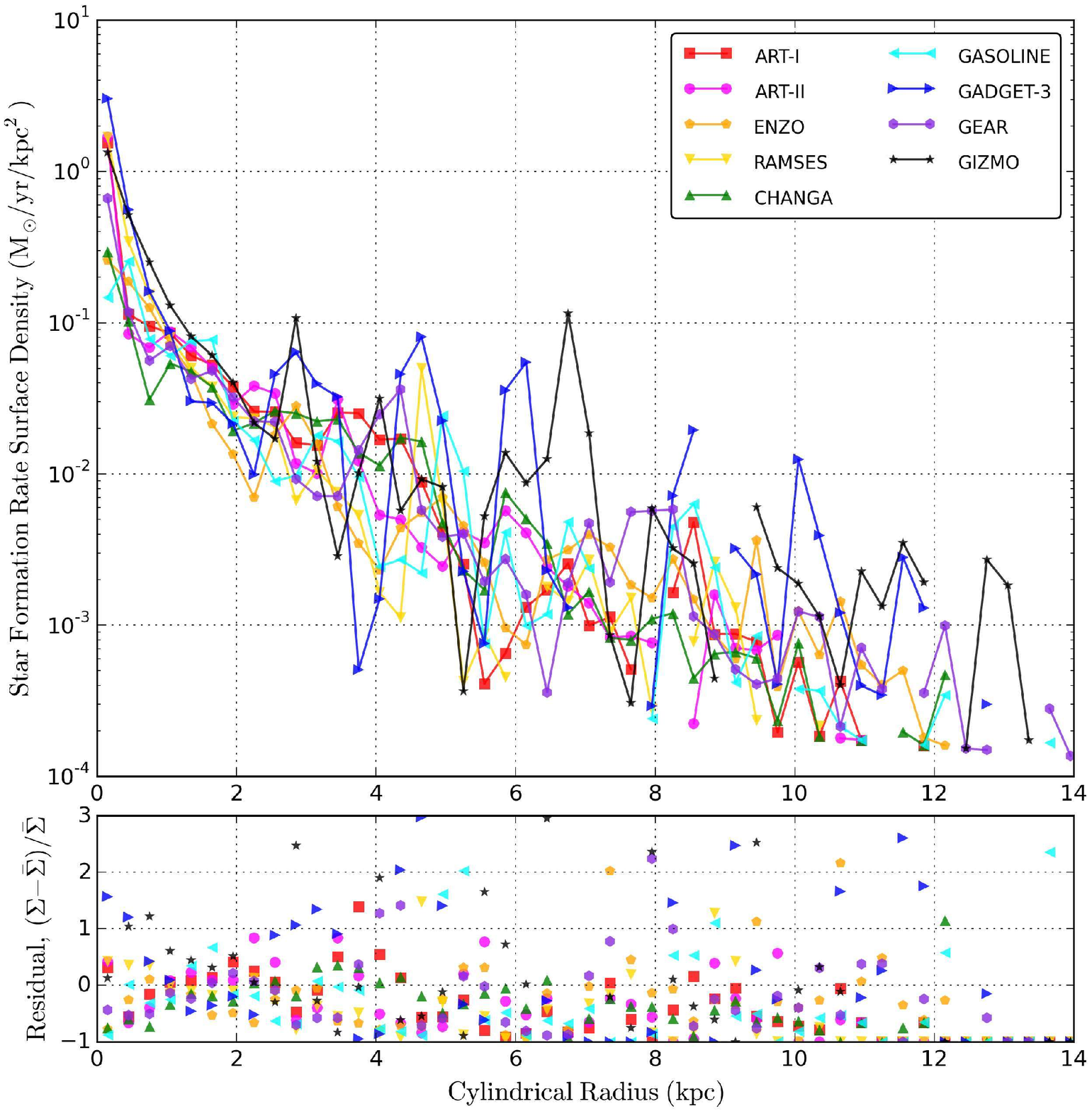}
    \caption{Cylindrically-binned star formation rate surface density profiles at 500 Myr for {\it Sim-SFF}.  Star formation rates are estimated using the newly-formed star particles that are younger than 20 Myr old.  Shown in the bottom panel is the fractional deviation from the mean of these profiles.  See Section \ref{results-sf} for a detailed explanation on how this figure is made.  Compare with Figures \ref{fig:gas_surface_density_radial_500_sim-sff} and \ref{fig:star_surface_density_radial_500_sim-sff}. 
\label{fig:sfr_surface_density_radial_500_sim-sff}}
\end{minipage}
\end{figure*}

Figure \ref{fig:star_surface_density_radial_500_sim-sff} depicts surface densities of newly-formed star particles -- excluding star particles present in the ICs -- for {\it Sim-SFF}, calculated in cylindrically symmetric radial bins.  
While there are differences up to a factor of a few among some codes (e.g., between {\sc Gadget-3} and {\sc Gasoline}, due to their different rates of galaxy-wide star formation shown in Figure \ref{fig:sfr_500_sim-sff}; averaged fractional deviation for 2 $< r <$ 10 kpc is 53.9\% or 0.187 dex), all the lines can be well fit by an exponential disk profile at radii $>$ 1.5 kpc. 
Occasional fluctuations visible in the profiles (e.g., at $\sim$ 6.5 kpc in {\sc Gizmo}) are due to dense  stellar clumps located at these particular radii. 

In order to compare the distribution of newly-formed star particles and the level of disk fragmentation between different codes, we identify clumps in the distribution of newly-formed star particles using the Friends-of-Friends (FOF) algorithm \citep{1985ApJS...57..241E}.\footnote{For the FOF machinery, it is implicitly assumed that all newly-formed star particles have the same mass.  The actual mass difference is no more than a factor of 2.  The FOF finder also implicitly demands that particles are located in a periodic box, whose size we manually set to $3.25^3$ Mpc (largest box size used by one of the groups).  Given the periodic boundary condition, we use a linking length equal to 0.25\% of the average inter-particle distance.}  
We only consider clumps with newly-formed stellar masses above $2.6 \times 10^6 \msun$  (equivalent to 30 times the mass of star particles present in the IC, $30\, m_{\rm \star,\,IC}$).  
The most massive clump found by FOF is excluded since it is always associated with the stellar bulge at the galactic center, but we do not explicitly remove gravitationally unbound clumps (which may have overestimated the number of clumps). 
Identified clumps are marked with circles in the bottom row of Figure \ref{fig:star_with_clumps_fof_500_sim-sff}, where the radius of each circle indicates the clump's virial radius.  
We also show the cumulative mass functions of newly-formed stellar clumps  in Figure \ref{fig:star_clump_stats_fof_500_sim-sff}.  
It is worth to note that very similar clump distributions and mass functions are discovered when we identify clumps using the HOP algorithm \citep{1998ApJ...498..137E} in lieu of FOF. 
The general trends for clumps discussed below are largely independent of the clump finder. 

In all codes, the majority of newly-formed stellar clumps have masses below $10^{7.5} \msun$, and there is a relatively sharp decline in the number of clumps towards higher masses.\footnote{These clumps are not to be confused with the giant clumps of masses between $10^8$ and $10^9 \msun$ observed in star-forming galaxies at redshift $z\sim2$.  These $z\sim2$ clumps are expected to form in disks with gas fractions of 40-50\%, much higher than the initial 20\% gas fraction in our experiment.} 
From these figures, it is also clear that nearly-formed stellar clumps are the most prevalent in the {\sc Gizmo} run, and less so in the {\sc Ramses} run than in other codes.   
The relatively large number of stellar clumps in {\sc Gizmo} is not a transient feature, but consistently observed across snapshots until 1 Gyr.  
This is related to the fact that {\sc Gizmo} produces the most clumpy gas disk among the codes even with star formation and feedback in {\it Sim-SFF} (see Figure \ref{fig:sigma_500_sim-sff}).
While preserving all the common elements in comparison (such as pressure floor or resolution; described in Sections \ref{physics} and \ref{params}), the {\sc Gizmo} group has carried out extensive tests with other simulation parameters to check what most dictates the level of fragmentation (e.g., smoothing kernel, $N_{\rm ngb}$, {\sc Grackle} version, slope limiter, dual energy formalism; but always within conventional norms), finding that the different parameter choices do not qualitatively alter the mass function.
However, we note that increasing the Jeans pressure floor by as little as 100\% -- well within the uncertainty in the geometry prefactor of Jeans pressure support equation, Eq. (\ref{eq:floor}) -- and reverting to {\sc Grackle} v2.0 -- where metal cooling rates are slightly lower compared to v2.1 because of the different solar metallicity definition -- entirely removes the discrepancy (see the black dashed line in Figure \ref{fig:star_clump_stats_fof_500_sim-sff} labeled {\sc Gizmo-ps2}).  
A possible explanation for the stronger fragmentation in {\sc Gizmo} is that the ``effective'' gravitational resolution in {\sc Gizmo} is slightly higher compared to other codes \citep[as observed by][]{2016MNRAS.460.4382F}, resulting from a combination of different choices in their implementation (e.g., slope limiter, gradient estimator, density estimator).\footnote{Careful readers may notice that the stronger fragmentation in {\sc Gizmo} conflicts what \cite{2016ApJ...830L..13M} found.  We caution that their setup is different from ours.  For example, \cite{2016ApJ...830L..13M} employed a different pressure floor prescription that depends on the kernel size, and more effective stellar feedback based on \cite{Stinson06}.  They modeled a low-metallicity massive gas rich galaxy without considering metal cooling, which was shown by the {\sc Gizmo} group to significantly affect the fragmentation. As discussed, a slight variation in {\sc Gizmo} parameters is enough to erase their discrepancy with other codes.  In \cite{2016ApJ...830L..13M}, this role may be played by the different subgrid models and the absence of metal cooling.}

\subsection{Stellar Disk Kinematics}\label{results-star-kin}

Following the analysis shown in Section \ref{results-gas-kin} for the gas disk kinematics, here we study the kinematics of stellar disks formed by each code in {\it Sim-SFF}.
In Figure \ref{fig:star_pos_vel_500_sim-sff} we show the rotation velocity curves for newly-formed star particles in {\it Sim-SFF}. 
As in gas rotation velocity curves of Figures \ref{fig:pos_vel_500_sim-sff}, each line represents a mass-weighted average of stellar rotational velocities as a function of cylindrical radius.  
As in Section \ref{results-star-morph}, only the newly-formed star particles are considered for these profiles, not the disk or bulge stars in the IC.  
Just as in the gas rotation curves, the stellar rotation velocities show a high degree of similarity among the different codes, as good as within a few percent at certain radii (averaged fractional deviation for 2 $< r <$ 10 kpc is 2.5\% or 0.011 dex).  
Disagreements seen at radius $>$ 12 kpc for some codes such as {\sc Art-II} and {\sc Gasoline} are attributed to a small number statistics near the edge of a stellar disk.  

In Figure \ref{fig:star_pos_disp_500_sim-sff} we analyze the velocity dispersion curves for newly-formed star particles in {\it Sim-SFF}.  
As in gas velocity dispersion curves of Figures \ref{fig:pos_disp_500_sim-sff}, this plot quantifies the residual velocity components other than the rotational velocity computed in Figure \ref{fig:star_pos_vel_500_sim-sff}.  
Once again, a good agreement is found that all codes lie within a few tens of percent from one another at radii $<$ 10 kpc (averaged fractional deviation for 2 $< r <$ 10 kpc is 11.2\% or 0.046 dex).  
When compared with Figures \ref{fig:pos_disp_500_sim-nosf} and \ref{fig:pos_disp_500_sim-sff}, the agreement is particularly better in the central region ($<$ 1.5 kpc).  
Note that when we plot the ratio of {\it vertical} velocity dispersion  to {\it total} velocity dispersion in the bottom panel of Figure \ref{fig:star_pos_disp_500_sim-sff}, the systematic discrepancy found in Figure \ref{fig:pos_disp_500_sim-sff} between mesh-based and particle-based codes at radii $<$ 1.5 kpc no longer exists.  
This confirms our assertion in Section \ref{results-gas-kin} that the said discrepancy in Figure \ref{fig:pos_disp_500_sim-sff} is due to vertical gas movement captured only in mesh-based codes.  

\begin{figure*}
    \centering
    \includegraphics[width=0.56\textwidth]{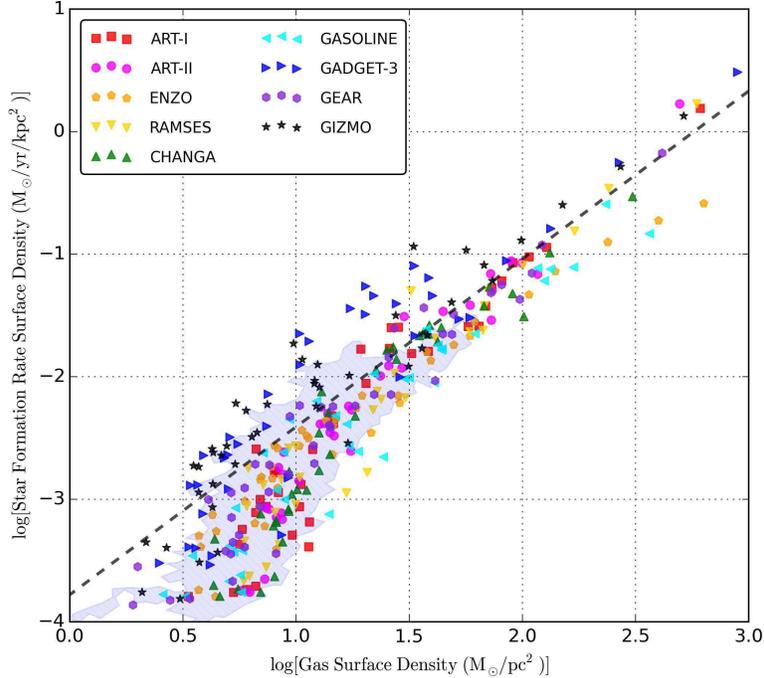}
    \caption{The Kennicutt-Schmidt relation for {\it Sim-SFF} at 500 Myr using the azimuthally-averaged gas surface densities (Figure \ref{fig:gas_surface_density_radial_500_sim-sff}) and SFR surface densities (Figure \ref{fig:sfr_surface_density_radial_500_sim-sff}).  The thick black dashed line denotes a best observational fit by \cite{2007ApJ...671..333K}.  The blue hatched contour marks the observed sub-kpc patches in nearby galaxies by \cite{Bigiel2008}, where their {\it hydrogen} surface density is multiplied by 1.36 to match the {\it total} gas surface density in our simulations.  See Section \ref{results-sf} for a detailed explanation on how this figure is made.  The full color version of this figure is available in the electronic edition.  
\label{fig:k-s_500_sim-sff}}
\end{figure*}

\subsection{Star Formation Relation}\label{results-sf}

In this section, we compare the star formation rates (SFR) of different codes in {\it Sim-SFF} and check whether we reproduce the observed relation between gas surface density and SFR surface density.
In the following discussion, SFRs are time-averaged over the past 20 Myrs and derived based on the ages of newly-formed star particles in the 500 Myr snapshots.

Figure \ref{fig:sfr_500_sim-sff} displays the evolution of the galaxy-wide SFR of each run by 500 Myr.  
For most codes, SFRs increase at early times, reach a maximum at 200-300 Myr after the start of the simulation, and then plateaus at later times. 
The SFRs evolve smoothly without evidence for strong bursts. 
The SFRs are within a factor of $\sim$3 from one another at all times (averaged fractional deviation for 50 $< t <$ 500 Myr is 32.8\% or 0.123 dex), and for most codes the values settle at 2-4 $\msun\,{\rm yr}^{-1}$ over most of the simulated time. 
Some differences are noteworthy.    
For example, {\sc Gasoline} and {\sc Enzo} predict somewhat lower SFRs, especially at intermediate times, while the SFR for {\sc Gizmo} never plateaus or begins to decline, but reaches a maximum of $\sim 6\,\,\msun\,{\rm yr}^{-1}$ at 500 Myr.
{\sc Gadget-3} produces the most stellar mass in this time period, but its SFR does not further grow after $\sim$ 300 Myr. 
The total stellar mass formed in 500 Myr ranges from $0.8 \times 10^9 \msun$ in {\sc Gasoline} to $2.4 \times 10^9 \msun$ in {\sc Gadget-3}. 
{\sc Gizmo}'s efficient and clumpy star formation is discussed in detail in Section \ref{results-star-morph} and Figure \ref{fig:star_clump_stats_fof_500_sim-sff}.   
We note again that increasing the Jeans pressure floor in {\sc Gizmo} by a factor of 2 -- well within the uncertainty in the geometry prefactor of Jeans pressure support equation, Eq. (\ref{eq:floor}) -- and reverting to {\sc Grackle} v2.0 entirely removes {\sc Gizmo}'s discrepancy (see the black dashed line in Figure \ref{fig:sfr_500_sim-sff} labeled {\sc Gizmo-ps2}; compare with Figure \ref{fig:star_clump_stats_fof_500_sim-sff}).  
In addition to the systematic tests the {\sc Gizmo} group performed, the {\sc Gadget-3} group has also tried a run with a slight variation in the treatment of supernova feedback that would increase the number of gas particles inside the ``shock radius'' (Section \ref{gadget}). 
We find that this slight variation indeed produces significantly lower SFR for {\sc Gadget-3}, closer to {\sc Enzo}'s value (results not shown here).  
Our tests strongly suggest that the SFR evolution is highly sensitive to the details of the numerical implementation of the common subgrid physics, including pressure floor and feedback prescriptions.

Now, to better understand the differences in SFR, we plot in Figure \ref{fig:sfr_surface_density_radial_500_sim-sff} the SFR surface densities as functions of cylindrical distance from the galactic center. 
Again, SFRs are estimated using the newly-formed star particles that are younger than 20 Myr old.
The agreement between the codes is generally encouraging, especially outside the central 0.5 kpc. 
We note, however, that the SFR within 0.5 kpc constitutes a large fraction of the total galactic SFR.
{\sc Enzo} and {\sc Gasoline} have significantly lower SFRs in the central region of the galaxy, thus explaining the difference seen in Figure \ref{fig:sfr_500_sim-sff}. 
In contrast, much of the excess star formation in {\sc Gadget-3} and {\sc Gizmo} takes place at large radii and is likely related to the formation of larger numbers of stellar clumps, as seen in Figures \ref{fig:star_with_clumps_fof_500_sim-sff} and \ref{fig:star_clump_stats_fof_500_sim-sff}.  
For example, a $\sim$ 6.5 kpc peak in {\sc Gizmo}'s SFR profile in Figure \ref{fig:sfr_surface_density_radial_500_sim-sff} coincides with a peak at the same radius in its stellar surface density profile in Figure \ref{fig:star_surface_density_radial_500_sim-sff}.

When measured on galactic scales ($\sim$ kpc), the gas surface density and SFR surface density are tightly correlated \citep[e.g.,][]{1959ApJ...129..243S, 1989ApJ...344..685K, 1998ApJ...498..541K, 2007ApJ...671..333K, Bigiel2008}. 
Consequently, the relation between these two observables, the Kennicutt-Schmidt (KS) relation, is frequently used to calibrate the modeling of star formation in galaxy-scale simulations. 
In Figure \ref{fig:k-s_500_sim-sff} we show the KS relation for {\it Sim-SFF}, where gas and SFR surface densities are measured within cylindrical annuli, as computed in Figures \ref{fig:gas_surface_density_radial_500_sim-sff} and \ref{fig:sfr_surface_density_radial_500_sim-sff}, respectively.  
Only annuli with nonzero averaged SFR surface densities are considered. 
The thick black dashed line denotes a best observational fit by Eq. (8) in \cite{2007ApJ...671..333K}, ${\rm log}\, \Sigma_{\rm SFR} = 1.37\,\, {\rm log}\, \Sigma_{\rm gas} - 3.78$, for a spatially-resolved patches in M51a.  
The blue hatched contour marks the observed sub-kpc patches in nearby galaxies taken from Figure 8 of \cite{Bigiel2008}, where their {\it hydrogen} surface density is multiplied by 1.36 to account for helium to match the {\it total} gas density in our simulations (see their Section 2.3.1).\footnote{We caution that galaxies in the \cite{Bigiel2008} sample are in nearby universe with relatively low gas fractions.  These are slightly different from the initial condition of our experiment that is modeled as a disk galaxy at $z\sim1$  with 20\% gas fraction (see Section \ref{IC}).}   

\begin{figure*}
    \centering
    \includegraphics[width=1.04\textwidth]{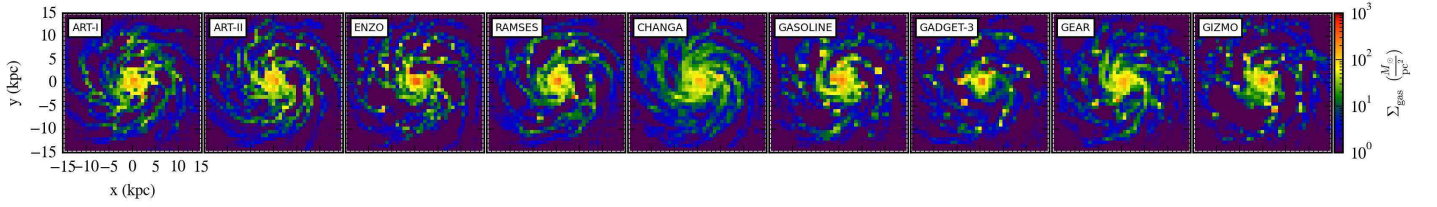}
    \caption{The 500 Myr snapshot of face-on gas surface densities for {\it Sim-SFF} at 750 pc resolution.  For particle-based codes, surface densities are estimated by depositing gas particles via the Cloud-In-Cell (CIC) scheme on to a 2-dimensional uniform grid with 750 pc resolution.  This image could be considered as a degraded version of Figure \ref{fig:sigma_500_sim-sff} although a different deposit algorithm is used for particle-based codes.  See Section \ref{results-sf} for a detailed explanation on how this figure is made. The full color version of this figure is available in the electronic edition.  
\label{fig:degraded_sigma_500_sim-sff}}
\end{figure*}

\begin{figure*}
    \centering
    \includegraphics[width=1.04\textwidth]{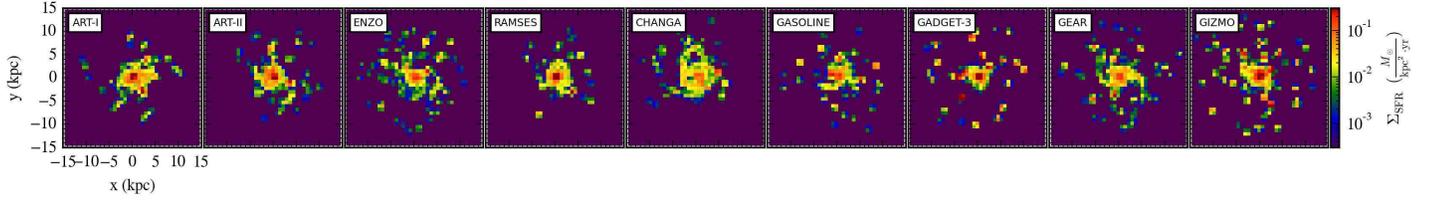}
    \caption{The 500 Myr snapshot of face-on star formation rate surface densities for {\it Sim-SFF} at 750 pc resolution.  SFR surface densities are estimated by depositing the newly-formed star particles that are younger than 20 Myr old on to a uniform grid with 750 pc resolution.  See Section \ref{results-sf} for a detailed explanation on how this figure is made.  The dynamic range of the color axis (3 orders of magnitude) are kept identical among Figures \ref{fig:degraded_sigma_500_sim-sff} and \ref{fig:degraded_sfr_map_500_sim-sff} to help see if the gas depletion times are similar among pixels.  
    \label{fig:degraded_sfr_map_500_sim-sff}}
\end{figure*}

\begin{figure*}
    \centering
    \includegraphics[width=0.56\textwidth]{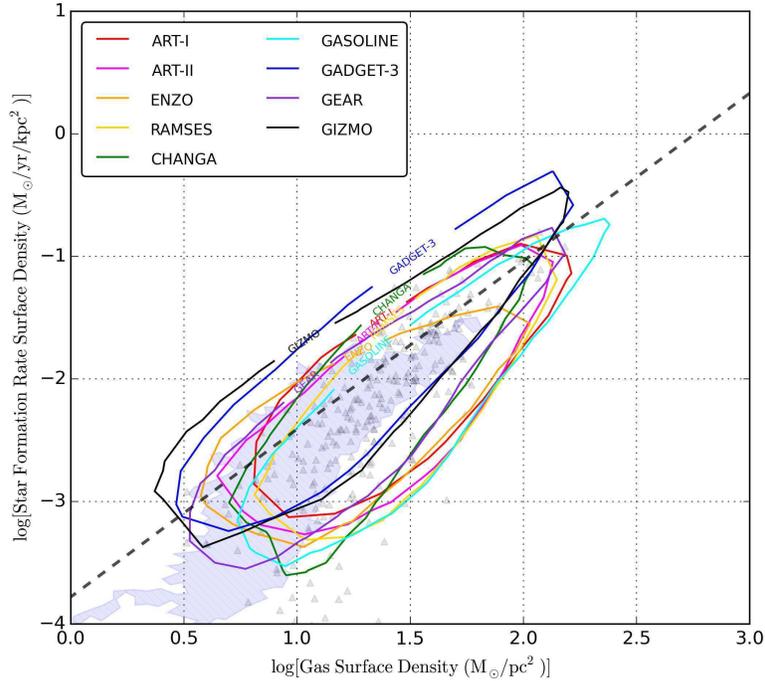}
    \caption{Same Kennicutt-Schmidt plane as Figure \ref{fig:k-s_500_sim-sff} but now with gas surface densities (Figure \ref{fig:degraded_sigma_500_sim-sff}) and SFR surface densities (Figure \ref{fig:degraded_sfr_map_500_sim-sff}) averaged in 750$\times$750 pc patches at 500 Myr.  As an example, shown as gray triangles are the patches with nonzero SFR surface density found in {\sc Changa}.  For all other codes, only 80\% percentile contours are drawn.   The thick black dashed line denotes a best observational fit by \cite{2007ApJ...671..333K}.  The blue hatched contour marks the observed sub-kpc patches in nearby galaxies by \cite{Bigiel2008}.  See Section \ref{results-sf} for a detailed explanation on how this figure is made.  The axes ranges are kept identical among Figures \ref{fig:k-s_500_sim-sff} and \ref{fig:k-s_local_500_sim-sff} for easier comparison.  The full color version of this figure is available in the electronic edition.  
\label{fig:k-s_local_500_sim-sff}}
\end{figure*}

As Figure \ref{fig:k-s_500_sim-sff} reveals, all participating codes predict a KS relation that agrees well with one another within a factor of a few, and with observed nearby disk galaxies in \cite{Bigiel2008}.
In particular, by combining star formation and thermal supernova feedback, most codes match both the normalization of the observed relation and the characteristic ``threshold'' value of the gas surface density ($\sim 10\, \msun\,{\rm pc}^{-2}$) below which star formation becomes less efficient.  
However, it should be noted that our simulations do not include multi-phase gas physics that explicitly models the transition between atomic and molecular hydrogen at $\sim 10\, \msun\,{\rm pc}^{-2}$ \citep[e.g.,][]{2009ApJ...699..850K}.  
More investigation may be needed to check how the apparent change in slope seen here is affected by our choice of subgrid physics (e.g., star formation efficiency $\epsilon_\star$, or thermal feedback energy budget; see Section \ref{physics-sim-sff}). 
Figure \ref{fig:k-s_500_sim-sff} also highlights that there are some differences between the codes. 
SFR surface densities of {\sc Gasoline} and {\sc Enzo} lie slightly below the other codes at a given gas surface density, while {\sc Gadget-3} and {\sc Gizmo} show higher SFRs than the rest of the codes. 
These differences are generally in line with what was observed in global SFR of Figure \ref{fig:sfr_500_sim-sff}. 

\begin{figure*}
    \centering
    \includegraphics[width=1.04\textwidth]{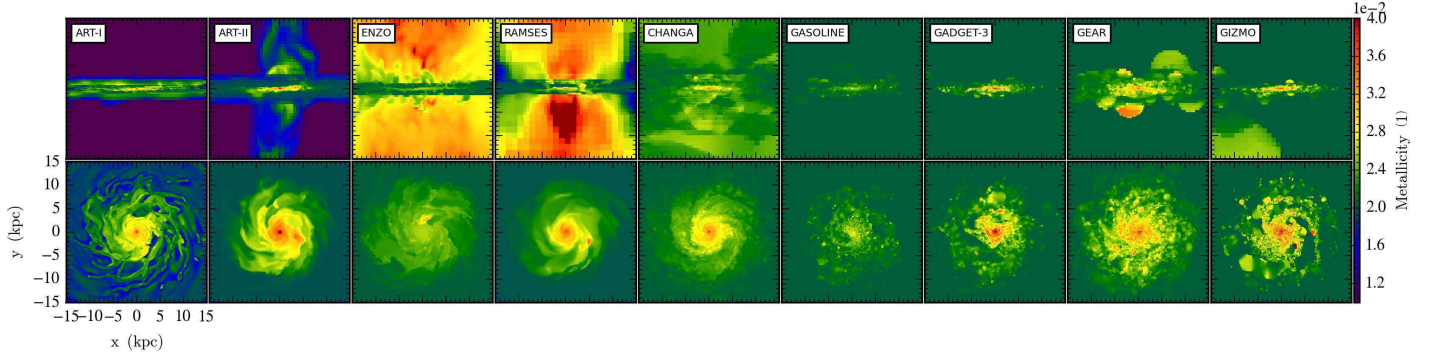}
    \caption{The 500 Myr composite of density-square-weighted gas metal fraction projections, edge-on {\it (top)} and face-on {\it (bottom)}, for {\it Sim-SFF} with star formation and feedback.  Colors represent the ratio of metal density to total gas density (not in units of $Z_\odot$).  The color axis spans from 0.01 to 0.04 in a linear scale.  See Section \ref{results-etc} for a detailed explanation of this figure.  Compare with Figures \ref{fig:sigma_500_sim-sff},  \ref{fig:temp_500_sim-sff}, and \ref{fig:star_with_clumps_fof_500_sim-sff}.  The full color version of this figure is available in the electronic edition.  The high-resolution versions of this figure and article are available at the Project website, http://www.AGORAsimulations.org/.
\label{fig:metal_500_sim-sff}}
\end{figure*}

\begin{figure*}
    \centering
    \includegraphics[width=0.75\textwidth]{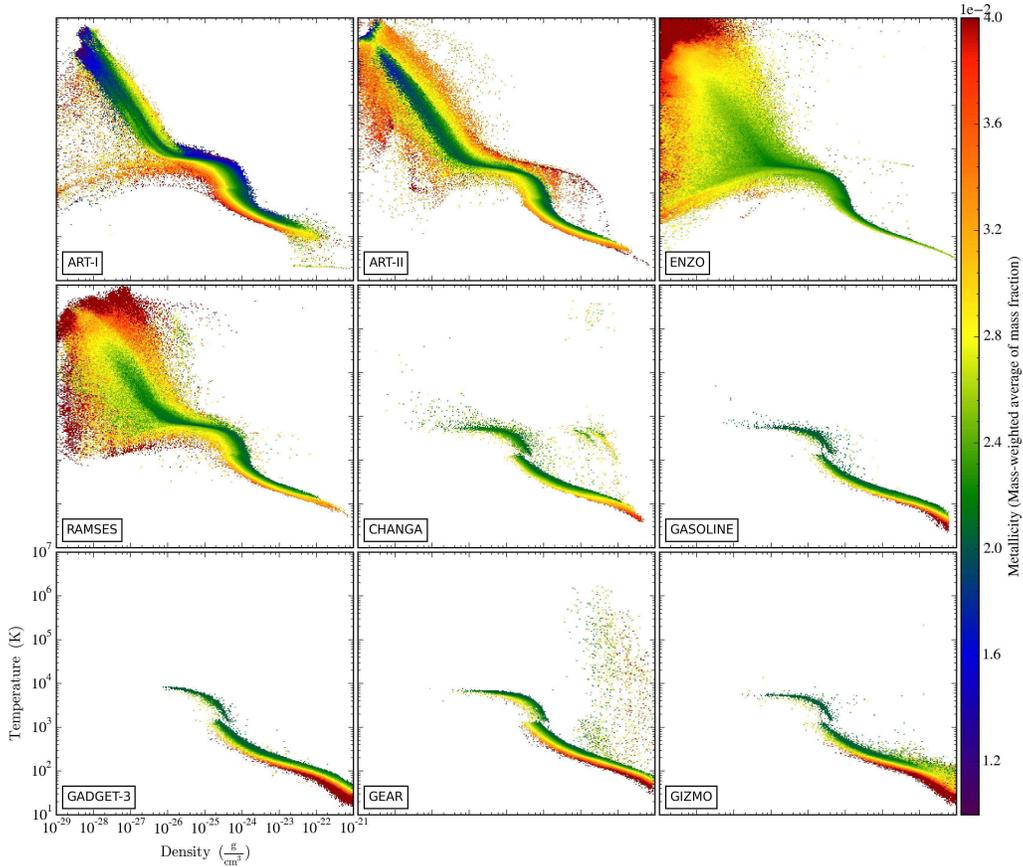}
    \caption{The 500 Myr composite of the mass-weighted averages of gas metal fraction on a density$-$temperature plane for the gas within 15 kpc from the galactic center in {\it Sim-SFF} with star formation and feedback.  The metal fraction is simply the ratio of metal density to total gas density (not in units of $Z_\odot$). Note that a gaseous halo  -- low density, high temperature gas in the upper left corner of each panel -- exists only in mesh-based codes, but not in particle-based codes (SPH codes or {\sc Gizmo}). Compare with Figure \ref{fig:pdf_500_sim-sff}.  See Section \ref{results-etc} for more information on this figure.  The full color version of this figure is available in the electronic edition.  The high-resolution versions of this figure and article are available at the Project website, http://www.AGORAsimulations.org/.
\label{fig:metal_pdf_500_sim-sff}}
\end{figure*}

In order to better match the observational technique by \cite{Bigiel2008}, one may consider using sub-kpc patches to generate the KS relation rather than cylindrical annuli.  
In Figures \ref{fig:degraded_sigma_500_sim-sff}  and \ref{fig:degraded_sfr_map_500_sim-sff} we present mock observations of gas surface densities and SFR surface densities for {\it Sim-SFF} at 500 Myr. 
For mesh-based codes in Figure \ref{fig:degraded_sigma_500_sim-sff}, their panels in Figure \ref{fig:sigma_500_sim-sff} are degraded to 750 pc resolution. 
For gas particles for particle-based codes in Figure \ref{fig:degraded_sigma_500_sim-sff} and young star particles of age less than 20 Myr in Figure \ref{fig:degraded_sfr_map_500_sim-sff}, we use the Cloud-In-Cell (CIC) scheme to deposit particle masses on to a uniform 2-dimensional grid with 750 pc resolution.\footnote{Thus, overall, Figure \ref{fig:degraded_sigma_500_sim-sff} could be considered as a ``degraded'' version of Figure \ref{fig:sigma_500_sim-sff} although, in Figure \ref{fig:sigma_500_sim-sff}, smoothing kernels -- not CIC -- are used to deposit gas particles in particle-based codes on to an octree.}
This resolution matches \cite{Bigiel2008}'s reported working resolution.
The dynamic range of the color axis are set to 3 orders of magnitude in both Figures \ref{fig:degraded_sigma_500_sim-sff} and \ref{fig:degraded_sfr_map_500_sim-sff}, in order to help readers to see if the gas depletion times are similar among pixels.  

\begin{figure*}
    \centering
    \includegraphics[width=1.04\textwidth]{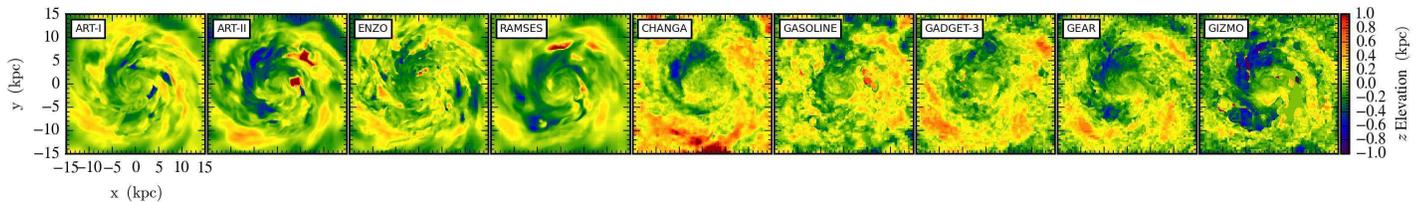}
    \caption{The 500 Myr face-on composite of density-weighted averages of gas elevations for {\it Sim-SFF} with star formation and feedback.  See Section \ref{results-etc} for a detailed explanation on how this figure is made.  Compare with Figures \ref{fig:sigma_500_sim-sff} and \ref{fig:temp_500_sim-sff}.  The full color version of this figure is available in the electronic edition.   The high-resolution versions of this figure and article are available at the Project website, http://www.AGORAsimulations.org/.
\label{fig:elevation_500_sim-sff}}
\end{figure*}

\begin{figure*}
    \centering
    \includegraphics[width=1.04\textwidth]{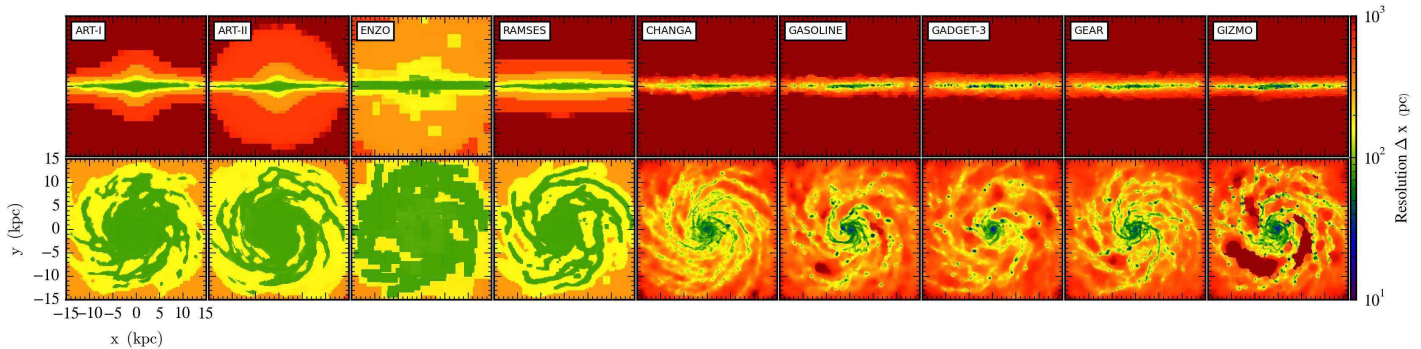}
    \caption{The 500 Myr composite of the size of resolution elements along different lines of sight, edge-on {\it (top)} and face-on {\it (bottom)}, for {\it Sim-SFF} with star formation and feedback.  The color axis spans from 10 to $10^3$ pc in a logarithmic scale, with highest resolution shown in dark blue.  See Section \ref{results-sf} for a detailed explanation on how this figure is made.  Compare with Figures \ref{fig:sigma_500_sim-sff}, \ref{fig:temp_500_sim-sff}, and \ref{fig:star_with_clumps_fof_500_sim-sff}.  The full color version of this figure is available in the electronic edition.  
\label{fig:resolution_500_sim-sff}}
\end{figure*}

Then we identify all 750$\times$750 pc patches with nonzero SFR surface density, and plot them in Figure \ref{fig:k-s_local_500_sim-sff} on the same KS plane as Figure \ref{fig:k-s_500_sim-sff}. 
As an example, all such patches found in the {\sc Changa} run are shown as gray triangles.
For all other codes, only 80\% percentile contours are drawn.  
Again, all participating codes reproduce the slope and normalization of the observed KS relation well.  
But slight differences in SFR surface densities exist.  
Contours for {\sc Gasoline} and {\sc Enzo} lie below the pack, while {\sc Gadget-3} and {\sc Gizmo}'s contours sit above all other codes at all gas surface densities.  
These findings are consistent with what we see in Figures \ref{fig:sfr_500_sim-sff} and \ref{fig:k-s_500_sim-sff}. 

\subsection{Other Comparisons: Metal Fraction, Disk Elevation, and Spatial Resolution}\label{results-etc}

In Figure \ref{fig:metal_500_sim-sff} we present the projections of density-square-weighted gas metal fraction for {\it Sim-SFF} with star formation and feedback.  
The metal fraction we show here is simply the ratio of metal density to total gas density, and it is not in units of $Z_\odot$, in order to minimize any confusion caused by {\sc Grackle} 2.0 versus 2.1 implementations (see footnotes \ref{grackle-zsun} and \ref{grackle-zsun-2}).  
The color axis spans from 0.01 to 0.04 in a linear scale.  
The edge-on views of mesh-based codes, particularly {\sc Enzo} and {\sc Ramses}, show high metallicity filaments flowing out of the disk, carrying metals into the embedding halo (see also Figure \ref{fig:temp_500_sim-sff}). 
However, as noted in Section \ref{IC}, a gaseous halo exists only in mesh-based codes, but not in particle-based codes (SPH codes and {\sc Gizmo}).  
Therefore, it is {\it not} the intended scope of this paper to compare the metal content of the halo which, by design, is captured only in mesh-based codes.   
In the face-on views within the disk we find qualitatively similar results across codes, with denser gas (corresponding to star forming regions) tending to be more metal-enriched.
Significant differences in the morphology of metal distribution exist, however.  
Differences in numerical implementations of the stellar feedback model are responsible for such discrepancies (see Section \ref{results-gas-thermal} for more discussion).\footnotemark[\getrefnumber{hemispherical-artifact}] 

Figure \ref{fig:metal_pdf_500_sim-sff} shows the mass-weighted averages of gas metal fraction in {\it Sim-SFF} on a 2-dimensional density$-$temperature plane for gas within 15 kpc of the galactic center.  
Raw particle fields are used for particle-based codes, not the interpolated or smoothed fields constructed in {\tt yt}.
Note again that a gaseous halo represented by low density, high temperature gas in the upper left corner of each panel exists only in mesh-based codes.  
For particle-based codes, high density, low temperature gas has higher metallicity because of its correspondingly higher star formation rate and thus metal enrichment.  
For mesh-based codes, on the other hand, high metal fraction is found in low density, high temperature gas as well, which is contaminated by hot metal-enriched materials dispersed by supernova feedback.\footnote{Readers may notice that {\sc Art-I} does not show strong outflows.  Rather, the continued inflow of gas from the halo provides zero metallicity gas mixed in to the disk (see the initial halo setup in Section \ref{IC}).  {\sc Art-I}'s discrepancies in Figures  \ref{fig:metal_500_sim-sff}-\ref{fig:metal_pdf_500_sim-sff} are manifestations of weaker stellar feedback than other mesh-based codes, which are exaggeratedly visible only because mesh-based codes include a large reservoir of zero metallicity gas around the disk.}  
These two observations are related to how well metals get mixed in each code.  
Readers may have noticed the difference between mesh-based and particle-based codes already in Figure \ref{fig:metal_500_sim-sff} by focusing on mixing of the metals. 
With neither the halo gas nor a sophisticated metal mixing scheme in place, metal-enriched gas in particle-based codes tends to stay near the dense star-forming sites that provided the metals.     

In Figure \ref{fig:elevation_500_sim-sff} is the density-weighted average of gas elevations from the $x-y$ disk plane for {\it Sim-SFF} with star formation and feedback.  
That is, averages of $(z_{\rm i} - z_{\rm center})$ such that a positive (negative) value indicates the gas along that line of sight is located above (below) the disk plane on average.  
For particle-based codes, we use the reconstructed density field from {\tt yt}'s in-memory octree on which gas particles are deposited using smoothing kernels (see Section \ref{results-gas-morph}).  
This figure helps to visualize and estimate the warping of the gas disk \citep[e.g.,][]{2006ApJ...643..881L}.
All participating codes produce largely flat disks, with vertical offsets less than $\pm 1$ kpc. 
Yet, it is also true that all codes show some levels of coherent warping or flaring along the disk plane.  
This strongly suggests that all these codes are able to resolve vertical instabilities.  

\begin{table*}
\renewcommand{\arraystretch}{1.35}
\caption{Relative Differences Between Codes in Main Observables 
(for {\it Sim-SFF} run)}
\centering
\begin{tabular}{c || c | c | c }
\hline\hline 
 & Level of agreement between codes & Figures & Relevant sections \\ 
\hline
Gas surface density (cylindrically-binned) & averaged fractional deviation for 2 $< r <$ 10 kpc $=$ 32.2\% (0.121 dex)\footnotemark[\getrefnumber{fractional_deviation_radius}] & Figure \ref{fig:gas_surface_density_radial_500_sim-sff} & Section \ref{results-gas-morph} \\
\hline 
Gas surface density (vertically-binned) & averaged fractional deviation for $z <$ 0.6 kpc $=$ 30.4\% (0.115 dex) & Figure \ref{fig:gas_surface_density_vertical_500_sim-sff} & Section \ref{results-gas-morph} \\
\hline 
Gas average vertical height  & averaged fractional deviation for 2 $< r <$ 10 kpc $=$ 19.1\% (0.076 dex) & Figure \ref{fig:rad_height_500_sim-sff} & Section \ref{results-gas-morph} \\
\hline 
Gas rotation velocity & averaged fractional deviation for 2 $< r <$ 10 kpc $=$ 2.8\% (0.012 dex) & Figure \ref{fig:pos_vel_500_sim-sff} & Section \ref{results-gas-kin} \\
\hline 
Gas velocity dispersion & averaged fractional deviation for 2 $< r <$ 10 kpc $=$ 17.8\% (0.071 dex) & Figure \ref{fig:pos_disp_500_sim-sff} & Section \ref{results-gas-kin} \\
\hline 
\multirow{2}{*}{Gas density probability distribution} & averaged fractional deviation in $10^{-25} < \rho < 10^{-22}\,{\rm g\,cm^{-3}}$  $=$ 28.6\% (0.109 dex), & \multirow{2}{*}{Figure \ref{fig:density_df_500_sim-sff}}  & \multirow{2}{*}{Section \ref{results-gas-thermal}} \\
& up to more than an order of magnitude difference at $\rho > 10^{-22}\,{\rm g\,cm^{-3}}$  & & \\ 
\hline 
Newly-formed stellar surface density & averaged fractional deviation for 2 $< r <$ 10 kpc $=$ 53.9\% (0.187 dex) & Figure \ref{fig:star_surface_density_radial_500_sim-sff} & Section \ref{results-star-morph} \\
\hline 
Newly-formed stellar clump mass function & all data points lie within a factor of $\sim$3 from the mean at each mass & Figure \ref{fig:star_clump_stats_fof_500_sim-sff} & Section \ref{results-star-morph} \\
\hline 
Newly-formed stellar rotation velocity & averaged fractional deviation for 2 $< r <$ 10 kpc $=$ 2.5\% (0.011 dex) & Figure \ref{fig:star_pos_vel_500_sim-sff} & Section \ref{results-star-kin} \\
\hline 
Newly-formed stellar velocity dispersion & averaged fractional deviation for 2 $< r <$ 10 kpc $=$ 11.2\% (0.046 dex) & Figure \ref{fig:star_pos_disp_500_sim-sff} & Section \ref{results-star-kin} \\
\hline 
Galaxy-wide star formation rate & averaged fractional deviation for 50 $< t <$ 500 Myr $=$ 32.8\% (0.123 dex) & Figure \ref{fig:sfr_500_sim-sff} & Section \ref{results-sf} \\
\hline 
KS relation (azimuthally-averaged) & all data points lie within a factor of $\sim$3 from the mean at each $\Sigma_{\rm gas}$ & Figure \ref{fig:k-s_500_sim-sff} & Section \ref{results-sf} \\
\hline 
KS relation (patch-averaged) & all data points lie within a factor of $\sim$3 from the mean at each $\Sigma_{\rm gas}$ & Figure \ref{fig:k-s_local_500_sim-sff} & Section \ref{results-sf} \\
\hline 
\end{tabular}
\label{table:summary}
\vspace*{5 mm}
\end{table*}

Finally, Figure \ref{fig:resolution_500_sim-sff} compares the sizes of spatial resolution elements each code imposes on the galactic disk of {\it Sim-SFF} at 500 Myr.
For mesh-based codes, this is a projection of gas cell sizes along different lines of sight, weighted by (cell volume)$^{-2}$ so that the maximal resolution element along that line of sight could stand out.    
For particle-based codes, this is a projection of gas particle sizes, defined as $(m_{\rm gas}/\rho_{\rm gas})^{1/3}$, smoothed on to {\tt yt}'s octree (the same octree {\tt yt} used in other edge-on/face-on visualizations; see Section \ref{results-gas-morph} for more information on the octree), weighted by (particle volume)$^{-2}$, defined as $(m_{\rm gas}/\rho_{\rm gas})^{-2}$.  
The color axis spans from 10 to $10^3$ pc in a logarithmic scale, with highest resolution shown in dark blue.  
The green color in mesh-based codes marks the finest mesh size permitted, 80 kpc, while the blue color in the spirals and clumps of particle-based codes can be associated with the minimum smoothing length permitted, $0.2 \times 80$ pc.  
Because of its Lagrangian nature, particle-based codes best demonstrate their strengths in dense clumps and spirals.
Meanwhile, with its flexible refinement strategy, mesh-based codes may best utilize their strengths in the contact regions with high density contrast, such as above and below the disk.  
For example, in the edge-on views of Figure \ref{fig:resolution_500_sim-sff}, the green-colored high resolution region covering the disk is thicker in mesh-based codes than in particle-based codes.\footnote{From the face-on views of Figure \ref{fig:resolution_500_sim-sff}, readers may distinguish the differences in grid construction machineries of mesh-based codes: octree structures in {\sc Art-I}, {\sc Art-II}, and {\sc Ramses}, versus block structures in {\sc Enzo}.}

\vspace{1mm}

\section{Discussion and Conclusion}\label{conclusion}

Using an isolated Milky Way-mass galaxy simulation, we compared results from 9 state-of-the-art gravito-hydrodynamics codes widely used in the numerical community (Section \ref{codes}).  
We utilized the infrastructure we have built for the {\it AGORA} High-resolution Galaxy Simulations Comparison Project.  
For the common initial conditions for these isolated galaxy simulations we used the ones generated by {\sc Makedisk} (Section \ref{IC}).\footnotemark[\getrefnumber{makedisk-website}]
We also adopted the common physics models (e.g., radiative cooling and UV background by the standardized package {\sc Grackle}; Section \ref{physics-sim-nosf})\footnotemark[\getrefnumber{grackle-website}] and common analysis toolkit {\tt yt},\footnotemark[\getrefnumber{yt-website}] both of which are publicly available.  
Subgrid physics such as pressure floor, star formation prescription, supernova feedback energy, and metal production have been meticulously constrained across participating codes (Sections \ref{physics-sim-nosf} and \ref{physics-sim-sff}).  
Strenuous efforts have also been made to ensure the consistency between the parameters that control resolutions of the codes (Section \ref{params}).  

With numerical accuracy that resolves the disk scale height -- high-order numerical methods in modern simulation codes combined with high spatial resolution -- we find that the codes overall agree well with one another in many dimensions, including: gas and stellar surface densities, gas and stellar rotation curves and velocity dispersions, gas density and temperature distribution functions, disk vertical heights, newly-formed stellar clumps, SFRs, and Kennicutt-Schmidt relations (Section \ref{results}).  
Quantities such as velocity dispersions are very robust (e.g., gas and newly-formed stellar velocity dispersions agree within a few tens of percent at all radii) while other measures like newly-formed stellar clump mass functions show more significant variation (differ by up to a factor of $\sim$3). 
In Table \ref{table:summary} we summarize the relative differences between codes for the main observables studied in this report.  
Some discrepancies can be understood as systematic differences between codes, for example, between mesh-based and particle-based codes in the low density region (Figures \ref{fig:sigma_500_sim-nosf}-\ref{fig:sigma_500_sim-sff} and Section \ref{results-gas-morph}), and between more diffusive and less diffusive schemes in the high density tail of the density distribution (Figure \ref{fig:density_df_500_sim-sff} and Section \ref{results-gas-thermal}).  
The latter translates into differences in clumping properties (Figure \ref{fig:star_clump_stats_fof_500_sim-sff}) and star formation rates (Figure \ref{fig:sfr_500_sim-sff}) of different codes.  
These intrinsic code differences are not as serious as some might have mistakenly extrapolated from previous code comparisons \citep[e.g.,][]{Aquila}, and are generally small compared to the variations in numerical implementations of the common subgrid physics such as supernovae feedback.
Our experiment reveals the remarkable level of agreement between different modern simulation tools despite their codebases having evolved largely independently for many years.  
It is also reassuring that our computational tools are more sensitive to input physics than to intrinsic  differences in numerical schemes, and that predictions made by the participating numerical codes are reproducible and likely reliable.  
If adequately designed in accordance with our proposed common parameters (e.g., cooling, metagalactic UV background, stellar physics, resolution; see Sections \ref{physics} and \ref{params}), results of a modern high-resolution galaxy  formation simulation are likely robust.  

It is worth briefly noting a few points about our study presented in this paper:  
(1) During the course of the present study, we have developed and field-tested important pieces of the {\it AGORA} infrastructure such as the common initial condition, common physics models, and common analysis toolkit.  
In particular, it should be noted that all the analyses in Section \ref{results} are carried out with common {\tt yt} scripts that are nearly independent of simulation codes.  
This common analysis platform approach has repeatedly proven its strength in {\it AGORA} comparisons including this study, significantly reducing the cost needed to hack the codes that any one researcher might not be familiar with, and allowing moving straight to {\it science-driven} comparisons of underlying physical properties. 
(2) While we find that the 500 Myr snapshot we used for the comparison is representative of each simulation, we caution that any similarity or discrepancy found here may not be universally the case at every single epoch.  
In an ongoing study using the same suite of simulations presented here, but multiple snapshots up to 2 Gyr, we are systematically checking if any conclusion drawn in this work is challenged by the fact that we compared a snapshot of a galaxy at a single epoch.\footnote{In fact, the evolution of the same {\it AGORA} isolated IC adopted in this work has been studied at different epochs already using many of the participating codes, albeit with different input physics and runtime parameters \citep[e.g.,][]{2013ApJ...770...25A, 2014MNRAS.442.3013K,  2015ApJ...814..131G, 2016ApJ...827...28G, 2016arXiv160907547A, 2016ApJ...826..200S}.} 
(3) Comparison studies in the {\it AGORA} Collaboration including this work are not intended to decide which numerical implementation is a ``correct'' one.   
The problem we are trying to numerically solve, i.e., galaxy formation, does not have a well-defined solution at a given resolution that every code is expected to converge to.  
Thus, it is never our intention to identify a ``correct'' or ``incorrect'' code, nor even a ``better'' or ``worse'' code. 
Instead, we aim to determine how much scatter one should expect among different numerical implementations in a particular problem of galaxy formation, given nominally similar physics and runtime parameters.   

We plan to further investigate our isolated disk galaxy simulations in other interesting dimensions, such as disk stability, bulge-to-disk decomposition, spiral and bar formation, and mass inflow and outflow, among others.  
As mentioned in Section \ref{physics-sim-sff}, we also intend to calibrate feedback schemes against observations (e.g., metrics such as galactic fountain, outflows, mass-loading, fraction of hot/warm/cold gas, and main sequence star formation) and against one another.  
While we complete analyses for these ongoing efforts, we are aiming to publicly release the 500 Myr snapshots used in the present article from all participating codes in January 2017 (tentatively) through the {\it AGORA} Project website.\footnotemark[\getrefnumber{agora-website}]
This is to allow any interested party in the numerical galaxy formation community to be able to compare their own simulations with the {\it AGORA} snapshots, using our publicly available common {\tt yt} script if needed.\footnotemark[\getrefnumber{agora-analysis-script-website}]

Finally, we emphasize the role the {\it AGORA} Project played in promoting {\it collaborative and reproducible research} in the numerical galaxy formation community.  
Over the past four years we have collaboratively formed a one-of-a-kind platform where members of the numerical community can work together and verify one another's work.  
Not only have we successfully built a common, publicly available infrastructure fully encompassing all the components to run galaxy-scale simulations in a reproducible manner -- initial conditions, physics packages, calibrated runtime parameters, analysis pipeline, and data storage -- but we also have founded an open forum where members could talk to and learn from one another.  
This Project has become a great experiment in itself in which it was continuously shown how beneficial a platform like this could be for any scientific community.    
Through workshops and teleconferences, and via common languages and infrastructure built together, Project participants were able to better understand other codes, and improve their own.  
Participants found an optimal set of simulation parameters that makes their code to be best compatible with others.  
We came to understand how seemingly identical parameters differ in their meanings in different codes, and how seemingly different parameters have in fact identical meanings.  
In some comparisons, numerical errors were discovered and fixed in participating codes.   
The {\it AGORA} framework, now tested with the common physics and subgrid models, are serving as a launchpad to initiate {\it astrophysically-motivated} comparisons aimed at raising the predictive power of galaxy simulations, especially as we run the zoom-in cosmological simulations outlined in our flagship paper \citep{2014agora}. 
In the coming years, we expect {\it AGORA} to continue to provide a sustainable and fertile platform on which numerical experiments are readily validated and cross-calibrated, and ambitious multi-platform collaborations are forged.  

\vspace{2mm}

The authors of this paper thank the members of the {\it AGORA} Collaboration who are not on the author list but have provided helpful suggestions throughout the progress of the paper, including John Forbes. 
We also thank Volker Springel for providing the original versions of {\sc Gadget-3} and {\sc Makedisk} to be used in the {\it AGORA} Project.  
We gratefully acknowledge the financial and logistical support from the University of California High-Performance AstroComputing Center (UC-HiPACC) during the annual {\it AGORA} Workshops held at the University of California Santa Cruz from 2012 to 2016.  
This research also used resources of the National Energy Research Scientific Computing Center (NERSC), a DOE Office of Science User Facility supported by the Office of Science of the U.S. Department of Energy under Contract No. DE-AC02-05CH11231.
The publicly available {\sc Enzo} and {\tt yt} codes used in this work are the products of collaborative efforts by many independent scientists from numerous institutions around the world.  
Their commitment to open science has helped make this work possible.  
Ji-hoon Kim acknowledges support from NASA through an Einstein Postdoctoral Fellowship, grant PF4-150147, and support from the Moore Center for Theoretical Cosmology and Physics at Caltech.  
A part of his computing time was provided by Extreme Science and Engineering Discovery Environment (XSEDE) allocation TG-AST140064.  
XSEDE is supported by National Science Foundation (NSF) grant No. ACI-1053575.
He is also grateful for the support from the computational team at SLAC National Accelerator Laboratory during the usage of the clusters for the simulation analysis.
Oscar Agertz acknowledges support from STFC consolidated grant ST/M000990/1 and the Swedish Research Council grant 2014-5791.
Daniel Ceverino acknowledges support from the European Research Council (ERC) via the ERC Advanced Grant STARLIGHT Project No. 339177.
Robert Feldmann acknowledges support in part by NASA through Hubble Fellowship grant HF2-51304.001-A awarded by the Space Telescope Science Institute, which is operated by the Association of Universities for Research in Astronomy, Inc., for NASA, under contract NAS 5-26555, in part by the Theoretical Astrophysics Center at UC Berkeley, and by NASA ATP grant 12-ATP-120183.
Alessandro Lupi acknowledges support by the ERC Project No. 267117 (PI J. Silk). 
Tom Quinn acknowledges partial support by the NSF through grants No. AST-1514868 and AST-1311956
and NASA Hubble grant HST-AR-13264.  
{\sc Changa} simulations where run on resources provided by XSEDE and NASA Pleiades.
Robert Thompson, Matthew Turk and Nathan Goldbaum acknowledge support by the Gordon and Betty Moore Foundation's Data-Driven Discovery Initiative through grant GBMF4561 (PI M. Turk), and by the NSF through grant No. ACI-1535651.
Tom Abel acknowledges partial support by the Kavli Foundation. 
Sukanya Chakrabarti acknowledges support by the NSF through grant No. AST-1517488.
Piero Madau acknowledges support by the NSF through grant No. AST-1229745 and by NASA through grant NNX12AF87G.
Kentaro Nagamine and Ikkoh Shimizu acknowledge support from the JSPS KAKENHI grant No. JP26247022.
Some of the {\sc Gadget-3} simulations were carried out on the XC30 machine at the Center for Computational Astrophysics, National Astronomical Observatory of Japan.
Brian O'Shea acknowledges support from NASA through grants NNX12AC98G, NNX15AP39G, and NASA Hubble theory grants HST-AR-13261.01-A and HST-AR-14315.001-A.  
He was also supported in part by the sabbatical visitor program at the Michigan Institute for Research in Astrophysics at the University of Michigan in Ann Arbor, and gratefully acknowledges their hospitality. 
Joel Primack acknowledges support from STScI through grant HST-GO-12060.12-A-004, and NASA Advanced Supercomputing for Pleiades time on which {\sc Art-I} simulations were run.  
Christine Simpson acknowledges support from the ERC under ERC-StG grant EXAGAL-308037 and from the Klaus Tschira Foundation.
John Wise acknowledges support by the NSF through grants No. AST-1333360 and AST-1614333 and NASA Hubble theory grants HST-AR-13895 and HST-AR-14326.

\begin{appendix}

\section{A. Pressure Floor Parameters in Selected Codes}\label{jeans-params}

We caution that the definition of $N_{\rm Jeans}$ in our Jeans pressure floor recommendation, Eq. (\ref{eq:floor}),
\begin{equation}
P_{\rm \,Jeans} = {1 \over \gamma \pi} N_{\rm Jeans}^2  G \rho_{\rm gas}^2  \Delta x^2 
\tag{\ref{eq:floor}}
\end{equation}
is different from another widely used formula in particle-based codes, Eq. (1) of \cite{2011MNRAS.417..950H},
\begin{equation}
P_{\rm \,Jeans} = {1.2  \over \gamma} N_{\rm Jeans}^{2/3}  G \rho_{\rm gas}^2  h_{\rm sml}^2.
\label{eq:floor-hopkins} 
\end{equation}
Our choice of $N_{\rm Jeans} = 4$ in Eq. (\ref{eq:floor}) (see Section \ref{physics-sim-nosf}) is equivalent to $N_{\rm Jeans} = 8.75$ in Eq. (\ref{eq:floor-hopkins}) if the smoothing length $h_{\rm sml}$ is replaced with a fixed number $\Delta x = 80$ pc.  
For {\sc Gear}, the runtime parameter {\tt JeansMassFactor} controls the Jeans pressure support and it equates to $N_{\rm Jeans}$ in Eq. (\ref{eq:floor-hopkins}), to be set to 8.75. 
For {\sc Changa} and {\sc Gasoline}, the relevant runtime parameter is {\tt dResolveJeans}, but it is not equal to $N_{\rm Jeans}$ in Eq. (\ref{eq:floor-hopkins}), but to the entire prefactor  $1.2 \,N_{\rm Jeans}^{2/3} \gamma^{-1} $.  
Therefore, to follow the {\it AGORA} recommendation in Section \ref{physics-sim-nosf}, {\tt dResolveJeans} should be set to $1.2 \times(8.75)^{2/3}\times (5/3)^{-1} = 3.06$, along with replacing $h_{\rm sml}$ with 80 pc.  
For some codes, one of the pressure floor formulae is simply hardcoded without a tunable runtime parameter: Eq. (\ref{eq:floor-hopkins}) in {\sc Gadget-3}, and Eq. (\ref{eq:floor}) in {\sc Gizmo}.  
The parameter {\tt MinimumPressureSupportParameter} in {\sc Enzo} refers to $N_{\rm Jeans}^2$ in Eq. (\ref{eq:floor}) above (see {\tt Grid$\_$SetMinimumSupport.C}).  
Thus, to follow the {\it AGORA} recommendation it should be set to 16.  
For pressure floor implementations using polytropes in {\sc Art-II} and {\sc Ramses}, see Sections \ref{art-ii} and \ref{ramses}, respectively.  

\section{B. Star Formation Efficiency Parameters in {\sc Changa} and {\sc Gasoline}}\label{sf-params}

Unlike in other codes, the parameter which controls the star formation efficiency in {\sc Changa} and {\sc Gasoline}, {\tt CStar} \citep[$c_\star$ defined in Eq. (3a) of][]{1992ApJ...391..502K}, is not equal to $\epsilon_\star$ in our Eq. (\ref{eq:KS}) of Section \ref{physics-sim-sff}.  
This is due to a difference in the denominator of the Schmidt formula:  $t_{\rm g} = (1/(4\pi G\rho_{\rm gas}))^{1/2}$ in Eq. (3a) of \cite{1992ApJ...391..502K}, versus $t_{\rm ff}=(3\pi/(32 G\rho_{\rm gas}))^{1/2}$ in our Eq. (\ref{eq:KS}).  
Therefore, in order to follow the {\it AGORA} recommendation in Section \ref{physics-sim-sff}, {\tt CStar} should be set to $\epsilon_\star (t_{\rm g} / t_{\rm ff}) = 0.01 \times (32/12\pi^2)^{1/2} = 0.0052$ in {\sc Changa} and {\sc Gasoline}.  

\section{C. Softening and Smoothing Parameters in Particle-Based Codes}\label{sph-params}

Throughout this paper we define the gravitational softening length $\epsilon_{\rm grav}$ (Section \ref{resolution}) as the equivalent Plummer softening length.  
This equates to the parameters {\tt Softening[Gas/Halo/Disk/Stars]} in {\sc Gadget-3}, {\sc Gear} and {\sc Gizmo}.  
It is however different from the typical definition of spline size $h$ beyond which the gravitational force becomes exactly Newtonian \citep[$h$ as defined in Eq. (4) of][]{gadget2}.  
For {\sc Gadget-3}, {\sc Gear} and {\sc Gizmo}, the spline size $h$ is equal to $2.8\, \epsilon_{\rm grav} = 2.8 \times 80 = 224$ pc (see {\tt ForceSoftening[i]} in {\tt gravtree.c}).  
Note that the parameter {\tt MinGasHsmlFractional} that controls the minimum hydrodynamical smoothing length (Section \ref{smoothing}) is defined as a fraction of $h$, not of $\epsilon_{\rm grav}$.  
Therefore, in order to set the minimum smoothing length to $0.2 \,\epsilon_{\rm grav}$, {\tt MinGasHsmlFractional} should be set to $0.2/2.8 = 0.0714$.  

By contrast, the spline size parameter {\tt dSoft} in {\sc Changa} and {\sc Gasoline} is not equal to the equivalent Plummer softening length, but to $h/2$ in the \cite{gadget2} definition of $h$ above.\footnote{It is worth to point out that {\tt dSoft} is indeed equal to $h$ defined in other papers, e.g., Eq. (14) of \cite{1996ApJS..105...19K} or Eq. (21) of the original \cite{1985A&A...149..135M}.}  
Therefore, in order to have $\epsilon_{\rm grav} = 80$ pc in  {\sc Changa} or {\sc Gasoline} runs, {\tt dSoft} should be set to $2.8\,\epsilon_{\rm grav}/2 = 224/2 = 112$ pc.  
One should keep in mind that the parameter which controls the minimum hydrodynamical smoothing length, {\tt dhMinOverSoft}, is defined as a fraction of {\tt dSoft}, not of $\epsilon_{\rm grav}$.  
Therefore, in order to set the minimum smoothing length to $0.2 \, \epsilon_{\rm grav}$, {\tt dhMinOverSoft} should be set to $0.2\times(2/2.8) = 0.143$.  

\end{appendix}

\end{document}